\documentclass[%
 aip,
 amsmath,amssymb,
 preprint,%
]{revtex4-1}

\usepackage[utf8]{inputenc}
\usepackage{graphicx}
\usepackage{dcolumn}
\usepackage{bm}
\usepackage[dvipsnames]{xcolor}
\usepackage{subcaption}
\usepackage{xcolor}
\usepackage[utf8]{inputenc}
\usepackage[T1]{fontenc}
\usepackage{mathptmx}

\begin{document}

\preprint{AIP/123-QED}

\title{Theory of Ion Aggregation and Gelation in Super-Concentrated Electrolytes}

\author{Michael McEldrew}
\affiliation{Department of Chemical Engineering, Massachusetts Institute of Technology, Cambridge, MA, USA}

\author{Zachary A. H. Goodwin}
\affiliation{Department of Physics, CDT Theory and Simulation of Materials, Imperial College of London, South Kensington Campus, London SW7 2AZ, UK}
\affiliation{Department of Chemistry, Imperial College of London, Molecular Sciences Research Hub, White City Campus, Wood Lane, London W12 0BZ, UK}
\affiliation{Thomas Young Centre for Theory and Simulation of Materials, Imperial College of London, South Kensington Campus, London SW7 2AZ, UK}

\author{Sheng Bi}
 \affiliation{ State Key Laboratory of Coal Combustion, School of Energy and Power Engineering, Huazhong University of Science and Technology (HUST), Wuhan, Hubei 430074, China}

\author{Martin Z. Bazant}
\affiliation{Department of Chemical Engineering, Massachusetts Institute of Technology, Cambridge, MA, USA}
\affiliation{Department of Mathematics, Massachusetts Institute of Technology, Cambridge, MA, USA}

\author{Alexei A. Kornyshev}
\affiliation{Department of Chemistry, Imperial College of London, Molecular Sciences Research Hub, White City Campus, Wood Lane, London W12 0BZ, UK}
\affiliation{Thomas Young Centre for Theory and Simulation of Materials, Imperial College of London, South Kensington Campus, London SW7 2AZ, UK}
\affiliation{Institute of Molecular Science and Engineering, Imperial College of London, South Kensington Campus, London SW7 2AZ, UK}

\date{\today}

\begin{abstract}
In concentrated electrolytes with asymmetric or irregular ions, such as ionic liquids and solvent-in-salt electrolytes, ion association is more complicated than simple ion-pairing. Large branched aggregates can form at significant concentrations at even moderate salt concentrations. When the extent of ion association reaches a certain threshold, a percolating ionic gel networks can form spontaneously. Gelation is a phenomenon that is well known in polymer physics, but it is practically unstudied in concentrated electrolytes. However, despite this fact, the ion-pairing description is often applied to these systems for the sake of simplicity. In this work, drawing strongly from established theories in polymer physics, we develop a simple thermodynamic model of reversible ionic aggregation and gelation in concentrated electrolytes accounting for the competition between ion solvation and ion association. Our model predicts the populations of ionic clusters of different sizes as a function of salt concentration, it captures the onset of ionic gelation and also the post-gel partitioning of ions into the gel. We discuss the applicability of our model, as well as the implications of its predictions on thermodynamic, transport, and rheological properties. 
\end{abstract}

\maketitle

\clearpage

\begin{table}[ht]
\centering
\caption{List of Variables}
\begin{tabular}{llll}
\hline
$N_{lmsq}$         & Number of $lmsq$ clusters    & $N^{gel}_i$     & Number of species $i$ in gel \\
$f_i$              & Functionality of species $i$ & $v_i$         & Volume of species $i$\\
$\xi_i$    & Scaled volume of species $i$ & $V$         & Total volume of mixture \\
$\Omega$     & Number of lattice sites      & $c_{lmsq}$       & Dimensionless concentration of cluster\\

$c^{gel}_{i}$    & Dimensionless concentration of & $c_{tot}$  & Total dimensionless concentration \\
&species $i$ in gel && of clusters\\
$\phi_i$         & Total volume fraction of species $i$ & $\phi_{\pm}$     & Volume fraction of salt\\
$\phi^{sol}_i$   & Volume fraction of species $i$ in sol & $\phi^{gel}_i$   & Volume fraction of species $i$ in gel\\
$\phi_{lmsq}$    & Volume fraction of an $lmsq$ cluster & $\psi_{i}$     & Concentration of association sites\\
& & &of species i\\
$\beta$ & Inverse thermal energy & $\Delta F$       & Free energy\\
$\Delta_{lmsq}$  & Free energy of formation of a rank & $\Delta_{lmsq}^{comb}$  & Combinatorial free energy of \\
&$lmsq$ cluster&& formation of a rank $lmsq$ cluster\\
$\Delta_{lmsq}^{bond}$  & Bonding free energy of & $\Delta_{lmqs}^{conf}$  & Configurational free energy of \\
&  formation of an $lmsq$ cluster &&formation of an $lmsq$ cluster\\
$\Delta_{lmsq}^{el}$  & Electrostatic free energy of &$\Delta^{gel}_{i}$ & Free energy change of species $i$\\
&formation of an lmsq cluster&&associating to the gel\\
$\mu_{lmsq}$    & Chemical potential of an $lmsq$ cluster & $\mu^{gel}_{i}$ & Chemical potential of species $i$\\
&&&in the gel \\
$K_{lmsq}$      & Equilibrium constant & $W_{lmsq}$      & Combinatorial enumeration \\
$\Delta u_{ij}$ & Association free energy  & $S_{lmsq}$      & Configurational entropy of a cluster\\
$S_{lmsq}$       & Configurational entropy of cluster & $Z$             & Coordination number of lattice\\
$\Lambda_{ij}$  & Association constant between $i$ and $j$ & $\tilde{\Lambda}$  & Association ratio\\
$p_{ij}$        & Association probabilities & $p^{sol}_{ij}$  & Association probabilities in the sol\\
$\zeta$         & Number of anion-cation associations & $\Gamma$        & Number of cation-solvent associations\\
$\Xi$        & Number of anion-solvent associations & $\alpha$        & Branching coefficient\\
$\bar{n}_w$     & Weight average of ionic aggregation & $\alpha_{lm}$   & Fraction of ions in $lm$ clusters \\
$\mathcal{K}$  & Cluster distribution constant     & $w^{gel}_i$     & Fraction of species in the gel\\
$w^{sol}_i$     & Fraction of species in the sol & $G_e$           & Equilibrium shear modulus\\
$R$             & Gas constant & $c$             & Molar concentration of salt\\
\hline
\end{tabular}
\label{tab:my_label}
\end{table}

\clearpage

\section{Introduction}

For most dilute electrolytes with high permittivity solvents, it is reasonable to assume that the salt is perfectly dissociated as confirmed by classical experiments~\cite{harned1959physical}. However, for moderately concentrated systems or dilute solutions with low permittivity solvents, incomplete dissociation of ions can be substantial~\cite{IBL}. Bjerrum popularized the concept of ion pairing, which was able to account for some deviations of experimental results from theoretical predictions \cite{bjerrum1926k}. In the Bjerrum theory of ion pairing, an ion pair is formed when the separation of oppositely charged ions is smaller than the length scale at which the Coulomb interaction is equivalent to thermal energy (known as the Bjerrum length). Many theoretical studies have focused on extending or modifying Bjerrum's treatment/definition of ions pairs, and we direct the readers to Ref.~\citenum{marcus2006ion} for an extensive review on the topic. Only a small fraction of studies considered ion aggregates larger than just simple ion pairs~\cite{kraus1933_1,fuoss1933_4,fuoss1933_9,barthel2000application}, but even those works only apply for moderate concentrations and model only simple ionic clusters.

In super-concentrated electrolytes, such as ionic liquids (ILs) or solvent-in-salt electrolytes (SiSEs) the picture is more complicated. With the recent explosion of interest in this regime for electrochemical applications~\cite{Suo2013,Sodeyama2014,Suo2015,Smith2015,Yamada2016,Wang2016,Gambou-Bosca2016,Sun2017,suo2017water,Dong2017,Diederichsen2017,yang2017a,Yang2017,wang2018hybrid,leonard2018,Yang2019,Dou2019}, a complete description of ion aggregation may be necessary for understanding the physicochemical, electrochemical, and thermodynamic properties of these concentrated mixtures. For ionic liquids, it has been useful to introduce the concept of free ions, without fully describing the nature of the associated species~\cite{Chen2017,goodwin2017underscreening}. These concepts have been applied to ILs to reproduce the temperature dependence of ionic conductivities~\cite{feng2019free} and differential capacitance~\cite{Chen2017}, although these simple pictures still cannot fully explain the so-called underscreening paradox in ILs ~\cite{Gebbie2013,Gebbie2015,Smith2016,goodwin2017mean,goodwin2017underscreening}. In SiSEs, as well as IL mixtures, there have been a multitude of molecular dynamics~\cite{kim2014ion,choi2014ion,choi2015ion,choi2017ion,borodin2017liquid,france2019,yu2020asymmetric} and experimental~\cite{borodin2017liquid,lim2018,lewis2020signatures} studies detailing complex ion association and hydration, often manifesting in highly asymmetric or even negative~\cite{molinari2019transport,molinari2019general} transference numbers. Although these molecular simulations and experimental studies provide valuable insight, it is often constrained to specific systems and is not readily transferable to new systems. 

For super-concentrated electrolytes it would therefore be beneficial to have a theoretical description of ion aggregates of arbitrary size, but to our knowledge, such a theory has not been reported in literature. Hence, in this article, we will formulate a thermodynamic model of ionic association beyond a simple description of ion pairing (or even triple and quadruple ions). Ultimately, we want our model to capture a distribution of aggregate sizes and even the formation of arbitrarily large ionic aggregates. 
In building such a model, we draw inspiration from polymer physics. In the early 1940's, Flory~\cite{flory1941molecular,flory1942constitution} and Stockmayer~\cite{stockmayer1943theory,stockmayer1944theory} derived expressions describing the most likely distribution of polymer molecular weights in a mixture. These expressions only require knowledge of the probability of the polymerization reaction, as well as the \emph{functionalities}, $f$, of the monomers. Functionalities refer to the number of bonds a monomer unit can make to extend the polymer. When $f=2$, then large linear chains can be formed, but when $f>2$, these aggregates will be branched and increasingly complex. Moreover, when $f>2$, Flory and Stockmayer were able to show that at a certain extent of reaction a percolating polymer network will be spontaneously formed in a process referred to as gelation. In the polymers community, this percolating network is referred to as a \emph{gel}, while the remaining finite species in mixture are referred to as the \emph{sol}. The gelation phenomenon outlined by Flory and Stockmayer turned out to be analogous to the percolation problem on a Bethe lattice~\cite{stauffer1994introduction}. 

The theories of Flory and Stockmayer were formulated to describe the largely irreversible covalent bond formation characteristic of condensation polymerization reactions, as opposed to the more reversible physical associations of ions. Starting in the late 1980's, Tanaka pioneered the theory of \emph{thermoreversible} polymer association and gelation \cite{tanaka1989,tanaka1990thermodynamic,tanaka1994,tanaka1995,ishida1997,tanaka1998,tanaka1999,tanaka2002}. In his work, Tanaka models the physical association between polymer strands within a thermodynamic framework that is able to capture the distribution of polymeric clusters, as well presence and breadth of gel networks. Of particular interest to us, is the two component case in which Tanaka describes a mixture of two types of polymer strands that associate heterogeneously in an alternating fashion~\cite{tanaka1998}. This is quite analogous to ion association in that ions will only associate to counterions. Thus, our theory of ion association and gelation in concentrated ionic systems will build upon that of Tanaka.

This paper is split into two main sections: Theory and Discussion. The Theory section is split into 5 subsections. First, we describe the stoichiometric definitions of our mixture, as well as its free energy of mixing. Then, we minimize that free energy yielding our pre-gel cluster distribution in terms of ``free" species volume fractions. In the third theory section we introduce ``association probabilities" that allow us to write the pre-gel cluster distributions in terms of experimentally accessible overall species volume fractions. In the fourth section, we describe the mechanism for gelation and derive the criterion for its onset. In the last theory section, we derive the post-gel relationships, yielding the post gel cluster distribution and the gel/sol partitioning. We end the paper, by discussing applicability of our model, and some of its implications on observable thermodynamic, transport, and rheological properties of the electrolyte solution, in particular those properties affected by the presence of ionic gel. At the start of this paper, we have a list of symbols in Tab.~\ref{tab:my_label}.

\section{Theory}

\begin{figure}[hbt!]
    \centering
    \includegraphics[width=0.35\textwidth]{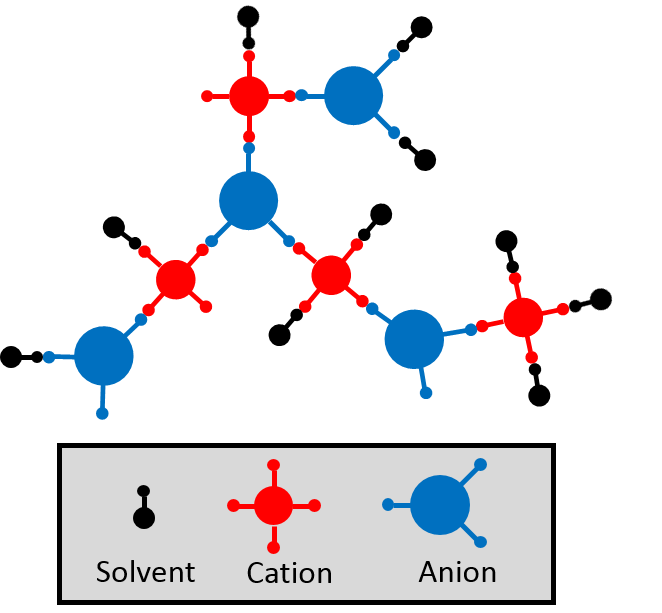}
    \caption{A cartoon example of cation/anion/solvent clusters that may be found with a certain probability in a model concentrated electrolyte. In this case, we have drawn a cluster in which $f_+=4$, $f_-=3$, $l=4$, $m=3$, $s=7$, and $q=3$.}
    \label{fig:ion_cluster}
\end{figure}

We consider a polydisperse mixture of $\sum_{lmsq}N_{lmsq}$ ionic clusters, each containing $l$ cations, $m$ anions, $s$ solvent molecules associated to cations, $q$ solvent molecules associated to anions ($lmsq$ cluster), and (if present) an interpenetrating gel network containing $N_+^{gel}$ cations, $N_-^{gel}$ anions, and $N_0^{gel}$ solvent molecules. We model the cations to have a functionality (defined as the number of associations that the species can make) of $f_+$, and anions to have a functionality of $f_-$. This means that a(n) cation (anion) is able to associate with $f_+$ ($f_-$) anions (cations) or solvent molecules. We also consider the ability of solvent molecules to coordinate to cations or anions with a functionality of 1. This, actually, means that we neglect the ability of solvent molecules to bridge ionic clusters through interactions with multiple ions, and thereby neglect the formation of any solvent-mediated clustering/gelation. This is obviously a simplification, justified by an assumption that the clusters that are not `glued' by direct ion--counter-ion interactions are more labile, and as such can be disregarded. A typical ion cluster consistent with our description is depicted in Fig.~\ref{fig:ion_cluster}. 

Following Tanaka, we account for molecular volumes by using a lattice model. We designate a single lattice site to have the volume of a single solvent molecule, $v_0$.  Thus the entire volume of the mixture, $V$, is divided into $\Omega = V/v_0$ lattice sites. Moreover, cations will occupy $\xi_+=v_+/v_0$ lattice sites, and anions will occupy $\xi_-=v_-/v_0$ lattice sites. Furthermore, when a gel is formed, then we distinguish between the volume fractions of gel (superscript $gel$) and sol (superscript $sol$). The volume fractions in the sol and gel constitutes the total volume fraction, $\phi_j$ of a given species, $j$, is given by
\begin{align}
    \phi_j=\phi_j^{sol}+\phi_j^{gel}    
\end{align}
in which the gel volume fractions is defined as $\phi_j^{gel}=\xi_j N_j^{gel}/\Omega$, with $N_j^{gel}$ as the mole number of species $j$ in the gel. The subscript $j=+,-,0$ corresponds to cation, anion, and solvent, respectively. The sol volume fraction of cations, anions, and solvent molecules have, respectively, the definitions
\begin{align}
    \phi^{sol}_+=\sum_{lmsq} \xi_+l c_{lmsq}
\end{align}
\begin{align}
    \phi^{sol}_-=\sum_{lmsq} \xi_-
    m c_{lmsq}
\end{align}
\begin{align}
    \phi^{sol}_0=\sum_{lmsq} (s+q) c_{lmsq}
\end{align}
where $c_{lmsq}=N_{lmsq}/\Omega$ is the dimensionless concentration of a $lmsq$ cluster (the number of $lmsq$ clusters per lattice site). Similarly, we define $\phi_\pm=\phi_++\phi_-$, which is the total volume fraction of the salt in solution. For simplicity the mixture is assumed to be incompressible, i.e.
\begin{align}
    1=\phi_\pm+\phi_0=\phi_++\phi_-+\phi_0
    \label{eq:incomp}
\end{align}
$\phi_+$ and $\phi_-$ are not independent owing to electroneutrality: $\phi_+/\xi_+=\phi_-/\xi_-$. The reduced volume of the mixture, $\Omega$, can also be expressed in terms of the mole number of each species/component due to the incompressibility constraint [Eq.~\eqref{eq:incomp}]
\begin{align}
    \Omega=\sum_{lmsq}(\xi_+l+\xi_-m+s+q)N_{lmsq}+\xi_+N^{gel}_++\xi_-N^{gel}_-+N^{gel}_0
    \label{eq:O}
\end{align}
This definition must be used when differentiating the free energy of mixture. Another important quantity that will be used abundantly later in the paper is the  dimensionless concentration of association sites (number of association sites per lattice site). We denote this quantity by $\psi_j$ and define it as the following
\begin{align}
    \psi_j=f_j \phi_j/\xi_j
\end{align}
Thus, $\psi_j$ is the number of $j$ association sites per lattice site. Note that for solvent molecules $\psi_0=\phi_0$.

\subsection{Free Energy}

We use a Flory-Huggins like free energy of mixing given in units of thermal energy, $\beta = 1/k_BT$,
\begin{align}
    \beta \Delta F &= \sum_{lmsq} \left[N_{lmsq}\ln \left( \phi_{lmsq} \right)+N_{lmsq}\Delta_{lmsq}^{\theta}\right]
    \nonumber \\
    &+ \sum_{lmsq} \left[N_{lmsq} \delta_{l,1}\delta_{m,0}(\ln \gamma^{DH}_++\Delta u_+^{born})+N_{lmsq}\delta_{m,1}\delta_{l,0}(\ln \gamma^{DH}_-+\Delta u_-^{born}) \right]
    \nonumber \\
    &+ \Delta^{gel}_+ N^{gel}_+ + \Delta^{gel}_- N^{gel}_- + \Delta^{gel}_0 N^{gel}_0 
    \label{eq:F}
\end{align}
where $\phi_{lmsq}=(\xi_+l+\xi_-m+s+q)N_{lmsq}/\Omega$ is the volume fraction of an $lmsq$ cluster, $\Delta^{\theta}_{lmsq}$ is the ideal free energy of formation of an $lmsq$ cluster from its unassociated constituents, $\gamma^{DH}_\pm$ is the Debye-Huckle ionic activity coefficient (defined later), $\Delta u^{Born}_\pm$ is the Born solvation free energy of an ion (defined later), $\delta_{i,j}$ is the Kroenecker delta, and $\Delta^{gel}_i$ is the free energy change of species, $i$, associating to the gel~\cite{flory1942thermodynamics,flory1953principles,tanaka1989}. We should note that Flory-Huggins type free energies typically contain regular solution interaction parameters between species in order to model phase separation, but we have omitted them here for the sake of simplicity. 


The free energy in Eq.~\eqref{eq:F} contains three essential pieces of physics: the entropy of mixing for a distribution of ion/solvent clusters and the gel, the association free energy corresponding to the formation of clusters or the gel, and finally the electrostatic non-idealities of free ions in solution. The entropy of mixing takes into account that species within specific clusters are not entropically independent, however the individual clusters are treated ideally. Additionally, $\phi_{lmsq}$ is constrained via the incompressibility condition [Eqs.~\eqref{eq:incomp} \& \eqref{eq:O}]. In the second line of Eq.~\eqref{eq:F}, we modify the chemical potential of unpaired or free ions by including terms to account for Debye-Huckel screening and Born solvation free energy  of free ions. 


Differentiating the free energy with respect to $N_{lmsq}$ yields the chemical potential of a cluster rank $lmsq$
\begin{align}
    \beta \mu_{lmsq}&=\ln \phi_{lmsq} + 1 - (l+m+s+q) c_{tot}+\Delta^{\theta}_{lmsq} \nonumber \\
    &+\delta_{l,1}\delta_{m,0}(\ln \gamma^{DH}_++\Delta u_+^{born})+\delta_{m,1}\delta_{l,0}(\ln \gamma^{DH}_-+\Delta u_-^{born})
    \label{eq:muclust}
\end{align}
where $c_{tot}=\sum_{lmsq}c_{lmsq}$ is the total reduced concentration. Note we have used Eq.~\eqref{eq:O} when differentiating the free energy. 
Additionally, we may define the chemical potential of species immersed in the gel
\begin{align}
    \beta \mu_+^{gel}=\Delta^{gel}_+  - c_{tot} 
\end{align}
\begin{align}
    \beta \mu_-^{gel}=\Delta^{gel}_--c_{tot}
\end{align}
\begin{align}
    \beta \mu_0^{gel}=\Delta^{gel}_0-c_{tot}. 
\end{align}


\subsection{Pre-gel Cluster Distribution}

The distribution of clusters can be derived by enforcing a chemical equilibrium between all of the clusters and their bare constituents (unassociated components)
\begin{align}
    l [\text{bare cation}]+m [\text{bare anion}]+(s+q) [\text{bare solvent}]\rightleftharpoons [lmsq\,\,\text{cluster}].
\end{align}
Chemical equilibrium requires that the chemical potentials of free species and those in clusters are equivalent
\begin{align}
    l \mu_{1000}+m \mu_{0100}+(s+q) \mu_{0010} = \mu_{lmsq} = l\mu^+_{lmsq} + m\mu^-_{lmsq} + (s+q)\mu^{0}_{lmsq}
    \label{eq:eqm}
\end{align}
Note that we may refer to free solvent molecules with either the index 0001 or 0010. For simplicity we will use the index 0010 to refer to free solvent molecules, for the remainder of the text. 
In  Eq.~\eqref{eq:eqm}, we have defined the chemical potential of a cation, anion or solvent molecule in an arbitrary cluster in the following manner
\begin{align}
    \mu^+_{lmsq}=\frac{\partial \mu_{lmsq}}{\partial l}=\mu_{1000}
\end{align}
\begin{align}
    \mu^-_{lmsq}=\frac{\partial \mu_{lmsq}}{\partial m}=\mu_{0100}
\end{align}
\begin{align}
    \mu^{0}_{lmsq}=\frac{\partial \mu_{lmsq}}{\partial s}=\frac{\partial \mu_{lmsq}}{\partial q}=\mu_{0010}
\end{align}

Solving Eq.~\eqref{eq:eqm} for an arbitrary $lmsq$ cluster obtains the following relation
\begin{align}
    \phi_{lmsq}=K_{lmsq}\phi_{1000}^l \phi_{0100}^m \phi_{0010}^{s+q}
    \label{eq:cluster_dist}
\end{align}
where $\phi_{1000}$, $\phi_{0100}$, and $\phi_{0010}$ are the bare species' volume fractions of cations, anions, and solvent molecules, respectively; and $K_{lmsq}$ is the equilibrium constant, given by
\begin{align}
    K_{lmsq}=\exp(l+m+s+q-1-\Delta^{\theta}_{lmsq}+\Delta_{lmsq}^{el})
\end{align}
where 
\begin{align}
    \Delta_{lmsq}^{el}=l(\delta_{l,1}\delta_{m,0}-1) (\ln \gamma_+^{DH}+\Delta u^{Born}_+) +m(\delta_{m,1}\delta_{l,0}-1)(\ln \gamma_-^{DH}+\Delta u^{Born}_-).
    \label{eq:delel1}
\end{align}
Thus, $\Delta_{lmsq}^{el}$ can be considered the electrostatic contribution to the free energy of formation of the cluster. It is convenient to employ the following definition: 
\begin{align}
    \Delta_{lmsq}=\Delta_{lmsq}^{\theta}+\Delta_{lmsq}^{el}
    \label{eq:delta2}
\end{align}
where $\Delta_{lmsq}$ is now the free energy of formation of an $lmsq$ cluster accounting for the electrostatic non-idealities of free ions, which we will discuss in more detail below. Thus, the partitioning of the species into clusters of different sizes is strongly governed by $\Delta_{lmsq}$. As such, this is where much of the physics of the ion/solvent association will be included.  $\Delta_{lmsq}$ contains four contributions
\begin{align}
    \Delta_{lmsq}=\Delta_{lmsq}^{comb}+\Delta_{lmsq}^{bond}+\Delta_{lmsq}^{conf}+\Delta_{lmsq}^{el}
    \label{eq:Delta}
\end{align}
where $\Delta_{lmsq}^{comb}$ is the \emph{combinatorial} (entropic) contribution, describing the multiplicity of clusters with the same number of constituents; $\Delta_{lmsq}^{bond}$ is the \emph{bonding} contribution, describing the association enthalpy of the constituents in the cluster; $\Delta_{lmsq}^{conf}$ is the \emph{configurational} contribution, describing the configurational entropy change upon forming a cluster from base constituents; and $\Delta_{lmsq}^{el}$ is the electrostatic contribution, accounting for the long range electrostatic interactions of free ions in the electrolyte. Note, the first three contributions are the same as included by Tanaka, however, the fourth contribution, $\Delta_{lmsq}^{el}$, is a necessary addition for modelling electrolytes due to the presence of free charges in solution.  

The entropy associated with the combinatorial enumeration, $W_{lmsq}$, of all of the possible ways a cluster with $l$ cations, $m$ anions, and $s+q$ solvent molecules can be formed is given by
\begin{align}
    \Delta_{lmsq}^{comb}=-\ln \left(W_{lmsq}\right)
    \label{eq:dcomb}
\end{align}
To derive $W_{lmsq}$ we use a two step procedure. First, we enumerate the number of ways to construct a network containing $l$ anions and $m$ cations, which are associated together in an alternating fashion, $W_{lm}$. This combinatorial problem is well known~\cite{stockmayer1952molecular}
\begin{align}
    W_{lm}=\frac{(f_+ l -l)!(f_- m -m)!}{l!m!(f_+l-l-m+1)!(f_-m-l-m+1)!}.
\end{align}
In the second step, we enumerate the number of ways $s+q$ solvent molecules can be placed on the cation-anion cluster. We know that we may only place the $s$ solvent molecules on the remaining $f_+l-l-m+1$ open cation sites. Thus $s$ must be less than or equal to $f_+l -l-m+1$. This enumeration is expressed via the binomial coefficient
\begin{align}
    \mathcal{C}^{f_+l-l-m+1}_s=\frac{(f_+l-l-m+1)!}{s!(f_+l-l-m-s+1)!}.
\end{align}
Similarly, we must place $q$ solvent molecules on the remaining $f_-m-m-l+1$ open anion sites, which can be enumerated via 
\begin{align}
    \mathcal{C}^{f_-m-m-l-q+1}_q=\frac{(f_-m-m-l+1)!}{q!(f_-m-m-l-q+1)!}.
\end{align}
Thus, we have 
\begin{align}
    W_{lmsq}&=W_{lm}\mathcal{C}^{f_+l-l-m+1}_s\mathcal{C}^{f_-m-m-l-q+1}_q \nonumber \\
    &=\frac{(f_+ l -l)!(f_- m-m)!}{l!m!s!q!(f_+l-l-m-s+1)!(f_-m-m-l-q+1)!}
\end{align}

Next, the bonding contribution, $\Delta^{bond}_{lmsq}$, can be described simply via the association free energies: $\Delta u_{ij}$ between species $i$ and $j$, where $i \neq j$ and $\Delta u_{ij} = \Delta u_{ji}$. Recall, that our model does not allow for solvent molecules to form clusters among themselves. For this reason, if a cluster contains 0 cations and anions, the cluster will necessarily only contain a single solvent molecule, corresponding to a free solvent molecule. Clearly, a free water molecule does not form associations and thus $\Delta^{bond}_{0010}=\Delta^{bond}_{0001}=0$. Overall, we can write $\Delta^{bond}_{lmsq}$ as
\begin{align}
    \Delta^{bond}_{lmsq}=\left[(l+m-1)\Delta u_{+-}+s\Delta u_{+0}+q\Delta u_{-0}\right][1-\delta_{l,0}\delta_{m,0}(\delta_{q,0}\delta_{s,1}+\delta_{q,1}\delta_{s,0})]
    \label{eq:delbond1}
\end{align}
where $\delta_{i,j}$ is Kroenecker delta function. For $l+m>0$, the association free energy for an $lmsq$ cluster is
\begin{align}
    \Delta^{bond}_{lmsq}=(l+m-1)\Delta u_{+-} + s\Delta u_{+0} + q\Delta u_{-0}
    \label{eq:dbond}
\end{align}
The coefficient in front of the cation-anion bond, $\Delta u_{+-}$, is due to the fact that there must be that many cation--anion associations to form a cluster with $l$ cations and $m$ anions. 

For the configurational contribution, $\Delta^{conf}_{lmsq}$, we use Flory's lattice theoretical expression for the entropy of disorientation \cite{flory1942thermodynamics,flory1953principles}. Tanaka adapted and modified Flory's expression for more complicated associating polymer mixtures in refs. \cite{tanaka1989,matsuyama1990theory,tanaka1999}, through a procedure outlined by Flory, involving the subsequent placement of lattice sized bits of molecules onto adjacent lattice sites. From this, we write the configurational entropy, $S_{lmsq}$, of an $lmsq$ cluster as
\begin{align}
    S_{lmsq}=-\ln \left( \frac{(\xi_+l+\xi_-m+s+q)Z(Z-1)^{\xi_+l+\xi_-m+s+q-2}}{\exp \left( \xi_+l+\xi_-m+s+q -1\right)} \right)
\end{align}
where $Z$ is the coordination number of the lattice. The configurational bit of $\Delta_{lmsq}$ is then
\begin{align}
    \Delta^{conf}_{lmsq}&=S_{lmsq}-lS_{1000}-mS_{0100}-(s+q)S_{0010} \nonumber \\
    &=-\ln\left(\frac{(\xi_+l+\xi_-m+s+q)\left[(Z-1)^2/Ze\right]^{l+m+s+q-1}}{\xi_+^l\xi_-^m}\right)
    \label{eq:dconf}
\end{align}

The last contribution to $\Delta_{lmsq}$ in Eq. \eqref{eq:Delta}, which Tanaka does not need to consider for his systems, is the electrostatic contribution, $\Delta^{el}_{lmsq}$. Note, we have already defined this quantity in Eq.~\eqref{eq:delel1} In determining it, we had to make the following simplifying assumption. Both in the limit high and low salt concentrations, the concentration of free ions will be small. Thus, we could describe the contribution to their free energy using simple Debye screening theory, as suggested by surface force data for ionic liquids \cite{gebbie2013ionic}. We neglected the contribution of charged clusters containing multiple ions, because their contribution to the ionic strength of the solution is expected to be small. However, we will take into account the effects of ionic clusters on the effective dielectric constant of the medium in which the free ions are dissolved. Such an approximation is expected to work effectively as an interpolation between the two limiting cases of low and high salt concentration. 

Hence, the electrostatic screening will be characterized by the Debye screening length, $\lambda_D$, 
\begin{align}
    \lambda_D^2= \frac{\varepsilon \varepsilon_0 k_BT}{e^2 I},
\end{align}
where $\varepsilon$ is the relative dielectric constant of the medium (affected by the degree of clustering), $\varepsilon_0$ is the vacuum permittivity, $e$ is the elementary charge, and $I$ is the ionic strength of the solution. In general, the ionic strength must take into account contributions from all the charged clusters:
\begin{align}
    I = \frac{1}{2}\sum\limits_{lmsq}(l-m)^2c_{lmsq}/v_0
    \label{eq:LI1}
\end{align}
where $c_{lmsq}$ is the number of clusters rank $lmsq$ per lattice site (dimensionless concentration). However, as previously mentioned, we will make the assumption that the free ions dominate the ionic strength, yielding the simplification
\begin{align}
    I = \frac{1}{2}\sum\limits_{i=+,-}\alpha_i\phi_i/\xi_iv_0
    \label{eq:LI2}
\end{align}
where $\alpha_+$ and $\alpha_-$ are the fraction of free cations and anions, respectively. In general, $\alpha_+$, $\alpha_-$, and $\varepsilon$ will depend on the composition of the electrolyte, and must be determined self-consistently as we will describe later. The DH formula for ionic activity [appearing in Eq.~\eqref{eq:delel1}] is given by 
\begin{align}
    \ln \gamma_{\pm}^{DH}=-\frac{e^2}{8 \pi \varepsilon \varepsilon_0 k_B T\lambda_D}\left(\frac{1}{1+a_\pm/\lambda_D}\right)
\end{align}
where $a_\pm$ is the radius of the free anion or cation~\cite{debye1923theory}. 

Additionally, the salt concentration is expected to change the dielectric permittivity of the fluid, which has a strong effect on ionic activity\cite{vincze2010nonmonotonic}, as first noted by Huckel\cite{huckel1925theorie}. The free energy of free ions is expected to change according to change in Born solvation energy, $\Delta u^{Born}_\pm$ [also appearing in Eq.~\eqref{eq:delel1}], which is written as 
\begin{align}
    \Delta u^{Born}_\pm= \frac{e^2}{8 \pi \varepsilon_0 k_B T a_\pm}\left(\frac{1}{\varepsilon}-\frac{1}{\varepsilon_s}\right)
    \label{eq:delel}
\end{align}
where $\varepsilon_s$ is the dielectric constant of the pure solvent~\cite{born1920volumen}. Note that the solvation energy here is defined with a positive sign. Thus if $\varepsilon$ decreases, the chemical potential of free ions will increase, weakening the propensity for ions to be free. For simplicity, hereafter, we will assume the free ion radius, $a_\pm$ to be equal for anions and cations $a_+=a_-=(v_0(\xi_++\xi_-)/2)^{1/3}$. In this way, the Debye-Huckel activities and Born solvation energies are made to be equivalent for anions and cations. The permittivity of the electrolyte is taken to change as a function of the electrolyte composition, through both the dielectric freezing of hydrating solvent molecules, and the degree ionic clustering. We employ the following phenomenological interpolation formula:
\begin{equation}
    \varepsilon=\varepsilon_s \alpha_0 (1-x)+ \varepsilon^{*}_s (1-\alpha_0) (1-x) +\varepsilon^{*}_\pm(1-\alpha_\pm)x,
\label{eq:perm}
\end{equation} 
where $x$ is the mole fraction of salt, $\alpha_0$, is the fraction of free solvent, $\varepsilon^{*}_s$  is the dielectric constant contribution of bound solvent, and $\varepsilon^{*}_\pm$ is the dielectric constant contribution of bound ions. Thus, $\varepsilon$ changes from $\varepsilon_s$ in the dilute regime to $\varepsilon_\pm^*$ as the ions become more and more bound in ionic clusters. Furthermore, this phenomenological expression will capture dielectric decrement via the decreasing fraction of free solvent molecules. However, this dielectric decrement will eventually level off as the free solvent disappears, in which case the dielectric constant would tend towards the lower value of a neat ionic liquid~\cite{weingartner2006static}. It is typical, when modelling dielectric decrement across wide concentration ranges, to employ nonlinear, empirical models, as in Ref. \citenum{shilov2015role}, but equation \eqref{eq:perm} will capture much of the same behavior, but with a more direct connection to the ion-association and ion-solvation modelled in this work.  

Thus, $\Delta_{lmsq}$ [written in Eq.~\eqref{eq:delel}] contains an electrostatic correction as a consequence of the chemical potential of free ions varying with electrolyte composition. The Debye-Huckel contribution stabilizes the free ions due to favorable electrostatic interactions with other free ions as concentration increases.  This results in a decreased affinity for ion association. However, the dielectric constant of the electrolyte decreases as a function of salt concentration, which will increase the Born solvation free energy of the free ions, ultimately resulting in an increasing affinity of ion association. These two effects (electrostatic interaction with screening cloud and Born Solvation) tend to counteract each other for a large majority of salt concentrations, and $\Delta^{el}_{lmsq}$ is roughly constant. However, when free ions are dilute (either at very low or very high salt fractions) the Debye-Huckel activities, and thus $\Delta^{el}_{lmsq}$, become strong functions of free ion concentration. 

Having defined each component of $\Delta_{lmsq}$, it is extremely useful to introduce the notion of the ``association constant", $\Lambda_{ij}$ for the association of species $i$ and $j$. The association constant characterizes the driving force or affinity--or more accurately the exponentiated driving force/affinity--for a specific type of association. It is written as the following
\begin{align}
    \Lambda_{+-}=\frac{(Z-1)^2}{Z}\gamma^{DH}_\pm \exp(-\Delta u_{+-}+\Delta u^{Born}_\pm)=\Lambda^0_{+-}\Lambda_{+-}^{el}
\end{align}
where we have defined a non-electrostatic ionic association constant, $\Lambda_{+-}^{\theta}$
\begin{align}
    \Lambda^{\theta}_{+-}=\frac{(Z-1)^2}{Z}\exp(-\Delta u_{+-})
\end{align} 
and the electrostatic association, $\Lambda_{+-}^{el}$
\begin{align}
    \Lambda_{+-}^{el}=\gamma_\pm^{DH}\exp \left(\Delta u^{Born}_\pm\right)
\end{align}
The ion-solvent association constant, $\Lambda_{\pm 0}$, contains only a non-electrostatic part the association constant:
\begin{align}
    \Lambda_{\pm 0}=\frac{(Z-1)^2}{Z}\exp \left( -\Delta u_{\pm 0} \right)
\end{align}
We then plug in each contribution of $\Delta_{lmsq}$ into Eq.~\eqref{eq:cluster_dist}. Due to the Kroenecker delta functions in Eqs. \eqref{eq:delbond1} and \eqref{eq:delel1} the distribution is most easily written separately for clusters with more than one ion, clusters containing a single ion, and clusters containing just solvent. First, for clusters containing more than one ion ($l+m>1$), we obtain the distribution 
\begin{align}
     c_{lmsq}=\frac{W_{lmsq}}{\lambda^{\theta}_{+-}}\left(\psi_{1000}\Lambda_{+-} \right)^{l} \left(\psi_{0100}\Lambda_{+-} \right)^{m}(\phi_{0010}\Lambda_{+0})^s(\phi_{0010}\Lambda_{-0})^q 
    \label{eq:dist}
\end{align}
where $\psi_{1000} = f_+\phi_{1000}/\xi_+$ and $\psi_{0100} = f_+\phi_{0100}/\xi_+$ are number of association sites  per lattice site for bare cations and free anions, respectively. For solvent-ion clusters containing only a single ion ($l+m=1$), the cluster may either contain a single cation:
\begin{align}
    c_{10s0}=W_{10s0}\psi_{1000}(\phi_{0010}\lambda_{+0})^s 
    \label{eq:dist2}
\end{align}
or a single anion:
\begin{align}
    c_{010q}=W_{010q}\psi_{0100}(\phi_{0010}\lambda_{-0})^q.
    \label{eq:dist3}
\end{align}
Note that when a cluster does not contain cations, $s$ must be 0. Similarly, if the cluster does not contain anions, $q$ must be 0.  Finally, within this model, for clusters not containing ions, the only non-zero component of the distribution corresponds to free solvent molecules:
\begin{align}
    c_{0010}=c_{0001}=\phi_{0010}
    \label{eq:dist4}
\end{align}

Equations \eqref{eq:dist}-\eqref{eq:dist4} give the thermodynamically consistent number distribution for clusters in the electrolyte mixture. It can readily give the volume fraction of a cluster of any size and makeup, if the volume fraction of the bare cations, anions, and solvent molecules are known. However, these bare species volume fractions are not experimentally accessible. Thus, we must write the volume fractions of the bare species in terms of the overall salt/solvent fractions, which are experimentally accessible.

\subsection{Association Probabilities}

Once again we follow Tanaka by introducing the association probabilities, $p_{ij}$. These probabilities are useful because we may write the bare species' volume fractions in terms of them. Formally, $p_{ij}$ is defined as the fraction of association sites of species, $i$, that are occupied with an association to species, $j$. Recall that cations, anions, and solvent molecules are said to have $f_+$, $f_-$, and 1 association sites per molecule, respectively. This implies that generally $p_{ij} \neq p_{ji}$, unless the functionalities and concentrations of species $i$ and $j$ are equivalent, as we will show below. We may write the bare cation volume fraction as
\begin{equation}
    \phi_{1000}=\phi_+(1-p_{+-}-p_{+0})^{f_+}
    \label{eq:x}
\end{equation}
The above equation arises because the probability that a given cation association site will be `dangling' (not participating in associations) will be $1-p_{+-}-p_{+0}$. Thus for all $f_+$ sites to be dangling is $(1-p_{+-}-p_{+0})^{f_+}$. Analogously, for the bare anions and solvent molecules we have
\begin{equation}
    \phi_{0100}=\phi_-(1-p_{-+}-p_{-0})^{f_-}
    \label{eq:y}
\end{equation}
\begin{equation}
    \phi_{0010}= \phi_0(1-p_{0+}-p_{0-})
    \label{eq:z}
\end{equation}

We may insert Eqs.~\eqref{eq:x}-\eqref{eq:z} into Eq.~\eqref{eq:dist}, obtaining a cluster distribution in terms of overall species volume fractions and the association probabilities, $p_{ij}$.  However, we now have six new variables, $p_{ij}$, which are unknown and a function of the overall species volume fractions. Thus, we need six equations to determine these six unknowns. We can obtain three equations straight away due to a conservation of each type of association. For cation-anion associations we have 
\begin{align}
    \psi_+ p_{+-}=\psi_- p_{-+}=\zeta
    \label{eq:sys1}
\end{align}
where $\zeta$ is the number of cation-anion associations per lattice site. For cation-solvent associations we have 
\begin{align}
    \psi_+ p_{+0}=\phi_0 p_{0+}=\Gamma
    \label{eq:sys2}
\end{align}
where $\Gamma$ is the number of cation-solvent associations per lattice site. Finally, for anion-solvent associations we have 
\begin{align}
    \psi_- p_{-0}=\phi_0 p_{0-}=\Xi
    \label{eq:sys3}
\end{align}
where $\Xi$ is the number of anion-solvent associations per lattice site.

We obtain the last three equations following Tanaka, by employing the law of mass action on the number of associations using the association constants $\Lambda_{+-}$, $\Lambda_{+0}$, and $\Lambda_{+0}$. For cation-anion associations we have
\begin{align}
    \Lambda_{+-}\zeta=\frac{p_{+-}p_{-+}}{(1-p_{+-}-p_{+0})(1-p_{-+}-p_{-0})}.
    \label{eq:sys4}
\end{align}
Similarly, for the cation-solvent associations we have
\begin{align}
    \Lambda_{+0}\Gamma=\frac{p_{+0}p_{0+}}{(1-p_{+-}-p_{+0})(1-p_{0+}-p_{0-})}.
    \label{eq:sys5}    
\end{align}
Finally, for the anion-solvent associations we have
\begin{align}
    \Lambda_{-0}\Xi=\frac{p_{-0}p_{0-}}{(1-p_{-+}-p_{-0})(1-p_{0+}-p_{0-})}.
    \label{eq:sys6}    
\end{align}
Here $\Lambda_{+-}$, $\Lambda_{+0}$, and $\Lambda_{-0}$ are treated as equilibrium constants for the individual associations made. Recall that $\Lambda_{+-}$ contains both an electrostatic factor ($\Lambda_{+-}^{el}$), and a non-electrostatic factor ($\Lambda_{+-}^\theta$). The non-electrostatic factor is a constant, but the electrostatic factor is a function of the overall electrolyte composition ($\phi_\pm$), as well as the fraction of free ions ($\alpha_+,\alpha_-$) and solvent ($\alpha_0$) via the Debye length, $\lambda_D$, and relative permittivity, $\varepsilon$. Thus, if we want to model the electrostatic contribution to ion association, we must additionally write $\alpha_i$ in terms of the association probabilities, $p_{ij}$. For $\alpha_+$, we have
\begin{align}
    \alpha_+=(1-p_{+-})^{f_+}
    \label{eq:freecat}
\end{align}
and $\alpha_-$ we have
\begin{align}
    \alpha_-=(1-p_{-+})^{f_-}
    \label{eq:freean}
\end{align}
Note that, for the fraction of ions contributing to the ionic strength we only require that the ion is not associated to a counter-ion; free ions can be hydrated by solvent in any capacity. For the fraction of free solvent we simply have 
\begin{align}
    \alpha_0=1-p_{0+}-p_{0-}.
\end{align}

Thus, Eqs.~\eqref{eq:sys1}-\eqref{eq:sys6} provide six equations from which we may solve for each $p_{ij}$ in terms of the overall species volume fractions. Without making approximations we cannot obtain an analytical solution to this system, but nonetheless we may solve it numerically. A useful approximation based on assumptions of ion symmetry and ``stickiness" permits an analytical solution of the association probabilities in terms of overall species volume fractions and is outlined in the Appendix. These association probabilities close the model, so that we may now obtain the full distributions of clusters as a function of the overall electrolyte composition. 
\begin{figure*}[hbt!]
    \centering
    \includegraphics[width=\textwidth]{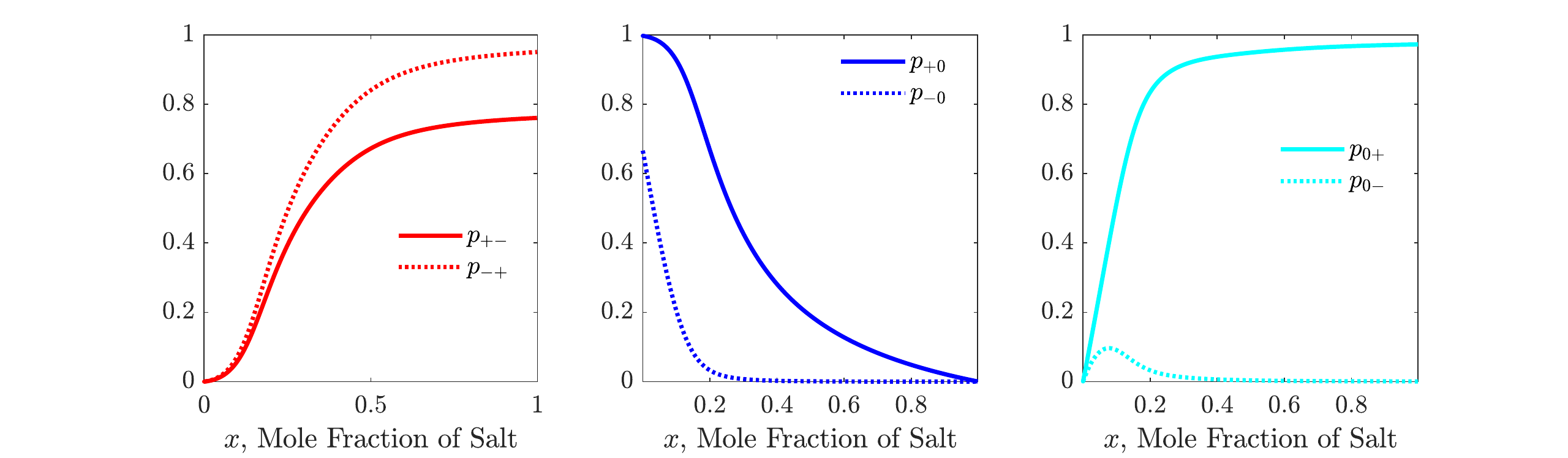}
    \caption{Association probabilities, $p_{ij}$, as a function of the mole fraction of salt for a model water-in-salt electrolyte. The ion--counter-ion association probabilities are plotted in the left panel ($p_{+-}$ and $p_{-+}$), the ion--solvent association probabilities are plotted in the middel panel ($p_{+0}$ and $p_{-0}$), and the solvent--ion associations are plotted in the right panel ($p_{0-}$ and $p_{0+}$). These curves are generated for $\xi_+=1$, $\xi_-=10$, $\Lambda_{+-}=50$, $\Lambda_{+0}=500$, $\Lambda_{-0}=2$, $f_+=5$, $f_-=4$, $v_0=25\text{A}^3$, $\varepsilon_s=80$, $\varepsilon^*_s=\varepsilon^*_\pm=10$.}
    \label{fig:prg_prob}
\end{figure*}

In Fig.~\ref{fig:prg_prob}, we plot sample curves of the concentration dependence of these association probabilities. The parameters detailed in the caption of Fig.~\ref{fig:prg_prob}, which will be used for the majority of this paper, were chosen to be representative of salts used in typical water-in-salt electrolytes (WiSEs), such as lithium bis(trifluoromethanesulfonyl)imide (LiTFSI)\cite{Suo2015}, sodium trifluoromethane sulfonate (NaOTF)\cite{suo2017water}, or even potassium containing analogues\cite{leonard2018}. Note that although these salts have extremely high solubility limits, they would likely precipitate from solution prior to the reaching the pure salt limit ($x=1$). Nevertheless, our figures will extend to the pure salt limit, in order to explore the behavior of the model in this regime. Furthermore, for different sets of parameters that are more representative of an ionic liquid salt, for example, the pure salt limit would be extremely relevant.  

Thus, the parameters used in most of our examples represent a model water-in-salt electrolyte. As would be expected for a LiTFSI-water or NaOTF-water system, the cation--solvent association constant ($\Lambda_{+0}=500$) is considerably larger than the anion--solvent association constant ($\Lambda_{-0}=2$). The anion is also made to be much larger ($\xi_-=10$) than the cation ($\xi_+=1$). Additionally, the cation has a larger functionality $f_+=5$ than the anion ($f_-=4$), to emphasize further cation/anion asymmetry.  

The ion--counter-ion association probabilities, $p_{\pm\mp}$ (left panel in Fig.~\ref{fig:prg_prob}), increase monotonically with salt volume fraction, and the difference between the solid and dotted blue curves in Fig.~\ref{fig:prg_prob} comes from the difference in cation and anion functionality; for a given total number of cation--anion associations, a lower fraction of cation association sites will be occupied with associations to anions. 

The ion--solvent association probabilities, $p_{\pm0}$ (middle panel in Fig.~\ref{fig:prg_prob}), both decrease monotonically with increasing ion concentration. This is expected because there is less water available to associate to ions, and more associations with counter-ions at high salt volume fractions. Again, the solvent is more likely to associate to cations because the association constants considered here dictate the solvent to interact stronger with cations than anions. 

The cation:anion asymmetry is manifested most clearly for the solvent--ion association probabilities, $p_{0\pm}$ (right panel in Fig.~\ref{fig:prg_prob}). The solvent-cation association probability increases monotonically with salt volume fraction due to the increasing concentration of cations and thus cationic association sites. However, the same argument does not hold for the solvent-anion association probability, which displays non-monotonic behavior. Initially, $p_{0-}$ increases due to increasing anion concentration, but then decreases because the cations monopolize the solvent association at high ion concentrations. The reason for this is that cations have more favorable association with the solvent ($\Lambda_{+0}>\Lambda_{-0}$), as well as having more open sites to accept solvent associations ($f_+>f_-$). 

 \begin{figure*}[hbt!]
    \centering
    \includegraphics[width=\textwidth]{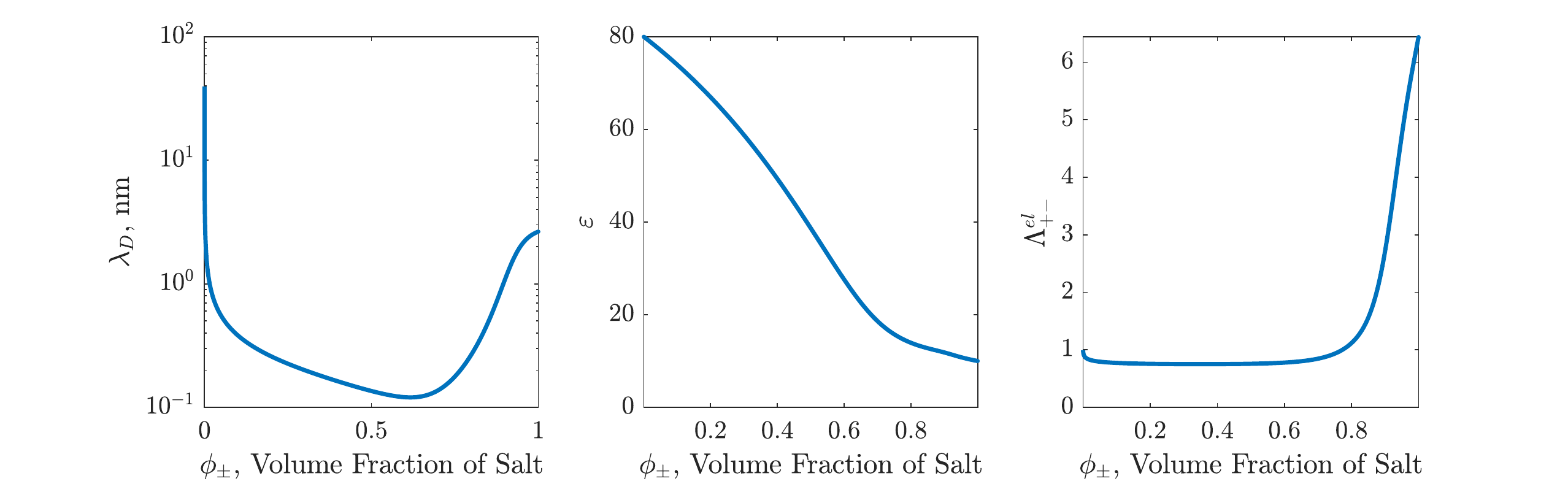}
    \caption{Debye screening length, $\lambda_D$ (left); dielectric constant, $\varepsilon$ (middle); and electrostatic ionic association constant, $\Lambda^{el}_{+-}$ (right) as a function of salt volume fraction for a model water-in-salt electrolyte. These curves are generated for $\xi_+=1$, $\xi_-=10$, $\Lambda_{+-}=50$, $\Lambda_{+0}=500$, $\Lambda_{-0}=2$, $f_+=5$, $f_-=4$, $v_0=25\text{A}^3$, $\varepsilon_s=80$, $\varepsilon^*_s=\varepsilon^*_\pm=10$.}
    \label{fig:elec}
\end{figure*}

Having solved for the association probabilities, we can compute the various quantities involved in the electrostatic portion of ion association. In Fig.~\ref{fig:elec}, the Debye screening length, $\lambda_D$, relative dielectric constant, $\varepsilon$, and the electrostatic ion association factor, $\Lambda^{el}_{+-}$ are plotted as functions of salt volume fraction. Interestingly, we see that $\lambda_D$ displays non-monotonic behavior as a function of $\phi_\pm$, with some qualitative similarities to the non-monotonic screening lengths observed in refs. \citenum{gebbie2013ionic} \& \citenum{smith2016electrostatic}. Although ion aggregation, as modelled here, likely plays a large role in phenomenon observed in refs. \citenum{gebbie2013ionic} \& \citenum{smith2016electrostatic}--since dubbed the ``underscreening paradox"--the full explanation of the underscreening paradox would likely involve a more comprehensive structural description of the electrolyte and its double layer.

The non-monotonicity in $\lambda_D$ is a direct result of the non-monotonicity of the ionic strength of the mixture. In this work, $\lambda_D$ was defined with an ionic strength that only accounts for ``free" ions. At low salt concentration ions remain largely unpaired, thus increasing salt concentration leads to an increase in ionic strength. At high concentrations, increasing salt concentration actually decreases the ionic strength of the mixture, leading to an increase in $\lambda_D$  The behavior of $\lambda_D$ also leads directly to the non-monotonic behavior of $\Lambda^{el}_{+-}$.

\subsection{Sol/Gel Transition}

\begin{figure}[hbt!]
    \centering
    \includegraphics[width=0.35\textwidth]{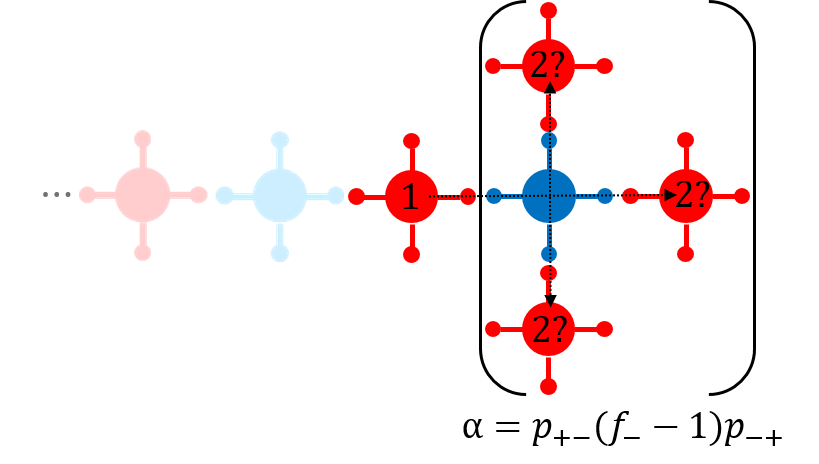}
    \caption{A schematic illustrating the concept of the branching coefficient, $\alpha$, which is an essential quantity in determining the criterion for gelation, Eq.~\eqref{eq:gel_crit}. Starting at the node labeled 1 (referring to a cation), we note that the cluster proceeds arbitrarily to the left. We then consider the probability ($\alpha$) of the cluster continuing to the right to the next cationic node (marked as 2). In order for the cluster to continue to the right the cationic node marked 1 must associate with an anion (with probability $p_{+-}$) and then one of the $f_--1$ remaining anionic association sites must associate with another cation (with probability $p_{-+}$).}
    \label{fig:thought}
\end{figure}

Since the functionalities, $f_\pm$, of anions and cations are both greater than two, the clusters have the potential to become infinitely large if the probabilities, $p_{+-}$ and $p_{-+}$ are large enough. The point at which this occurs (i.e. the gelation point) can be determined in the following manner with the help of Fig.~\ref{fig:thought}. Consider for example, that we traverse along a specific branch of the cluster until we stop arbitrarily at a cation, labeled as `1' in Fig.~\ref{fig:thought}. The cation contains $f_+-1$ sites in addition to the site that was traversed to arrive at the cation. In order for the cluster to proceed infinitely--thus forming a gel--one of the additional $f_+-1$ sites must continue the chain with a probability of unity~\cite{tanaka2011polymer}:
\begin{align}
    (f_+-1)\alpha^*=1
    \label{eq:gel_crit}
\end{align}
where $\alpha$ (not to be confused with the fraction of free species, $\alpha_+,\, \alpha_-$, or $\alpha_0$) is known as the branching coefficient with a ``$*$" denoting its critical value for gelation, and the factor of $f_+-1$ arises because there are $f_+-1$ additional branches on the cation capable of extending the cluster. The same criteria arises for mean-field percolation on a Bethe lattice with coordination number of $f_+$ \cite{stauffer1994introduction}. In our case, though, $\alpha$ refers to the probability that cation 1 continues to a subsequent cationic node (labeled as 2 in Fig.~\ref{fig:thought}) along any available branch, as depicted by the dotted arrows in Fig.~\ref{fig:thought}. In order to get from one cationic node to the next cationic node, we require that one of the cation sites associates with an anion with probability, $p_{+-}$, and that one of the $f_--1$ remaining anionic sites reacts with a cation with probability, $p_{-+}$. Thus,
\begin{align}
    \alpha = p_{+-}(f_--1)p_{-+}
\end{align}
The criterion for gelation is then 
\begin{align}
    (f_+-1)p^*_{+-}(f_--1)p^*_{-+}=1
    \label{eq:gel_crit2}
\end{align}

If this criterion is met, then we expect a macroscopic ionic gel network to spontaneously form and percolate through the electrolyte. Thus, if we know the probabilities, $p_{+-}$ and $p_{-+}$, as functions of concentration, then we may predict the critical concentration at which gelation will occur using Eq.~\eqref{eq:gel_crit2}.

We can also see this criterion arise when analyzing the weight averaged degree of ionic aggregation, $\Bar{n}_w$ (the average sized cluster of which an ion is a part), which is defined by the following formula:
\begin{align}
    \Bar{n}_w&=\frac{\sum_{lmsq}(l+m)^2c_{lmsq}}{\sum_{lmsq}(l+m)c_{lmsq}}
\end{align}
We can then plug in Eq.~\eqref{eq:dist}, and perform the sum over $s$ and $q$ by invoking the binomial theorem obtaining
\begin{align}
\Bar{n}_w = \sum_{lm}(l+m)\alpha_{lm}
\label{eq:nw1}
\end{align}
where $\alpha_{lm}$ is the fraction of total ions in clusters containing $l$ cations and $m$ anions. For clusters containing more than one ion, $\alpha_{lm}$ is given by
\begin{align}
    \alpha_{lm}=\frac{\Lambda^{el}_{+-}\mathcal{K}}{2}(l+m)W_{lm}\left(\frac{p_{-+}}{1-p_{-+}}(1-p_{+-})^{f_+-1}\right)^{l}\left(\frac{p_{+-}}{1-p_{+-}}(1-p_{-+})^{f_--1}\right)^{m}
    \label{eq:clust_frac}
\end{align}
where $\mathcal{K}=f_+ (1-p_{+-})(1-p_{-+})/p_{-+}$ (analogously defined by Stockmayer in Ref.~\citenum{stockmayer1952molecular}). Note that Eq.~\eqref{eq:clust_frac} will not reduce to $\alpha_+$ or $\alpha_-$ for single ion clusters (free ions), because the cluster distribution is slightly modified for single ion clusters [recall Eqs.~\eqref{eq:dist2} and \eqref{eq:dist3}]. 

We can write the sum in Eq.~\eqref{eq:nw1} in a closed form with the help of Stockmayer (Ref.~\citenum{stockmayer1952molecular}), or with the methods developed within Ref.~\citenum{macosko1976}:
\begin{align}
    \Bar{n}_w&=\Lambda^{el}_{+-}\left(1+\frac{p_{+-}p_{-+}\left((f_+-1)p_{+-}+(f_--1)p_{-+}+2\right)}{\left(\frac{p_{+-}}{f_-}+\frac{p_{-+}}{f_+}\right)\left(1-(f_+-1)(f_--1)p_{+-}p_{-+}\right)}\right) \nonumber \\
    &+\frac{1}{2}(1-\Lambda_{+-}^{el})\left(\alpha_++\alpha_-\right)
    \label{eq:nw2}
\end{align}
Note that Eq.~$\eqref{eq:nw2}$ will reduce to Stockmayer's result in Ref.~\citenum{stockmayer1952molecular} for $\Lambda_{+-}^{el}=1$. Interestingly, Eq.~\eqref{eq:nw2} predicts that $\Bar{n}_w$ diverges when $(f_+-1)p_{+-}(f_--1)p_{-+}=1$, which is the exact condition we previously derived for gelation. 

\begin{figure}[hbt!]
    \centering
    \includegraphics[width=0.49\textwidth]{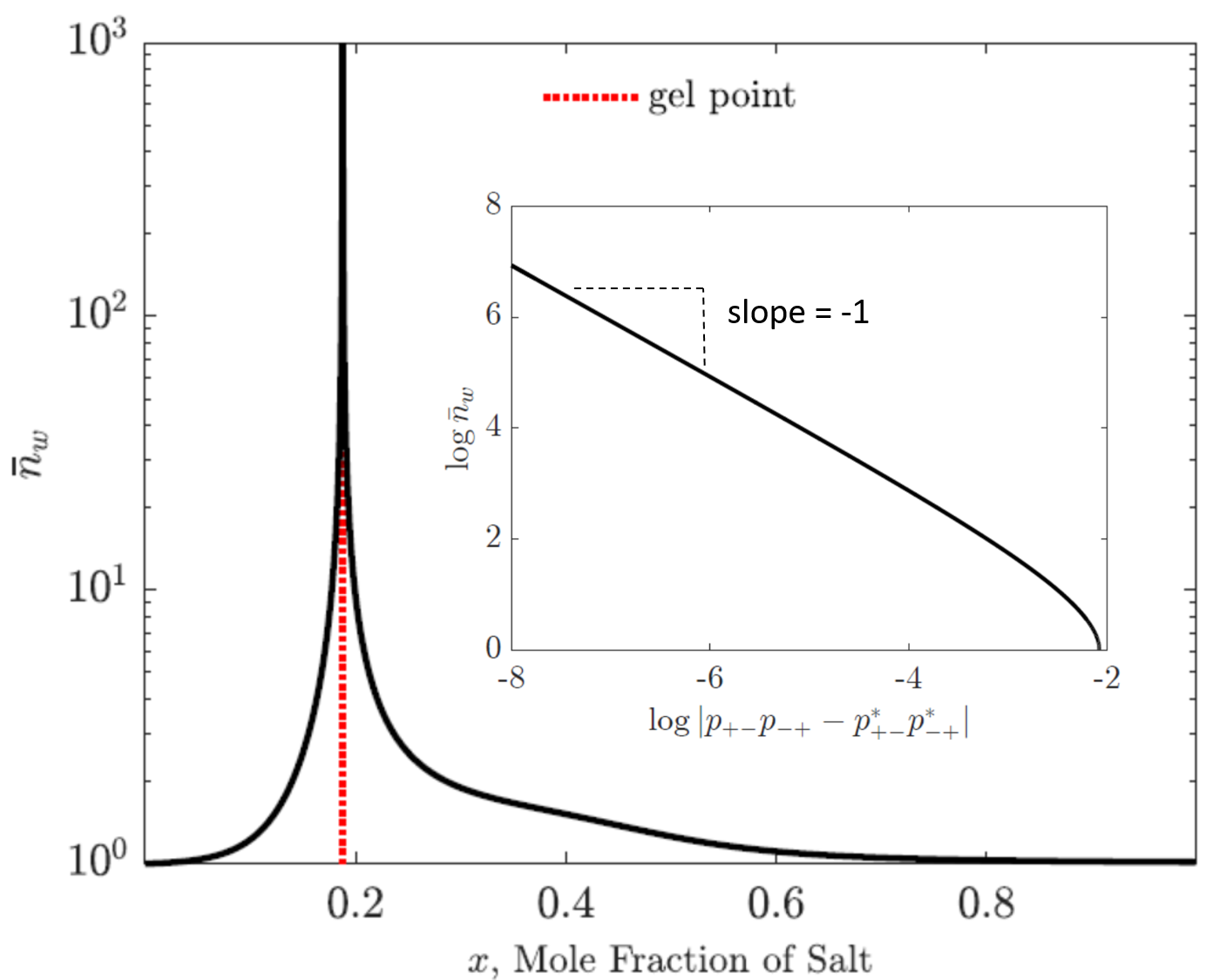}
    \caption{The weight averaged degree of ion aggregation, $\bar{n}_w$ plotted against the volume fraction of salt, $\phi_\pm$, using Eq.~\eqref{eq:nw2} with probabilities for association restricted to the sol (excluding the gel). In the inset we plot (on a log-log scale) the weight averaged degree of ion aggregation, $\bar{n}_w$, against the deviation from the gel point, $|p_{+-}p_{-+}-p^*_{+-}p^*_{-+}|$, showing a critical exponent of -1. This curve was generated for $\xi_+=1$, $\xi_-=10$, $\Lambda_{+-}=50$, $\Lambda_{+0}=500$, $\Lambda_{-0}=2$, $f_+=5$, $f_-=4$, $v_0=25\text{A}^3$, $\varepsilon_s=80$, $\varepsilon^*_s=\varepsilon^*_\pm=10$.}
    \label{fig:nwbar}
\end{figure}

As an example, we plot the weight averaged degree of aggregation as a function of concentration in Fig.~\ref{fig:nwbar} using Eq.~\eqref{eq:nw2} with the model parameters listed in the caption, corresponding to the aforementioned fictitious water-in-salt electrolyte. As can be seen, the weight average degree of aggregation diverges at the gelation point. In the inset of Fig.~\ref{fig:nwbar}, we display a log-log plot of the weight average degree of aggregation as a function of deviation in $p_{+-}p_{-+}$ from the critical value, yielding a linear curve with a slope of -1. Thus, $\bar{n}_w$ diverges at the gel point with a critical exponent of -1. This type of behavior is expected, when considering the direct analogy of our gelation model with percolation on a Bethe lattice. 

Interestingly in Fig.~\ref{fig:nwbar}, beyond the gel point, $\bar{n}_w$ rapidly decreases. This is because we are plotting the weight averaged degree of aggregation for species in the sol only, excluding the gel. After the gel forms the vast majority of ion associations are contributing to the gel, as opposed to finite clusters in the sol. As we approach the no-solvent limit (ionic liquid/crystal limit), the degree of aggregation in the sol is essentially 1, implying that at large salt fractions, the electrolyte looks like a simple mixture of dilute free ions immersed in an ionic gel. 

\subsection{Post-Gel Regime}

For salt concentrations beyond the critical concentration, we expect a gel to be present in the electrolyte containing an increasing fraction of the electrolyte's ions. Thus, we must quantify the fraction of the species in the gel and in the sol. We employ Flory's treatment of the post-gel regime in which the volume fraction of free species can be written equivalently in terms of overall association probabilities, $p_{ij}$, and association probabilities taking into account only the species residing in the sol, $p^{sol}_{ij}$. 

\begin{align}
    \phi_+(1-p_{+-}-p_{+0})^{f_+}=\phi_+^{sol}(1-p^{sol}_{+-}-p^{sol}_{+0})^{f_+}
    \label{eq:gel1}
\end{align}
\begin{align}
    \phi_-(1-p_{-+}-p_{-0})^{f_-}=\phi_-^{sol}(1-p^{sol}_{-+}-p^{sol}_{-0})^{f_-}
    \label{eq:gel2}
\end{align}
\begin{align}
    \phi_0(1-p_{0+}-p_{0-})=\phi_0^{sol}(1-p^{sol}_{0+}-p^{sol}_{0-})
    \label{eq:gel3}
\end{align}
Where $\phi_i^{sol}$ is the volume fraction of species, $i$ remaining in the sol. We may determine each of the three unknown $\phi_i^{sol}$ variables, as well as the six unknown sol association probabilities, $p_{ij}^{sol}$, using \eqref{eq:gel1}-\eqref{eq:gel3} in addition to Eqs.~\eqref{eq:sys1}-\eqref{eq:sys6}, however in this case we use sol-specific quantities.

Thus, we have nine equations and nine unknowns (six sol association probabilities and three sol species volume fractions). The fraction of species, $i$, in the gel is simply given by 
\begin{align}
    w_i^{gel}=1-\phi^{sol}_i/\phi_i
\end{align}
Note that prior to the critical gel concentration, we have the trivial solution that $p_{ij}=p^{sol}_{ij}$ and $\phi_i=\phi^{sol}_i$, yielding a gel fraction of $w_i^{gel}=0$. However, beyond the gel point, there is a non-trivial solution yielding $w_i^{gel}>0$.

\begin{figure*}[hbt!]
    \centering
    \includegraphics[width=\textwidth]{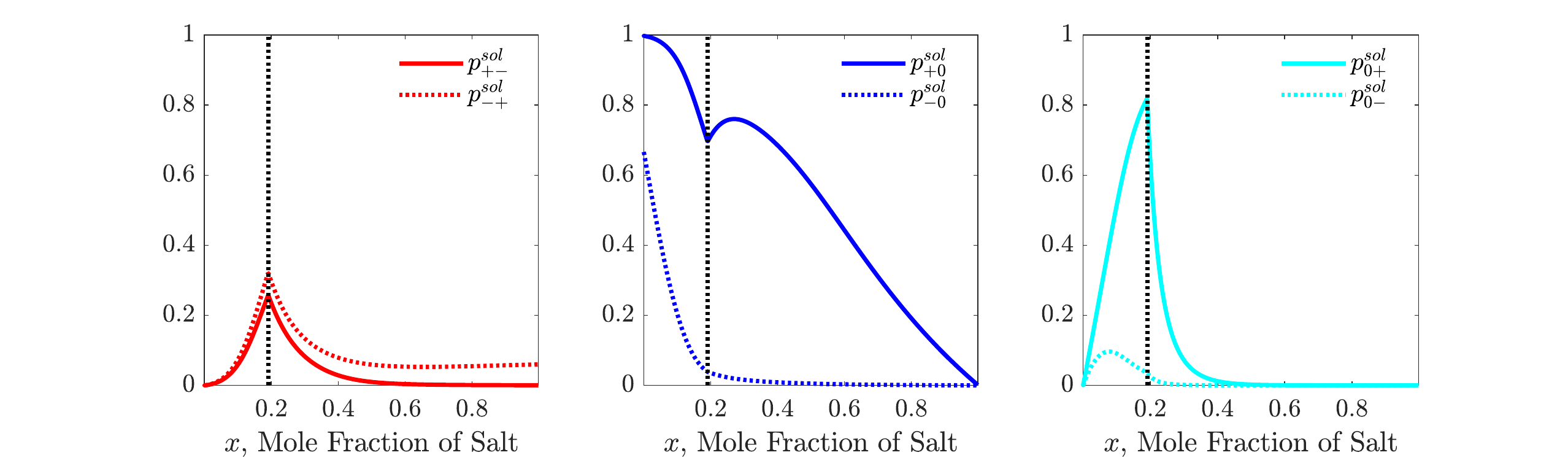}
    \caption{The ``sol" association probabilities, $p^{sol}_{ij}$, are plotted against the mole fraction of salt. The ion--counter-ion association probabilities ($p^{sol}_{+-}$ and $p^{sol}_{-+}$) are plotted in the left panel, the ion--solvent association probabilities ($p^{sol}_{+0}$ and $p^{sol}_{-0}$) are plotted in the middle panel, and the solvent--ion associations ($p^{sol}_{0-}$ and $p^{sol}_{0+}$) are plotted in the right panel. These curves are generated for $\xi_+=1$, $\xi_-=10$, $\Lambda_{+-}=50$, $\Lambda_{+0}=500$, $\Lambda_{-0}=2$, $f_+=5$, $f_-=4$, $v_0=25\AA^3$, $\varepsilon_s=80$,$\varepsilon^*_s=\varepsilon^*_\pm=10$.The black dotted line corresponds to the gel point.}
    \label{fig:pog_prob}
\end{figure*}

As an example, we plot the ``sol" association probabilities in Fig.~\ref{fig:pog_prob} using the model parameters listed in the caption, corresponding to the aforementioned fictitious water-in-salt electrolyte. As expected, we observe distinct cusps in the ``sol" association probabilities at the gel point. These cusps are the result of a bifurcation point for solutions to the equations. One solution branch belongs to the overall association probabilities that transition smoothly through the gel point, and the other solution branch belongs to the ``sol" association probabilities that bifurcate from the gel point. Typically, beyond the gel point all of the sol association probabilities decrease, because the majority of associations are consumed by the gel.

\begin{figure*}[hbt!]
    \centering
    \includegraphics[width=\textwidth]{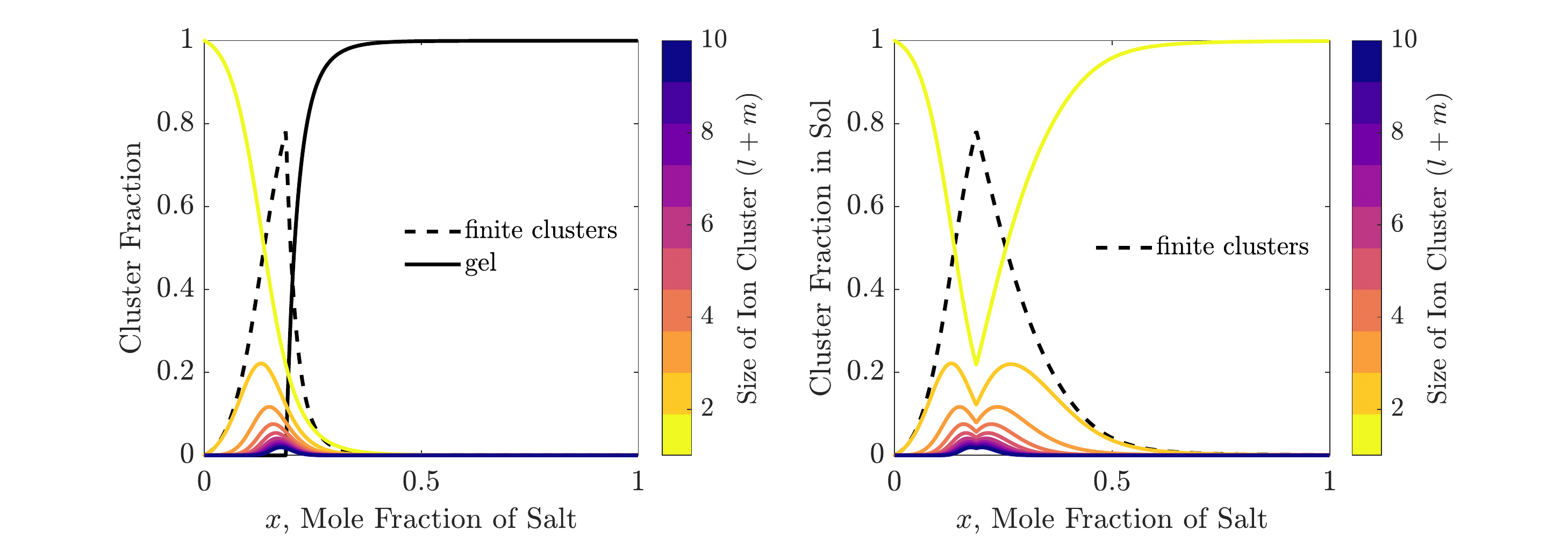}
    \caption{Fraction of ion clusters of different size ($l+m$), including both total finite cluster fraction ( $l+m\geq2$) and the gel fraction as a function of salt volume fraction. The plot on the left corresponds to cluster fractions considering all ions in the electrolyte, while the right corresponds to cluster fractions for ions in the sol only. These curves are generated for $\xi_+=1$, $\xi_-=10$, $\Lambda_{+-}=50$, $\Lambda_{+0}=500$, $\Lambda_{-0}=2$, $f_+=5$, $f_-=4$, $v_0=25\text{A}^3$, $\varepsilon_s=80$, $\varepsilon^*_s=\varepsilon^*_\pm=10$.}
    \label{fig:cluster_curves}
\end{figure*}

In the left panel of Fig.~\ref{fig:cluster_curves}, we plot the concentration dependence of a various ion clusters of different sizes ($1\leq l+m\leq10$ and the ionic gel). As expected, we see that the fraction of free ions ($l+m=1$) decreases monotonically as a function of salt volume fraction due to the increasing ionic association probability. Interestingly, all other finite ion clusters behave non-monotonically with salt fraction. In general, ion clusters with $l+m\geq2$ first increase with salt concentration due to the increasing ion association probability. However, as salt concentration increases further, more and more associations are directed towards the formation of higher order clusters, and eventually the ionic gel. Fig.~\ref{fig:cluster_curves} also defines three distinct ``regimes" in the solution. In the low concentration regime ($0<\phi_\pm\leq 0.15$), free ions are the major ionic species in the electrolyte. For $0.2<\phi_\pm\leq0.3$, finite ion clusters dominate the electrolyte. Finally, for high salt concentrations ($\phi_{\pm}>0.3$) the electrolyte is majorly comprised of the ionic gel. 

In the right panel of Fig.~\ref{fig:cluster_curves}, we plot the same cluster fractions, but consider only the ions that remain in the sol. The curves are identical to those in the left plot of Fig.~\ref{fig:cluster_curves} prior to the gel point. Beyond the gel point, the cluster fractions behave in a very peculiar manner. The fraction of free ions in the sol actually increases as a function of concentration. This is due to the fact that the ion association probabilities for ions in the sol actually decreases after the gel point. Thus, the sol is looks more and more like ``dilute" electrolyte as we increase the overall salt concentration. For the parameters chosen in Fig.~\ref{fig:cluster_curves}, we see that nearly all of the ions in the sol are free as we approach the pure salt limit. Though, this is actually a very small amount of free ions overall, because the electrolyte is nearly all gel. For model parameters more akin to an ionic liquid salt, we might expect a much larger fraction of free ions in the pure salt limit \cite{feng2019free}.  

\begin{figure*}[hbt!]
    \centering
    \includegraphics[width=\textwidth]{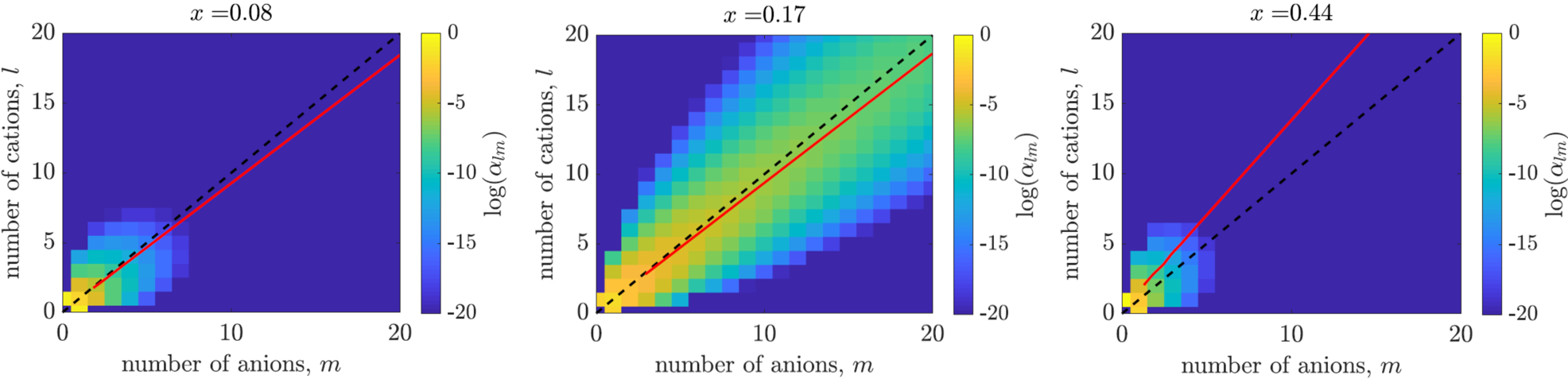}
    \caption{Probability distribution of ion clusters rank $lm$ (containing $l$ cations and $m$ anions), for various mole fractions of salt. (left) The cluster distribution for a pre-gel salt mole fraction of $x=0.09$. (middle) The cluster distribution for a near-gel salt mole fraction of $x=0.19$. (Right) The cluster distribution for a post-gel salt mole fraction of $x=0.47$. In each plot, the black curve corresponds to a 1:1 anion:cation ratio, and the red curve corresponds to the most probable cluster of total rank $l+m$. Note that the probabilities are plotted on a log scale to better visualize the distribution. These plots are generated for $\xi_+=1$, $\xi_-=10$, $\Lambda_{+-}=50$, $\Lambda_{+0}=500$, $\Lambda_{-0}=2$, $f_+=5$, $f_-=4$, $v_0=25\text{A}^3$, $\varepsilon_s=80$, $\varepsilon^*_s=\varepsilon^*_\pm=10$.}
    \label{fig:clust_hist}
\end{figure*}

Figure~\ref{fig:cluster_curves} informs us as to the probabilities of seeing clusters containing a total amount of  ions. However, it does not tell us specifically how many anions or cations compose those clusters. The full bivariate probability distribution of clusters, $\alpha_{lm}$, defined in Eq.~\eqref{eq:clust_frac}, is plotted for various salt fractions in Fig.~\ref{fig:clust_hist}. We have chosen three different mole fractions of salt for plotting the distribution: $x=0.08$ (pre-gel), $x=0.17$ (near-gel), and $x = 0.44$ (post-gel). Both the pre-gel and near-gel distributions are skewed below the neutral cluster line (black dashed line), centered around the red solid line (denoting the most probable cluster of rank $l+m$). This indicates that clustered ions have a slight tendency to be negatively charged, containing more anions than cations. This effect is expected when the functionalities for ions are different. In this case, because the cations have a larger functionality than anions, each cation can accept more ion associations than each anion. Thus, there will be a tendency for there to be more anions in each cluster than cations.  Additionally, the cluster distribution is pushed towards larger clusters as the mole fraction is increased from 0.08 to 0.17, due to the increasing ionic association probability. However, as the mole fraction is increased to 0.44 (well above the gel point) the distribution is both pushed towards smaller clusters than at $x=0.44$, as well as being skewed above the neutral cluster line, indicating that the finite clusters will on average more likely to be positively charged. When the gel is formed it absorbs many of the large negative clusters, and is overall negatively charged. Therefore, the sol will have a net positive charge, leading to positively skewed cluster distribution. 

We may probe the effect of solvent or salt type by tuning the different association constants, $\Lambda_{ij}$. If we assume that ion association sites are never empty (either occupied by solvent or counter-ions), and that the ions have equal functionality, we may use the ``sticky symmetric ion approximation," which is outlined in the Appendix. If we operate within the sticky symmetric ion approximation, we are left with one primary variable to manipulate: $\tilde{\Lambda}=\Lambda_{+-}/\Lambda_{+0}\Lambda_{-0}$. By varying $\tilde{\Lambda}$ we are tuning the ``strength" of the electrolyte: weak electrolytes have $\tilde{\Lambda}\gg 1$ and strong electrolytes have  $\tilde{\Lambda} \ll 1$. 
\begin{figure}[hbt!]
    \centering
    \includegraphics[width=0.49\textwidth]{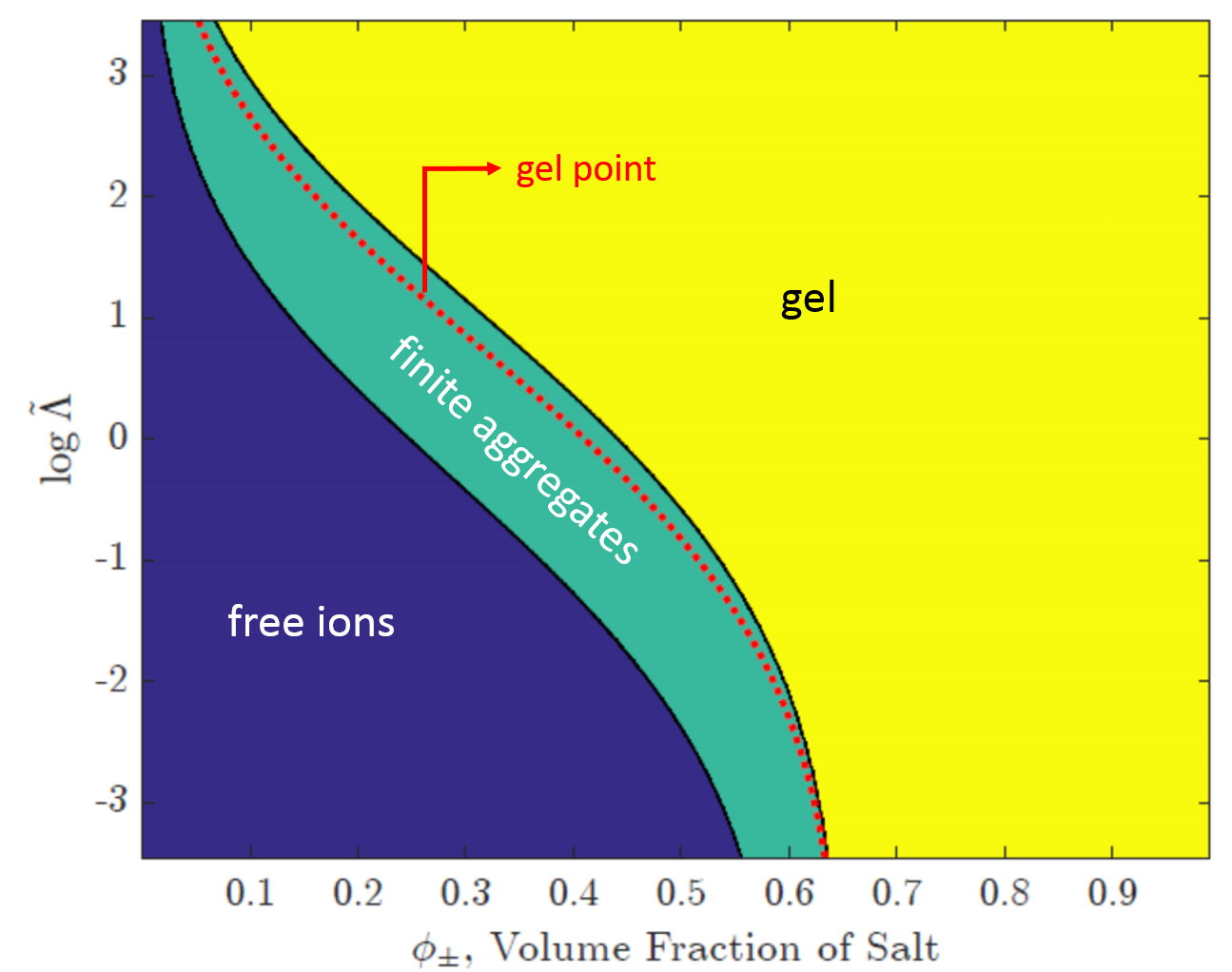}
    \caption{A psuedo-phase diagram of the most probable ionic ``state" (either free, in a finite cluster, or in the gel) as a function of $\tilde{\Lambda}$ and $\phi_\pm$. The Red dotted line denotes the critical gel boundary. The diagram was generated within the sticky symmetric ion approximation (see Appendix) for $\xi_+=\xi_-=5$, and $f_+=f_-=4$. }
    \label{fig:frac_maps}
\end{figure}
In Fig.~\ref{fig:frac_maps} we display a psuedo-phase diagram of the most probable ionic ``state" (either free, in a finite cluster, or in the ionic gel) of an ion as a function of $\tilde{\Lambda}$ and $\phi_\pm$. Note that Fig.~\ref{fig:frac_maps} is generated within the sticky symmetric ion approximation. As was noted in Fig.~\ref{fig:cluster_curves}(left), free ions dominate at low salt fractions and gel dominates at moderate-high salt fractions, with a narrow region of phase space where finite aggregates dominate. The critical gel boundary is denoted by the red dotted line, which generally resides within the finite aggregate region of the phase diagram, because at the along the gel boundary, the fraction of ions within gel will be infinitesimal. However, the gel tends to grow rapidly beyond the gel point by consuming the larger ion clusters. Thus, the gel dominates the mixture soon after crossing the gel boundary. For $\ln(\tilde{\Lambda}) > 0$, the strength of the ion-ion attraction more favorable than the ion-solvent interaction, which results in the onset of gelation occurring at smaller salt fractions. Whereas, for $\ln(\tilde{\Lambda}) < 0$, the favourable ion-solvent interaction tends to ``pull" free ions out of finite aggregates and gel, which pushes out the onset of gelation to larger salt fractions.

\section{Discussion}
Within the accuracy of our various assumptions, our developed model can be applied to the entire range of salt concentrations from dilute to pure IL. In the dilute regime, our model recovers Debye-Huckel behavior \cite{debye1923theory}; this regime is not of much interest in terms of aggregation and gelation. Rather, the more interesting regime occurs for super-concentrated solvent-in-salt electrolytes (including IL solvent mixtures, hydrate melts etc..) and ILs, which are highly relevant for battery or super-capacitor applications. Often SiSEs and ILs contain bulky or asymmetric ions that leads to high solubility or low melting points of the salts. Moreover, the ion aggregates that are formed in these systems tend to be irregular and disordered, which is quite consistent with the approximation of Cayley tree-like ion aggregates. Thus, the physics included in our model should be highly relevant for SiSEs and Ils in particular. More typical salts, such as NaCl for example, form aggregates that may be ordered and semi-crystalline, as opposed to the branched structures that are characteristic of Cayely trees. Ordered aggregates nucleate, phase separate, and induce the precipitation of crystalline salt, without forming a gel. In these types of system, the physics of ion gelation would probably not be as relevant, and our description of ion aggregation would be somewhat flawed. Nonetheless, we expect that our model is well-equipped for capturing the ion association, solvation, and gelation in super-concentrated SiSEs and ILs.

\subsection{Thermodynamic Implications}
Our theory can also be used to predict some important thermodynamic quantities, such as the activity coefficients of species in the mixture. In Eq.~\eqref{eq:muclust}, we wrote the chemical potential of a cluster of rank $lmsq$. The equilibrium condition [Eq.~\eqref{eq:eqm}] implies that the chemical potential of species in the cluster will be equal to their bare counterparts. Thus, we may write the chemical potential of an ion or solvent molecule as simply the chemical potential of a bare ion or solvent molecule: 
\begin{align}
    \beta \Delta \mu_{+}=\beta \Delta \mu_{1000}=\ln \left( \phi_{1000} \gamma_+^{DH} \right) + \Delta u^{Born}_{+} + 1-c_{tot}
\end{align}
\begin{align}
    \beta \Delta \mu_{-}=\beta \Delta \mu_{0100}=\ln \left( \phi_{0100} \gamma_-^{DH} \right) + \Delta u^{Born}_{-}+1-c_{tot}
\end{align}
\begin{align}
    \beta \Delta \mu_{0}=\beta \Delta \mu_{0010}=\ln \phi_{0010}+1-c_{tot}
\end{align}

We may derive ionic activity coefficients (with respect to a dilute solution reference state) by obtaining the excess part of the chemical potential. First, we must subtract off ideal entropy of mixing terms ($\ln \phi_i$). Then, we must subtract off the excess part of the chemical potential of the bare ions in the dilute limit obtaining
\begin{align}
    \ln \gamma_\pm &= \beta \Delta \mu_\pm - \ln \left\{\phi_{+,-}(1-p^\circ_{\pm\mp}-p^\circ_{\pm0})^{f_\pm}\right\}
\end{align}
where the ``$\circ$" superscript denotes the association probabilities in the dilute limit (as salt concentration approaches 0), $\phi_{+,-}$ denotes $\phi_+$ or $\phi_-$ (not to be confused with $\phi_\pm$, the volume fraction of salt). The limiting ionic association probabilities, $p^{\circ}_{\pm\mp}$ tend towards 0. However, the limiting ion-solvent association probabilities, $p^{\circ}_{\pm0}$, tend toward $\Lambda^{\theta}_{\pm0}/(\Lambda^{\theta}_{\pm0}+1)$. Thus, if $\Lambda^{\theta}_{\pm0}\gg1$, we would expect ions to be fully associated with water in the dilute limit. We can then write the ionic activity coefficient as 
\begin{align}
   \ln \gamma_\pm =  \ln \gamma_\pm^{DH} +\Delta u_\pm^{Born} + f_\pm \ln \left\{(1-p_{\pm\mp}-p_{\pm0})(1+\Lambda^{\theta}_{\pm0})\right\} +1-c_{tot}
\end{align}
Similarly, we may write the activity coefficient of solvent molecules as 
\begin{align}
    \ln \gamma_0 = \ln (1-p_{0+}-p_{0-}) +1-c_{tot} 
\end{align}
It is also useful to define a mean ionic activity coefficient, $\bar{\gamma}_\pm=(\gamma_+\gamma_-)^{1/2}$, which is the more experimentally accessible quantity. 
\begin{figure}[hbt!]
    \centering
    \includegraphics[width=0.49\textwidth]{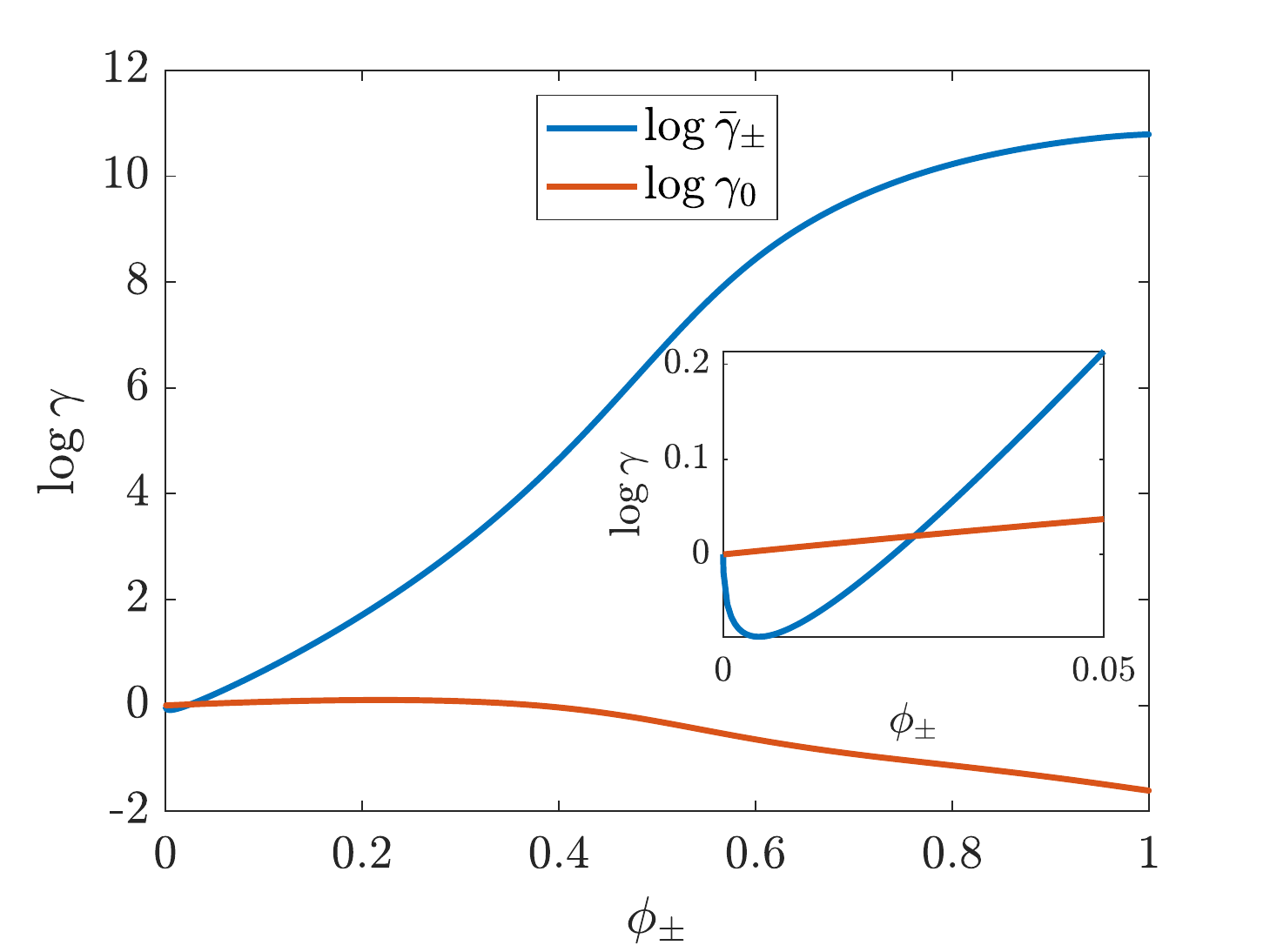}
    \caption{The activity coefficients of the salt and solvent are plotted against the mole fraction of salt. Within the inset of the figure, we zoom in on the dilute region where the model recovers Debye-H\"uckel behavior for salt activity. These curves are generated for $\xi_+=1$, $\xi_-=10$, $\Lambda_{+-}=50$, $\Lambda_{+0}=500$, $\Lambda_{-0}=2$, $f_+=5$, $f_-=4$, $v_0=25\text{A}^3$, $\varepsilon_s=80$, $\varepsilon^*_s=\varepsilon^*_\pm=10$.}
    \label{fig:loggam}
\end{figure}

In Fig.~\ref{fig:loggam}, we plot the mean ionic activity coefficient, as well as that of the solvent, as a function of the volume fraction of salt. A fairly general prediction of our model, which can be seen in Fig.~\ref{fig:loggam}, is that the activity of the salt tends to rise extraordinarily as a function of concentration, while that of the solvent simultaneously decreases. The salt activity increases for two primary reasons. First, the magnitude of Born solvation energy of free ions decreases due to the decreasing dielectric constant of the electrolyte--free ions become more active in lower dielectric constant fluids. Second, ions become more paired with counter-ions as opposed to solvent, which is unfavorable entropically, as well as enthalpically for very strongly hydrating solvents. At the same time the solvent activity tends to decrease at high salt concentrations, due to the increasing fraction of solvent that is favorably-incorporated within the hydration shell of ions. These trends in salt and solvent activity are interesting, because one of the primary reasons water-in-salt electrolytes (WiSEs), in particular, have garnered so much interest is their ability to form a passivating solid-electrolyte interface (SEI) at the negative electrode. This SEI layer suppresses the deleterious hydrogen evolution reaction, which prevents the use of more dilute aqueous electrolytes. The SEI layer on an anode in contact with a WiSE would consist of reduction products involving the salt. By raising the activity of the salt and lowering the activity of the solvent, the reduction potential of the salt is increased, while that of the solvent is decreased. Thus, by increasing salt concentration, the affinity to form an SEI layer is expected to increase, and that to evolve hydrogen is expected decrease, as observed experimentally~\cite{Suo2015}. At some salt concentration, there must be a crossover, where it becomes more favorable to form an SEI layer, than to evolve hydrogen. Because our model can capture the trends in activity for both ions and the solvent, it could potentially help predict when this crossover might occur, and how it might change for different electrolyte materials.

\subsection{Transport Implications}
Although our model does not include any dynamics, we can begin to speculate on how certain transport properties, such as conductivity or ion transference numbers, may be influenced by ion association in the super-concentrated regime. For transport in multi-component, concentrated mixtures, it is often necessary to consider coupled diffusive fluxes \cite{de2013non,krishna1997maxwell,deen1998analysis}, which are related to the vector of species chemical potential gradients through the Onsager linear-response tensor, or, after transformation to concentration gradients, the Stefan-Maxwell diffusivity tensor.  This mathematical framework is the basis for concentrated solution theories of electrolyte transport \cite{newman2012electrochemical}, which have been widely applied to batteries \cite{smith2017multiphase,thomas2002advances} and fitted to experiments\cite{valoen2005transport,nyman2008electrochemical,lundgren2015electrochemical} and molecular simulations \cite{wheeler2004molecular}. The Stefan-Maxell formulation has also been extended to charged electrolytes in double layers\cite{psaltis2011comparing,balu2018role}. Even for moderately concentrated electrolytes, however, the diffusivity tensor and ionic activity coefficients are  fitted to experimental data with little theoretical guidance, and complex many-body interactions with solvent at high concentration are neglected.  Our statistical model could provide a detailed, microscopic basis to model coupled fluxes in superconcentrated electrolytes as originating from the presence of ionic clusters.

Remarkably, as a result of the ion clustering predicted by our model,  superconcentrated electrolytes may behave more like dilute electrolytes in that low concentrations of mobile charge carriers drift and diffuse with nearly independent fluxes. As such,for an associative mixture of ions Ref.~\citenum{france2019} proposed a modified Nernst-Einstein equation for conductivity, $\sigma$,
\begin{align}
    \sigma = \frac{e^2c_{salt}}{k_B T} \sum_{lm} (l-m)^{2} \alpha_{lm} D_{lm}
    \label{eq:sig1}
\end{align}
where $D_{lm}$ is the diffusivity of a cluster of rank $lm$, and the factor of $(l-m)^2$ arises because $l-m$ is the valence charge of a cluster of rank $lm$. Our model is able to predict the cluster fractions, $\alpha_{lm}$, (as in Fig.~\ref{fig:clust_hist}) for different electrolyte compositions and temperatures, which could be extremely helpful when designing more conductive electrolytes. However, the cluster diffusivities, $D_{lm}$, would still be unknown, though, they would undoubtedly decrease with increasing cluster size. As detailed in Refs.~\citenum{france2019,molinari2019transport,molinari2019general}, the contribution of clusters to the ionic current may be largely responsible for the very interesting observation of negative transference numbers for species in ionic liquid mixtures and solid-state electrolytes [See Eq. (5) in Ref.~\citenum{france2019}]. Though, for binary liquid electrolytes, we would not expect such exotic observations in ion transference numbers. Along the same vein, observations of negative Stefan-Maxwell diffusion coefficients\cite{kraaijeveld1993negative,wesselingh1995exploring} have been reported for ion transport of concentrated electrolytes through membranes, which may be due to ion clustering.  

\begin{figure*}[hbt!]
    \centering
    \includegraphics[width=\textwidth]{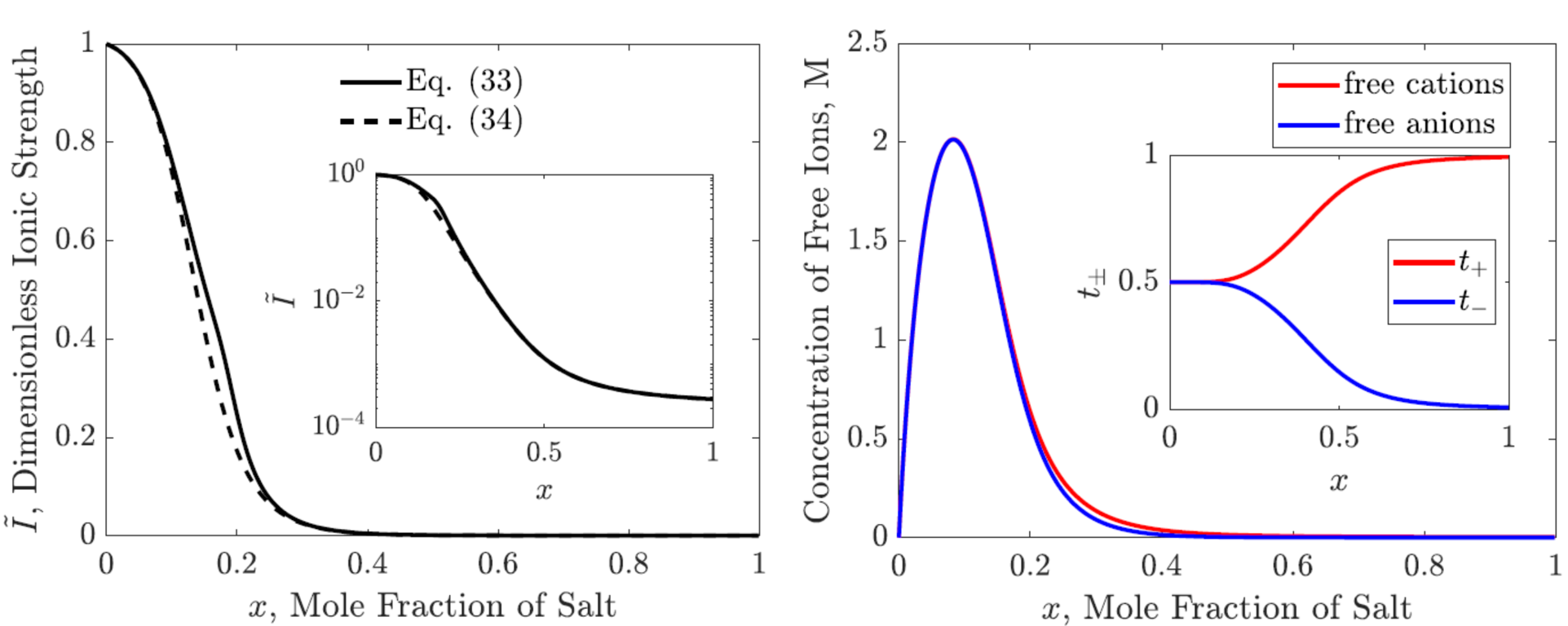}

    \caption{(left) A plot of the dimensionless ionic strength as a function of the mole fraction of salt when account for all charged cluster [solid line, Eq.~\eqref{eq:LI1}] or just free ions [dashed line, Eq.~\eqref{eq:LI2}]. The inset of the left panel displays the same curves on a log-log plot to better visualize the high mole fraction regime. (right) The concentration of free anions and cations are plotted against the mole fraction of salt, displaying non-monotonic salt concentration dependence. Within the inset of the figure, the transference number of anions ($t_-$) and cations ($t_+$) are plotted against the mole fraction of salt according to Eq.~\eqref{eq:t}. These curves are generated for $\xi_+=1$, $\xi_-=10$, $\Lambda_{+-}=50$, $\Lambda_{+0}=500$, $\Lambda_{-0}=2$, $f_+=5$, $f_-=4$, $v_0=25 \text{A}^3$, $\varepsilon_s=80$, $\varepsilon^*_s=\varepsilon^*_\pm=10$.}
    \label{fig:trans}
\end{figure*}

Although there are likely systems in which ion clusters play a large role in conducting ionic current, recent work in Ref.~\citenum{feng2019free} found that free ions ($l+m=1$) are the major contributor to ionic current in neat ionic liquids. In that case, conductivity obeys an even simpler equation 
\begin{align}
    \sigma=\frac{e^2c_{salt}}{k_B T}\left(\alpha_+ D_+ + \alpha_- D_-  \right)
    \label{eq:sig2}
\end{align}
where $D_\pm$ is the diffusivity of the free cation or anion. The ability  to use eq. \eqref{eq:sig2} instead of \eqref{eq:sig1} depends on if we can neglect the cluster contribution to the ionic strength of the electrolyte (Eq.\eqref{eq:LI1} vs. \eqref{eq:LI2}). Our model allows us to predict the ionic strength, and decompose the respective contributions from free ions and clusters. In the left panel of Fig. \ref{fig:trans}, we plot the dimensionless ionic strength (non-dimensionalized by the overall salt concentration) using both Eq. \eqref{eq:LI1} and \eqref{eq:LI2}. The dashed line in  Fig. \ref{fig:trans}, represents the free ion contribution to the ionic strength, while the solid curve represents the total ionic strength. It is apparent that free ions dominate the ionic strength of the electrolyte no matter the salt concentration, at least for the model parameters given in caption of the figure. There is a small region where there is a perceptible contribution of ion clusters to the ionic strength, which corresponds to concentrations very close to the gel point of the electrolyte ($x=0.18$). Nonetheless, it appears as if Eq.~\eqref{eq:sig2} could suffice for modelling the conductivity of our fictional electrolyte.

Within our model, the concentration of free ions can display nonlinear or even non-monotonic behavior as a function of overall salt concentration. At high concentrations adding more salt can actually decrease the amount of free ions in solution. This can be seen in Fig.~\ref{fig:trans}, where we have plotted the concentration of free ions as a function of the mole fraction of salt. Here, we have used the parameters listed in the figure caption to generate the curves, which are the same parameters that have been used for the majority of the paper. The non-monotonic concentration of free ions is likely largely responsible for the non-monotonic ionic conductivity that have been widely observed for concentrated electrolytes \cite{lobo1989handbook} or ionic liquid solvent mixtures \cite{stoppa2009conductivities,li2007effect,chaban2012acetonitrile}. Though we must note that $D_\pm$ is also expected to have a large role in the concentration dependence of ionic conductivity.

One interesting aspect of this model, is that for asymmetrically associating ions, we obtain different fractions of free anions and cations, as seen in Fig.~\ref{fig:trans}. If the free anions and cations have equivalent diffusivities, then we can write the transference number as:
\begin{align}
    t_\pm=\frac{\alpha_\pm}{\alpha_++\alpha_-}
    \label{eq:t}
\end{align}
Thus, assuming free ions are the dominant carrier of charge, our model would predict asymmetric transference numbers ($t_{\pm} \neq 0.5$) for salts with ions that do not have equivalent functionalities, as seen in the inset of Fig.~\ref{fig:trans}. In general for binary mixtures of monovalent salts, the ion with more association sites will have a higher fraction of free ions than the ion with less association sites. The reason for this is quite subtle when examining the expressions for $\alpha_+$ and $\alpha_-$ [Eqs.~\eqref{eq:freecat} \& \eqref{eq:freean}].  Ultimately, when $f_+>f_-$, for a fixed number of ion--counter-ion associations, the cations need less molecules to form those associations than anions. Thus, more cations will be free than anions, and we would observe that $t_+>0.5$ and $t_-<0.5$.

\subsection{Rheological Implications}
Gel-forming electrolytes should display intriguing viscoelastic properties. In polymers, typically the presence of gel is detected by probing the rheology of the mixture. At the gel point, the viscosity is expected to diverge and the equilibrium shear modulus is expected to become finite~\cite{flory1953principles}. Because our gel is composed of reversible physical associations between ions, we do not expect the viscosity to formally diverge. Nonetheless, thermoreversible gels should display a finite shear modulus. Flory related the equilibrium shear modulus, $G_e$ to the fraction of gel in the mixture for tetrafunctional associating polymer strands \cite{flory1953principles}. This was later extended for any $f$ functional associating polymer strand by Nijenhuis. This extension would be applicable for our case of ion gels if the ions have the equal functionalities, $f$:
\begin{align}
    G_e=-2cRT\left( \frac{\ln w^{sol}_\pm}{1-w^{sol}_\pm}\cdot \frac{1-(w^{sol}_\pm)^{f/2}}{1-(w^{sol}_\pm)^{f/2-1}}\cdot \frac{f-2}{f}+1\right)(1-w^{sol}_\pm)
    \label{eq:Geq}
\end{align}
where $c$ is the molar concentration of salt, and $R$ is the gas constant. Eq.~\eqref{eq:Geq} predicts, as expected, that $G_e$ will be zero prior to the formation of gel, and then increase with increasing gel fraction. If we again operate within the sticky symmetric ion approximation, then we can see how the equilibrium shear modulus is modulated by the electrolyte concentration (via $\phi_\pm$) and strength (via $\tilde{\Lambda}$) 
\begin{figure}[hbt!]
    \centering
    \includegraphics[width=0.49\textwidth]{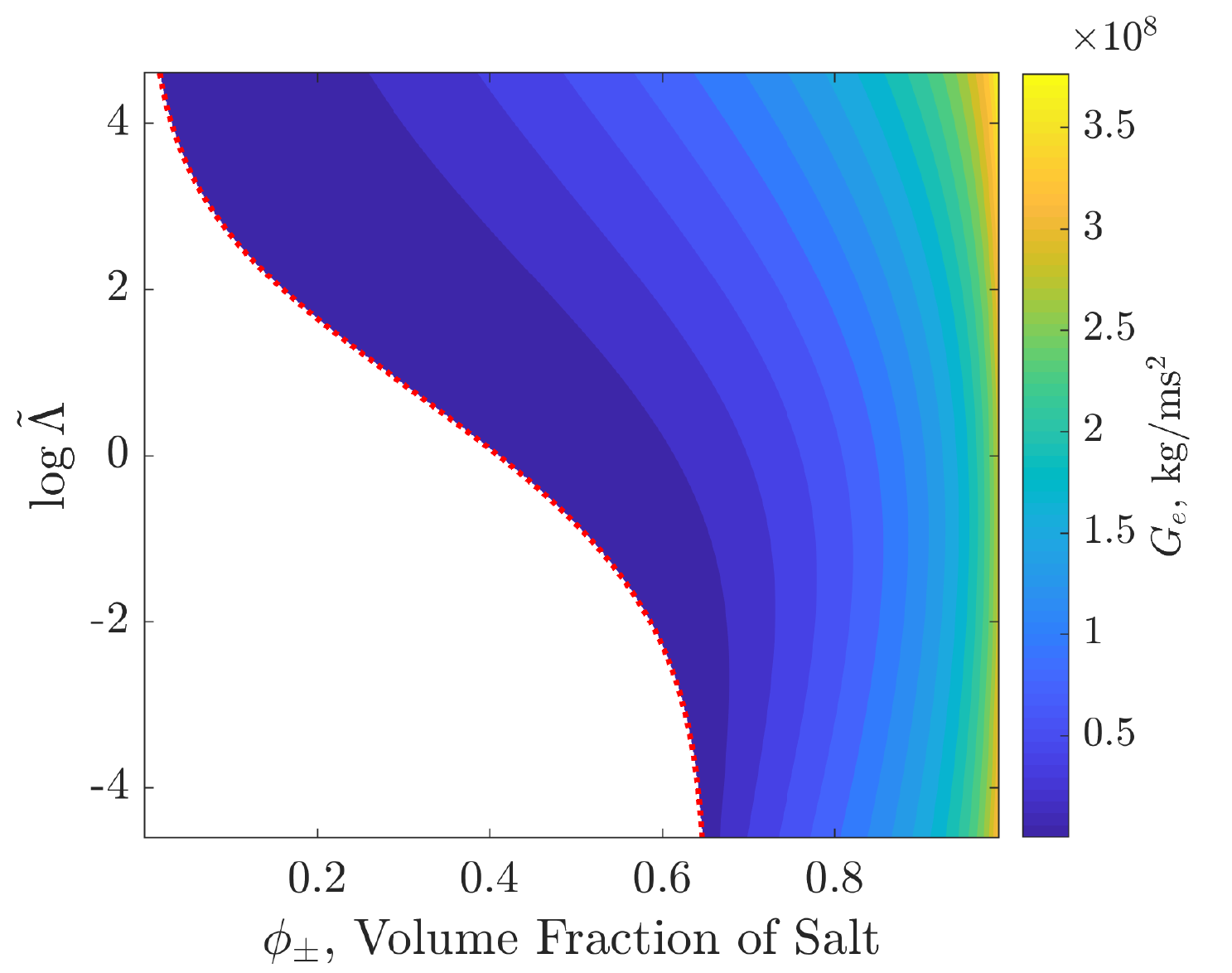}
    \caption{A Contour map of the equilibrium shear modulus, $G_e$, as a function of $\tilde{\Lambda}$ and $\phi_\pm$. The Red dotted line denotes the critical gel boundary. The region of white denotes the pre-gel region, where the equilibrium shear modulus is exactly 0. The diagram was generated within the sticky symmetric ion approximation (see Appendix) for $\xi_+=\xi_-=5$, and $f_+=f_-=4$.}
    \label{fig:eqmshear}
\end{figure}

In Fig.~\ref{fig:eqmshear}, we display a contour map of the equilibrium shear modulus using Eq.~\eqref{eq:Geq} as functions of $\phi_\pm$ and $\log \tilde{\Lambda}$. The shear modulus is predictably zero (white region), when there is no ionic gel present in the electrolyte, and becomes finite beyond the gel point. Additionally, the shear modulus increases monotonically with increasing gel fraction. As such, it increases with concentration, but tends to decrease as the electrolyte becomes weaker ($\log \Lambda$ decreases). There is a subtlety to this statement, as can be seen by the non-monotonicity in the contours of $G_e$ at high salt concentrations and low $\tilde{\Lambda}$. For very strong electrolytes ($\log \tilde{\Lambda}<2$) and for a given volume fraction of salt that is beyond the gel boundary, by increasing $\tilde{\Lambda}$ (increasing the affinity for ion association), the gel fraction actually decreases. This is quite counter-intuitive because we would expect more gel when the affinity for ion association is stronger. However, within the gel, the model allows for intramolecular loops. For very weakly associating salts, we would expect the gel to contain minimal intramolecular loops. Increasing the affinity for ion association, would induce more intramolecular loops, which would actually free up ions from the gel. The gel would simultaneously contain more ion--counter-ion associations, with less ions. Thus, in this regime, increasing the affinity for ion association actually decreases the shear modulus. 

This, subtlety should not obscure the result that when an ionic gel is present, the mixture may display viscoelastic properties. Interestingly, viscoelastic properties have been indeed observed experimentally for some common imidizolium-based ionic liquids~\cite{makino2008viscoelastic}. In that work, the equilibrium shear modulus of elasticity decreases as a function of temperature, which would be consistent with the melting and destruction of an ionic gel. 

There is limited literature on this topic, however. Furthermore, Ref.~\citenum{makino2008viscoelastic} does not actually attempt to compute a gel point. Perhaps the most reliable method for determining the exact gel point was introduced by Winter and Chambon~\cite{winter1986analysis}. They determined the gel point to occur at the intersection of the dynamic loss and the storage moduli for an oscillatory shear experiment. This could be a route to experimentally probe gelation in concentrated electrolytes. 

\section{Conclusion}

Here we have cast the mean-field theory of thermoreversible association and gelation from polymer physics into the context of electrolytes. The presented theory allows complicated, branched ionic aggregates to be included in models of concentrated electrolytes. Previously, ion pairs have only (typically) been included in models  of ionic association for concentrated electrolytes. However, these simple models break down when the system becomes sufficiently concentrated, which motivated the presented theory. More specifically, we developed a model for aggregation and gelation between cations, anions and solvent molecules, with alternating cation-anion aggregates/gel and solvent molecules decorating this ``ionic backbone". The theory can describe the composition of an electrolyte as a function of salt concentration and temperature, where different ionic states (free, aggregated, or gelled) dominate depending on the conditions. Higher salt concentrations favor the formation of a percolating gel, while smaller salt concentrations tend to have only free ions or small aggregates; between these extremes exists a narrow domain where finite aggregates dominate in the electrolyte. Note that the developed theory is best applied to electrolytes with ``complicated" ions, such as ionic liquids and water in salt electrolytes, where crystaline solids cannot precipitate out. Moreover, since model is a mean-field theory that neglects any loops in ionic clusters, it cannot describe the strongly correlated “spin glass” ordering recently discovered in simulations of ionic liquids, which transitions to long-range order in ionic crystals for “simple” ions \cite{levy2019spin}.  
Nevertheless, motivated by the success of mean-field theories from polymer physics, we expect that our model will have implications for the bulk thermodynamic, transport, and rheological properties of super-concentrated electrolytes, which can be probed experimentally and used guide the design of these dense ionic fluids. 

It is possible to extend our approach to interfacial properties as well. Specifically, it has already been shown that  understanding the partitioning of ions\cite{goodwin2017mean,goodwin2017underscreening,avni2020charge} and solvent\cite{mceldrew2018} between free and bound states has already been shown to be extremely enlightening in modelling the electrical double layer (EDL) of ionic liquids and WiSE's. Our model provides a more detailed and generalized  picture of the states of of ions or solvent, which may be leveraged to develop more accurate and general models of the EDL. EDL properties will also influence electrokinetic phenomena and may help to explain many puzzling observations, such as flow reversals in concentrated electrolytes \cite{Bazant2009a,storey2012effects}.  As with polymers under confinement, it will also be interesting to extend our model to nanopores, where cluster sizes are influenced by geometrical constraints. 

\begin{acknowledgments}
All authors would like to acknowledge the Imperial College-MIT seed fund. MM and MXB acknowledge support  from a Amar G. Bose Research Grant. 
ZG was supported through a studentship in the Centre for Doctoral Training on Theory and Simulation of Materials at Imperial College London funded by the EPSRC (EP/L015579/1) and from the Thomas Young Centre under grant number TYC-101. S.B. was also supported by the National Natural Science Foundation of China (51876072) and the financial support from the China Scholarship Council. A.K. would like to acknowledge the research grant by the Leverhulme Trust (RPG-2016- 223). 
\end{acknowledgments}

\appendix

\section{Sticky Symmetric Ions}

An analytical solution for the association probabilities is possible, if we make three primary assumptions. First, we must assume the ions to have equal number of association sites ($f_+=f_-=f$). Second, we assume that the electrostatic contribution to ion association is negligible. This implies that $\Lambda_{+-}^{el}=1$ or equivalently that $\Lambda_{+-}=\Lambda^{\theta}_{+-}$. This approximation is motivated by that $\Lambda_{+-}^{el}$ is mostly on the order of 1 (recall the right panel in Fig. \ref{fig:elec}), and only changes significantly with very dilute free ion concentrations. For our final assumption, we require that the ions do not contain any open association sites: they are either associated to counter-ions or solvent molecules. The cluster distributions is then limited to aggregates containing $fl-l-m+1$ solvent molecules attached to cations and $fm-m-l+1$ solvent molecules attached to anions. Thus, we have the cluster distribution
\begin{align}
    c_{lm}=W_{lm}\Lambda_{+-}^{l+m-1}\Lambda_{+0}^{fl-l+m+1}\Lambda_{-0}^{fm-m+l+1}\psi_{100}^l\psi_{010}^m\phi_{001}^{f(m+l)+2} + \Phi_{001}
    \label{eq:distss}
\end{align}
with a slightly modified correction for when $l=m=0$
\begin{align}
    \Phi_{001} =  \phi_{001}\left[1-\phi_{001}/\tilde{\Lambda} \right]\delta_{l,0}\delta_{m,0}
\end{align}
where $\tilde{\Lambda}=\Lambda_{+-}/\Lambda_{+0}\Lambda_{-0}$. We can rewrite Eq. \eqref{eq:distss} in the following manner:
\begin{align}
    c_{lm}=\frac{\mathcal{K}}{2}W_{lm}\left(\frac{p_{-+}}{1-p_{-+}}(1-p_{+-})^{f_+-1}\right)^{l}\left(\frac{p_{+-}}{1-p_{+-}}(1-p_{-+})^{f_--1}\right)^{m}
\end{align}
with an identical definition for $\mathcal{K}$ as previously written in the main text. The association probabilities are governed by the following equations:
\begin{align}
    p_{+-}=1-p_{+0}
\end{align}
\begin{align}
    p_{-+}=1-p_{-0}
\end{align}
\begin{align}
    p_{+-}=p_{-+}
\end{align}
\begin{align}
     \psi_{+} p_{+0}=\phi_0 p_{0+}
\end{align}
\begin{align}
    \psi_- p_{-0}=\phi_0 p_{0-}
\end{align}
\begin{align}
    \frac{\tilde{\Lambda}}{\psi_{+}}=\frac{p_{+-}(1-p_{0+}-p_{0-})}{p_{0+}p_{0-}}
\end{align}
where $\psi_{\pm}=f \phi_+/\xi_+=f \phi_-/\xi_-$ is the concentration of cationic or anionic association sites. Solving this system of equations we obtain the solution:
\begin{align}
    p_{+-}=p_{-+}=\frac{2 \tilde{\Lambda} \psi_{\pm} \phi_0\left(\phi_0-2\psi_{\pm}-\sqrt{4 \tilde{\Lambda}\psi_{\pm}+(\phi_0-2\psi_\pm)^2} \right)}{2 \psi_{\pm}(\tilde{\Lambda}-2\phi_0)}
\end{align}
\begin{align}
    p_{+0}=p_{-0}=\frac{2 \psi_{\pm}}{2 \psi_{\pm}+\phi_0+\sqrt{4 \tilde{\Lambda}\psi_{\pm}+(\phi_0-2\psi_\pm)^2}}
\end{align}
\begin{align}
    p_{0+}=p_{0-}=\frac{\phi_0-2\psi_{\pm}-\sqrt{4 \tilde{\Lambda}\psi_{\pm}+(\phi_0-2\psi_\pm)^2} }{2 \psi_{\pm}(\tilde{\Lambda}-2\phi_0)}
\end{align}
The approximations we have made yield association probabilities that do not distinguish between anions and cations. We should note that taking the limit of this approximation for a salt volume fraction of 1 (ionic liquid/solid limit) yields the trivial solution that $p_{\pm\mp}=1$. Thus, we have a fully connected alternating ion network, somewhat resembling an ionic crystal. Thus, there will be no finite ion clusters and certainly no free ions that can conduct ionic current. This actually consistent with behavior we would expect for many salts, which do not conduct charge without solvent present to induce dissociation. Thus, we ionic liquid salts would not be captured with this sticky symmetric ion approximation. 

The gelation criterion for systems within the sticky symmetric ion approximation is identical to that of the general theory, except the symmetry of the ion allows for simplified expression:
\begin{align}
    p_{\pm\mp}=1/(1-f)
\end{align}
 The post-gel relations will be slightly different to that for the general theory. We can write the fraction of free ions equivalently with overall probabilities and sol probabilities:
 \begin{align}
     \phi_+(1-p_{+-})^f=\phi^{sol}_+(1-p^{sol}_{+-})^f
 \end{align}
 \begin{align}
     \phi_-(1-p_{-+})^f=\phi^{sol}_-(1-p^{sol}_{-+})^f
 \end{align}
Similarly for free solvent molecules we have
\begin{align}
     \phi_0(1-p_{0+}-p_{0-})=\phi^{sol}_0(1-p^{sol}_{0+}-p^{sol}_{0-})
\end{align}
The sticky symmetric ion assumptions are also valid for sol probabilities
\begin{align}
    p^{sol}_{\pm0}=1-p^{sol}_{\pm\mp}
\end{align}
\begin{align}
    \frac{\tilde{\Lambda}}{\psi^{sol}_{+}}=\frac{p^{sol}_{+-}(1-p^{sol}_{0+}-p^{sol}_{0-})}{p^{sol}_{0+}p^{sol}_{0-}}
\end{align}
And finally we have the conservation of the associations made in the sol 
 \begin{align}
     \psi^{sol}_{+} p^{sol}_{+0}=\phi^{sol}_0 p^{sol}_{0+}
\end{align}
\begin{align}
    \psi^{sol}_- p^{sol}_{-0}=\phi^{sol}_0 p^{sol}_{0-}
\end{align}
\begin{align}
    p^{sol}_{+-}=p^{sol}_{-+}
\end{align}

Thus, we have 9 equations and 9 unknowns, exactly analogous to the general case. One thing to note is that the symmetry of the system implies that many of these equations will redundant. For sticky symmetric ions, $p^{sol}_{+-}=p^{sol}_{-+}$, $p^{sol}_{+0}=p^{sol}_{-0}$, $p^{sol}_{0+}=p^{sol}_{0-}$,and $\phi^{sol}_{+}=\phi^{sol}_{-}$.

\bibliography{gel_paper.bib}

\begin{thebibliography}{95}%
\makeatletter
\providecommand \@ifxundefined [1]{%
 \@ifx{#1\undefined}
}%
\providecommand \@ifnum [1]{%
 \ifnum #1\expandafter \@firstoftwo
 \else \expandafter \@secondoftwo
 \fi
}%
\providecommand \@ifx [1]{%
 \ifx #1\expandafter \@firstoftwo
 \else \expandafter \@secondoftwo
 \fi
}%
\providecommand \natexlab [1]{#1}%
\providecommand \enquote  [1]{``#1''}%
\providecommand \bibnamefont  [1]{#1}%
\providecommand \bibfnamefont [1]{#1}%
\providecommand \citenamefont [1]{#1}%
\providecommand \href@noop [0]{\@secondoftwo}%
\providecommand \href [0]{\begingroup \@sanitize@url \@href}%
\providecommand \@href[1]{\@@startlink{#1}\@@href}%
\providecommand \@@href[1]{\endgroup#1\@@endlink}%
\providecommand \@sanitize@url [0]{\catcode `\\12\catcode `\$12\catcode
  `\&12\catcode `\#12\catcode `\^12\catcode `\_12\catcode `\%12\relax}%
\providecommand \@@startlink[1]{}%
\providecommand \@@endlink[0]{}%
\providecommand \url  [0]{\begingroup\@sanitize@url \@url }%
\providecommand \@url [1]{\endgroup\@href {#1}{\urlprefix }}%
\providecommand \urlprefix  [0]{URL }%
\providecommand \Eprint [0]{\href }%
\providecommand \doibase [0]{http://dx.doi.org/}%
\providecommand \selectlanguage [0]{\@gobble}%
\providecommand \bibinfo  [0]{\@secondoftwo}%
\providecommand \bibfield  [0]{\@secondoftwo}%
\providecommand \translation [1]{[#1]}%
\providecommand \BibitemOpen [0]{}%
\providecommand \bibitemStop [0]{}%
\providecommand \bibitemNoStop [0]{.\EOS\space}%
\providecommand \EOS [0]{\spacefactor3000\relax}%
\providecommand \BibitemShut  [1]{\csname bibitem#1\endcsname}%
\let\auto@bib@innerbib\@empty
\bibitem [{\citenamefont {Harned}, \citenamefont {Owen},\ and\ \citenamefont
  {King}(1959)}]{harned1959physical}%
  \BibitemOpen
  \bibfield  {author} {\bibinfo {author} {\bibfnamefont {H.~S.}\ \bibnamefont
  {Harned}}, \bibinfo {author} {\bibfnamefont {B.~B.}\ \bibnamefont {Owen}}, \
  and\ \bibinfo {author} {\bibfnamefont {C.}~\bibnamefont {King}},\ }\bibfield
  {title} {\enquote {\bibinfo {title} {The physical chemistry of electrolytic
  solutions},}\ }\href@noop {} {\bibfield  {journal} {\bibinfo  {journal}
  {Journal of The Electrochemical Society}\ }\textbf {\bibinfo {volume}
  {106}},\ \bibinfo {pages} {15C--15C} (\bibinfo {year} {1959})}\BibitemShut
  {NoStop}%
\bibitem [{\citenamefont {Zwanikken}\ and\ \citenamefont {van
  Roij}(2009)}]{IBL}%
  \BibitemOpen
  \bibfield  {author} {\bibinfo {author} {\bibfnamefont {J.}~\bibnamefont
  {Zwanikken}}\ and\ \bibinfo {author} {\bibfnamefont {R.}~\bibnamefont {van
  Roij}},\ }\bibfield  {title} {\enquote {\bibinfo {title} {Inflation of the
  screening length induced by bjerrum pairs},}\ }\href@noop {} {\bibfield
  {journal} {\bibinfo  {journal} {J. Phys. Condens. Matter}\ }\textbf {\bibinfo
  {volume} {41}},\ \bibinfo {pages} {424102} (\bibinfo {year}
  {2009})}\BibitemShut {NoStop}%
\bibitem [{\citenamefont {Bjerrum}(1926)}]{bjerrum1926k}%
  \BibitemOpen
  \bibfield  {author} {\bibinfo {author} {\bibfnamefont {N.}~\bibnamefont
  {Bjerrum}},\ }\href@noop {} {\enquote {\bibinfo {title} {K. danske vidensk.
  selsk.}}\ } (\bibinfo {year} {1926})\BibitemShut {NoStop}%
\bibitem [{\citenamefont {Marcus}\ and\ \citenamefont
  {Hefter}(2006)}]{marcus2006ion}%
  \BibitemOpen
  \bibfield  {author} {\bibinfo {author} {\bibfnamefont {Y.}~\bibnamefont
  {Marcus}}\ and\ \bibinfo {author} {\bibfnamefont {G.}~\bibnamefont
  {Hefter}},\ }\bibfield  {title} {\enquote {\bibinfo {title} {Ion pairing},}\
  }\href@noop {} {\bibfield  {journal} {\bibinfo  {journal} {Chemical reviews}\
  }\textbf {\bibinfo {volume} {106}},\ \bibinfo {pages} {4585--4621} (\bibinfo
  {year} {2006})}\BibitemShut {NoStop}%
\bibitem [{\citenamefont {Kraus}\ and\ \citenamefont
  {Fuoss}(1933)}]{kraus1933_1}%
  \BibitemOpen
  \bibfield  {author} {\bibinfo {author} {\bibfnamefont {C.~A.}\ \bibnamefont
  {Kraus}}\ and\ \bibinfo {author} {\bibfnamefont {R.~M.}\ \bibnamefont
  {Fuoss}},\ }\bibfield  {title} {\enquote {\bibinfo {title} {Properties of
  electrolytic solutions. i. conductance as influenced by the dielectric
  constant of the solvent medium1},}\ }\href@noop {} {\bibfield  {journal}
  {\bibinfo  {journal} {Journal of the American Chemical Society}\ }\textbf
  {\bibinfo {volume} {55}},\ \bibinfo {pages} {21--36} (\bibinfo {year}
  {1933})}\BibitemShut {NoStop}%
\bibitem [{\citenamefont {Fuoss}\ and\ \citenamefont
  {Kraus}(1933{\natexlab{a}})}]{fuoss1933_4}%
  \BibitemOpen
  \bibfield  {author} {\bibinfo {author} {\bibfnamefont {R.~M.}\ \bibnamefont
  {Fuoss}}\ and\ \bibinfo {author} {\bibfnamefont {C.~A.}\ \bibnamefont
  {Kraus}},\ }\bibfield  {title} {\enquote {\bibinfo {title} {Properties of
  electrolytic solutions. iv. the conductance minimum and the formation of
  triple ions due to the action of coulomb forces1},}\ }\href@noop {}
  {\bibfield  {journal} {\bibinfo  {journal} {Journal of the American Chemical
  Society}\ }\textbf {\bibinfo {volume} {55}},\ \bibinfo {pages} {2387--2399}
  (\bibinfo {year} {1933}{\natexlab{a}})}\BibitemShut {NoStop}%
\bibitem [{\citenamefont {Fuoss}\ and\ \citenamefont
  {Kraus}(1933{\natexlab{b}})}]{fuoss1933_9}%
  \BibitemOpen
  \bibfield  {author} {\bibinfo {author} {\bibfnamefont {R.~M.}\ \bibnamefont
  {Fuoss}}\ and\ \bibinfo {author} {\bibfnamefont {C.~A.}\ \bibnamefont
  {Kraus}},\ }\bibfield  {title} {\enquote {\bibinfo {title} {Properties of
  electrolytic solutions. ix. conductance of some salts in benzene},}\
  }\href@noop {} {\bibfield  {journal} {\bibinfo  {journal} {Journal of the
  American Chemical Society}\ }\textbf {\bibinfo {volume} {55}},\ \bibinfo
  {pages} {3614--3620} (\bibinfo {year} {1933}{\natexlab{b}})}\BibitemShut
  {NoStop}%
\bibitem [{\citenamefont {Barthel}\ \emph {et~al.}(2000)\citenamefont
  {Barthel}, \citenamefont {Krienke}, \citenamefont {Holovko}, \citenamefont
  {Kapko},\ and\ \citenamefont {Protsykevich}}]{barthel2000application}%
  \BibitemOpen
  \bibfield  {author} {\bibinfo {author} {\bibfnamefont {J.}~\bibnamefont
  {Barthel}}, \bibinfo {author} {\bibfnamefont {H.}~\bibnamefont {Krienke}},
  \bibinfo {author} {\bibfnamefont {M.}~\bibnamefont {Holovko}}, \bibinfo
  {author} {\bibfnamefont {V.}~\bibnamefont {Kapko}}, \ and\ \bibinfo {author}
  {\bibfnamefont {I.}~\bibnamefont {Protsykevich}},\ }\bibfield  {title}
  {\enquote {\bibinfo {title} {The application of the associative mean
  spherical approximation in the theory of nonaqueous electrolyte solutions},}\
  }\href@noop {} {\bibfield  {journal} {\bibinfo  {journal} {Condensed Matter
  Physics}\ } (\bibinfo {year} {2000})}\BibitemShut {NoStop}%
\bibitem [{\citenamefont {Suo}\ \emph {et~al.}(2013)\citenamefont {Suo},
  \citenamefont {Hu}, \citenamefont {Li}, \citenamefont {Armand},\ and\
  \citenamefont {Chen}}]{Suo2013}%
  \BibitemOpen
  \bibfield  {author} {\bibinfo {author} {\bibfnamefont {L.}~\bibnamefont
  {Suo}}, \bibinfo {author} {\bibfnamefont {Y.-S.}\ \bibnamefont {Hu}},
  \bibinfo {author} {\bibfnamefont {H.}~\bibnamefont {Li}}, \bibinfo {author}
  {\bibfnamefont {M.}~\bibnamefont {Armand}}, \ and\ \bibinfo {author}
  {\bibfnamefont {L.}~\bibnamefont {Chen}},\ }\bibfield  {title} {\enquote
  {\bibinfo {title} {{A new class of Solvent-in-Salt electrolyte for
  high-energy rechargeable metallic lithium batteries}},}\ }\href {\doibase
  10.1038/ncomms2513} {\bibfield  {journal} {\bibinfo  {journal} {Nat.
  Commun.}\ }\textbf {\bibinfo {volume} {4}},\ \bibinfo {pages} {1481}
  (\bibinfo {year} {2013})}\BibitemShut {NoStop}%
\bibitem [{\citenamefont {Sodeyama}\ \emph {et~al.}(2014)\citenamefont
  {Sodeyama}, \citenamefont {Yamada}, \citenamefont {Aikawa}, \citenamefont
  {Yamada},\ and\ \citenamefont {Tateyama}}]{Sodeyama2014}%
  \BibitemOpen
  \bibfield  {author} {\bibinfo {author} {\bibfnamefont {K.}~\bibnamefont
  {Sodeyama}}, \bibinfo {author} {\bibfnamefont {Y.}~\bibnamefont {Yamada}},
  \bibinfo {author} {\bibfnamefont {K.}~\bibnamefont {Aikawa}}, \bibinfo
  {author} {\bibfnamefont {A.}~\bibnamefont {Yamada}}, \ and\ \bibinfo {author}
  {\bibfnamefont {Y.}~\bibnamefont {Tateyama}},\ }\bibfield  {title} {\enquote
  {\bibinfo {title} {{Sacrificial Anion Reduction Mechanism for Electrochemical
  Stability Improvement in Highly Concentrated Li-Salt Electrolyte}},}\ }\href
  {\doibase 10.1021/jp501178n} {\bibfield  {journal} {\bibinfo  {journal} {J.
  Phys. Chem. C}\ }\textbf {\bibinfo {volume} {118}},\ \bibinfo {pages}
  {14091--14097} (\bibinfo {year} {2014})}\BibitemShut {NoStop}%
\bibitem [{\citenamefont {Suo}\ \emph {et~al.}(2015)\citenamefont {Suo},
  \citenamefont {Borodin}, \citenamefont {Gao}, \citenamefont {Olguin},
  \citenamefont {Ho}, \citenamefont {Fan}, \citenamefont {Luo}, \citenamefont
  {Wang},\ and\ \citenamefont {Xu}}]{Suo2015}%
  \BibitemOpen
  \bibfield  {author} {\bibinfo {author} {\bibfnamefont {L.}~\bibnamefont
  {Suo}}, \bibinfo {author} {\bibfnamefont {O.}~\bibnamefont {Borodin}},
  \bibinfo {author} {\bibfnamefont {T.}~\bibnamefont {Gao}}, \bibinfo {author}
  {\bibfnamefont {M.}~\bibnamefont {Olguin}}, \bibinfo {author} {\bibfnamefont
  {J.}~\bibnamefont {Ho}}, \bibinfo {author} {\bibfnamefont {X.}~\bibnamefont
  {Fan}}, \bibinfo {author} {\bibfnamefont {C.}~\bibnamefont {Luo}}, \bibinfo
  {author} {\bibfnamefont {C.}~\bibnamefont {Wang}}, \ and\ \bibinfo {author}
  {\bibfnamefont {K.}~\bibnamefont {Xu}},\ }\bibfield  {title} {\enquote
  {\bibinfo {title} {{"Water-in-salt" electrolyte enables high-voltage aqueous
  lithium-ion chemistries.}}}\ }\href {\doibase 10.1126/science.aab1595}
  {\bibfield  {journal} {\bibinfo  {journal} {Science}\ }\textbf {\bibinfo
  {volume} {350}},\ \bibinfo {pages} {938--43} (\bibinfo {year}
  {2015})}\BibitemShut {NoStop}%
\bibitem [{\citenamefont {Smith}\ and\ \citenamefont {Dunn}(2015)}]{Smith2015}%
  \BibitemOpen
  \bibfield  {author} {\bibinfo {author} {\bibfnamefont {L.}~\bibnamefont
  {Smith}}\ and\ \bibinfo {author} {\bibfnamefont {B.}~\bibnamefont {Dunn}},\
  }\bibfield  {title} {\enquote {\bibinfo {title} {{Opening the window for
  aqueous electrolytes}},}\ }\href {\doibase 10.1126/science.aad5575}
  {\bibfield  {journal} {\bibinfo  {journal} {Science}\ }\textbf {\bibinfo
  {volume} {350}},\ \bibinfo {pages} {918--918} (\bibinfo {year}
  {2015})}\BibitemShut {NoStop}%
\bibitem [{\citenamefont {Yamada}\ \emph {et~al.}(2016)\citenamefont {Yamada},
  \citenamefont {Usui}, \citenamefont {Sodeyama}, \citenamefont {Ko},
  \citenamefont {Tateyama},\ and\ \citenamefont {Yamada}}]{Yamada2016}%
  \BibitemOpen
  \bibfield  {author} {\bibinfo {author} {\bibfnamefont {Y.}~\bibnamefont
  {Yamada}}, \bibinfo {author} {\bibfnamefont {K.}~\bibnamefont {Usui}},
  \bibinfo {author} {\bibfnamefont {K.}~\bibnamefont {Sodeyama}}, \bibinfo
  {author} {\bibfnamefont {S.}~\bibnamefont {Ko}}, \bibinfo {author}
  {\bibfnamefont {Y.}~\bibnamefont {Tateyama}}, \ and\ \bibinfo {author}
  {\bibfnamefont {A.}~\bibnamefont {Yamada}},\ }\bibfield  {title} {\enquote
  {\bibinfo {title} {{Hydrate-melt electrolytes for high-energy-density aqueous
  batteries}},}\ }\href {\doibase 10.1038/nenergy.2016.129} {\bibfield
  {journal} {\bibinfo  {journal} {Nat. Energy}\ }\textbf {\bibinfo {volume}
  {1}},\ \bibinfo {pages} {16129} (\bibinfo {year} {2016})}\BibitemShut
  {NoStop}%
\bibitem [{\citenamefont {Wang}\ \emph {et~al.}(2016)\citenamefont {Wang},
  \citenamefont {Yamada}, \citenamefont {Sodeyama}, \citenamefont {Chiang},
  \citenamefont {Tateyama},\ and\ \citenamefont {Yamada}}]{Wang2016}%
  \BibitemOpen
  \bibfield  {author} {\bibinfo {author} {\bibfnamefont {J.}~\bibnamefont
  {Wang}}, \bibinfo {author} {\bibfnamefont {Y.}~\bibnamefont {Yamada}},
  \bibinfo {author} {\bibfnamefont {K.}~\bibnamefont {Sodeyama}}, \bibinfo
  {author} {\bibfnamefont {C.~H.}\ \bibnamefont {Chiang}}, \bibinfo {author}
  {\bibfnamefont {Y.}~\bibnamefont {Tateyama}}, \ and\ \bibinfo {author}
  {\bibfnamefont {A.}~\bibnamefont {Yamada}},\ }\bibfield  {title} {\enquote
  {\bibinfo {title} {{Superconcentrated electrolytes for a high-voltage
  lithium-ion battery}},}\ }\href {\doibase 10.1038/ncomms12032} {\bibfield
  {journal} {\bibinfo  {journal} {Nat. Commun.}\ }\textbf {\bibinfo {volume}
  {7}},\ \bibinfo {pages} {12032} (\bibinfo {year} {2016})}\BibitemShut
  {NoStop}%
\bibitem [{\citenamefont {Gambou-Bosca}\ and\ \citenamefont
  {B{\'{e}}langer}(2016)}]{Gambou-Bosca2016}%
  \BibitemOpen
  \bibfield  {author} {\bibinfo {author} {\bibfnamefont {A.}~\bibnamefont
  {Gambou-Bosca}}\ and\ \bibinfo {author} {\bibfnamefont {D.}~\bibnamefont
  {B{\'{e}}langer}},\ }\bibfield  {title} {\enquote {\bibinfo {title}
  {{Electrochemical characterization of MnO2-based composite in the presence of
  salt-in-water and water-in-salt electrolytes as electrode for electrochemical
  capacitors}},}\ }\href {\doibase 10.1016/j.jpowsour.2016.04.088} {\bibfield
  {journal} {\bibinfo  {journal} {Journal of Power Sources}\ }\textbf {\bibinfo
  {volume} {326}} (\bibinfo {year} {2016}),\
  10.1016/j.jpowsour.2016.04.088}\BibitemShut {NoStop}%
\bibitem [{\citenamefont {Sun}\ \emph {et~al.}(2017)\citenamefont {Sun},
  \citenamefont {Suo}, \citenamefont {Wang}, \citenamefont {Eidson},
  \citenamefont {Yang}, \citenamefont {Han}, \citenamefont {Ma}, \citenamefont
  {Gao}, \citenamefont {Zhu},\ and\ \citenamefont {Wang}}]{Sun2017}%
  \BibitemOpen
  \bibfield  {author} {\bibinfo {author} {\bibfnamefont {W.}~\bibnamefont
  {Sun}}, \bibinfo {author} {\bibfnamefont {L.}~\bibnamefont {Suo}}, \bibinfo
  {author} {\bibfnamefont {F.}~\bibnamefont {Wang}}, \bibinfo {author}
  {\bibfnamefont {N.}~\bibnamefont {Eidson}}, \bibinfo {author} {\bibfnamefont
  {C.}~\bibnamefont {Yang}}, \bibinfo {author} {\bibfnamefont {F.}~\bibnamefont
  {Han}}, \bibinfo {author} {\bibfnamefont {Z.}~\bibnamefont {Ma}}, \bibinfo
  {author} {\bibfnamefont {T.}~\bibnamefont {Gao}}, \bibinfo {author}
  {\bibfnamefont {M.}~\bibnamefont {Zhu}}, \ and\ \bibinfo {author}
  {\bibfnamefont {C.}~\bibnamefont {Wang}},\ }\bibfield  {title} {\enquote
  {\bibinfo {title} {{“Water-in-Salt” electrolyte enabled LiMn2O4/TiS2
  Lithium-ion batteries}},}\ }\href {\doibase 10.1016/J.ELECOM.2017.07.016}
  {\bibfield  {journal} {\bibinfo  {journal} {Electrochem. Commun.}\ }\textbf
  {\bibinfo {volume} {82}},\ \bibinfo {pages} {71--74} (\bibinfo {year}
  {2017})}\BibitemShut {NoStop}%
\bibitem [{\citenamefont {Suo}\ \emph {et~al.}(2017)\citenamefont {Suo},
  \citenamefont {Borodin}, \citenamefont {Wang}, \citenamefont {Rong},
  \citenamefont {Sun}, \citenamefont {Fan}, \citenamefont {Xu}, \citenamefont
  {Schroeder}, \citenamefont {Cresce}, \citenamefont {Wang} \emph
  {et~al.}}]{suo2017water}%
  \BibitemOpen
  \bibfield  {author} {\bibinfo {author} {\bibfnamefont {L.}~\bibnamefont
  {Suo}}, \bibinfo {author} {\bibfnamefont {O.}~\bibnamefont {Borodin}},
  \bibinfo {author} {\bibfnamefont {Y.}~\bibnamefont {Wang}}, \bibinfo {author}
  {\bibfnamefont {X.}~\bibnamefont {Rong}}, \bibinfo {author} {\bibfnamefont
  {W.}~\bibnamefont {Sun}}, \bibinfo {author} {\bibfnamefont {X.}~\bibnamefont
  {Fan}}, \bibinfo {author} {\bibfnamefont {S.}~\bibnamefont {Xu}}, \bibinfo
  {author} {\bibfnamefont {M.~A.}\ \bibnamefont {Schroeder}}, \bibinfo {author}
  {\bibfnamefont {A.~V.}\ \bibnamefont {Cresce}}, \bibinfo {author}
  {\bibfnamefont {F.}~\bibnamefont {Wang}},  \emph {et~al.},\ }\bibfield
  {title} {\enquote {\bibinfo {title} {“water-in-salt” electrolyte makes
  aqueous sodium-ion battery safe, green, and long-lasting},}\ }\href@noop {}
  {\bibfield  {journal} {\bibinfo  {journal} {Advanced Energy Materials}\
  }\textbf {\bibinfo {volume} {7}},\ \bibinfo {pages} {1701189} (\bibinfo
  {year} {2017})}\BibitemShut {NoStop}%
\bibitem [{\citenamefont {Dong}\ \emph {et~al.}(2017)\citenamefont {Dong},
  \citenamefont {Yu}, \citenamefont {Ma}, \citenamefont {Bao}, \citenamefont
  {Truhlar}, \citenamefont {Wang},\ and\ \citenamefont {Xia}}]{Dong2017}%
  \BibitemOpen
  \bibfield  {author} {\bibinfo {author} {\bibfnamefont {X.}~\bibnamefont
  {Dong}}, \bibinfo {author} {\bibfnamefont {H.}~\bibnamefont {Yu}}, \bibinfo
  {author} {\bibfnamefont {Y.}~\bibnamefont {Ma}}, \bibinfo {author}
  {\bibfnamefont {J.~L.}\ \bibnamefont {Bao}}, \bibinfo {author} {\bibfnamefont
  {D.~G.}\ \bibnamefont {Truhlar}}, \bibinfo {author} {\bibfnamefont
  {Y.}~\bibnamefont {Wang}}, \ and\ \bibinfo {author} {\bibfnamefont
  {Y.}~\bibnamefont {Xia}},\ }\bibfield  {title} {\enquote {\bibinfo {title}
  {{All-Organic Rechargeable Battery with Reversibility Supported by
  “Water-in-Salt” Electrolyte}},}\ }\href {\doibase 10.1002/chem.201700063}
  {\bibfield  {journal} {\bibinfo  {journal} {Chemistry - A European Journal}\
  }\textbf {\bibinfo {volume} {23}} (\bibinfo {year} {2017}),\
  10.1002/chem.201700063}\BibitemShut {NoStop}%
\bibitem [{\citenamefont {Diederichsen}, \citenamefont {McShane},\ and\
  \citenamefont {McCloskey}(2017)}]{Diederichsen2017}%
  \BibitemOpen
  \bibfield  {author} {\bibinfo {author} {\bibfnamefont {K.~M.}\ \bibnamefont
  {Diederichsen}}, \bibinfo {author} {\bibfnamefont {E.~J.}\ \bibnamefont
  {McShane}}, \ and\ \bibinfo {author} {\bibfnamefont {B.~D.}\ \bibnamefont
  {McCloskey}},\ }\bibfield  {title} {\enquote {\bibinfo {title} {{Promising
  Routes to a High Li+ Transference Number Electrolyte for Lithium Ion
  Batteries}},}\ }\href {\doibase 10.1021/acsenergylett.7b00792} {\bibfield
  {journal} {\bibinfo  {journal} {ACS Energy Letters}\ }\textbf {\bibinfo
  {volume} {2}} (\bibinfo {year} {2017}),\
  10.1021/acsenergylett.7b00792}\BibitemShut {NoStop}%
\bibitem [{\citenamefont {Yang}\ \emph
  {et~al.}(2017{\natexlab{a}})\citenamefont {Yang}, \citenamefont {Chen},
  \citenamefont {Qing}, \citenamefont {Fan}, \citenamefont {Sun}, \citenamefont
  {von Cresce}, \citenamefont {Ding}, \citenamefont {Borodin}, \citenamefont
  {Vatamanu}, \citenamefont {Schroeder} \emph {et~al.}}]{yang2017a}%
  \BibitemOpen
  \bibfield  {author} {\bibinfo {author} {\bibfnamefont {C.}~\bibnamefont
  {Yang}}, \bibinfo {author} {\bibfnamefont {J.}~\bibnamefont {Chen}}, \bibinfo
  {author} {\bibfnamefont {T.}~\bibnamefont {Qing}}, \bibinfo {author}
  {\bibfnamefont {X.}~\bibnamefont {Fan}}, \bibinfo {author} {\bibfnamefont
  {W.}~\bibnamefont {Sun}}, \bibinfo {author} {\bibfnamefont {A.}~\bibnamefont
  {von Cresce}}, \bibinfo {author} {\bibfnamefont {M.~S.}\ \bibnamefont
  {Ding}}, \bibinfo {author} {\bibfnamefont {O.}~\bibnamefont {Borodin}},
  \bibinfo {author} {\bibfnamefont {J.}~\bibnamefont {Vatamanu}}, \bibinfo
  {author} {\bibfnamefont {M.~A.}\ \bibnamefont {Schroeder}},  \emph {et~al.},\
  }\bibfield  {title} {\enquote {\bibinfo {title} {4.0 v aqueous li-ion
  batteries},}\ }\href@noop {} {\bibfield  {journal} {\bibinfo  {journal}
  {Joule}\ }\textbf {\bibinfo {volume} {1}},\ \bibinfo {pages} {122--132}
  (\bibinfo {year} {2017}{\natexlab{a}})}\BibitemShut {NoStop}%
\bibitem [{\citenamefont {Yang}\ \emph
  {et~al.}(2017{\natexlab{b}})\citenamefont {Yang}, \citenamefont {Suo},
  \citenamefont {Borodin}, \citenamefont {Wang}, \citenamefont {Sun},
  \citenamefont {Gao}, \citenamefont {Fan}, \citenamefont {Hou}, \citenamefont
  {Ma}, \citenamefont {Amine}, \citenamefont {Xu},\ and\ \citenamefont
  {Wang}}]{Yang2017}%
  \BibitemOpen
  \bibfield  {author} {\bibinfo {author} {\bibfnamefont {C.}~\bibnamefont
  {Yang}}, \bibinfo {author} {\bibfnamefont {L.}~\bibnamefont {Suo}}, \bibinfo
  {author} {\bibfnamefont {O.}~\bibnamefont {Borodin}}, \bibinfo {author}
  {\bibfnamefont {F.}~\bibnamefont {Wang}}, \bibinfo {author} {\bibfnamefont
  {W.}~\bibnamefont {Sun}}, \bibinfo {author} {\bibfnamefont {T.}~\bibnamefont
  {Gao}}, \bibinfo {author} {\bibfnamefont {X.}~\bibnamefont {Fan}}, \bibinfo
  {author} {\bibfnamefont {S.}~\bibnamefont {Hou}}, \bibinfo {author}
  {\bibfnamefont {Z.}~\bibnamefont {Ma}}, \bibinfo {author} {\bibfnamefont
  {K.}~\bibnamefont {Amine}}, \bibinfo {author} {\bibfnamefont
  {K.}~\bibnamefont {Xu}}, \ and\ \bibinfo {author} {\bibfnamefont
  {C.}~\bibnamefont {Wang}},\ }\bibfield  {title} {\enquote {\bibinfo {title}
  {{Unique aqueous Li-ion/sulfur chemistry with high energy density and
  reversibility}},}\ }\href {\doibase 10.1073/pnas.1703937114} {\bibfield
  {journal} {\bibinfo  {journal} {Proceedings of the National Academy of
  Sciences of the United States of America}\ }\textbf {\bibinfo {volume} {114}}
  (\bibinfo {year} {2017}{\natexlab{b}}),\ 10.1073/pnas.1703937114}\BibitemShut
  {NoStop}%
\bibitem [{\citenamefont {Wang}\ \emph {et~al.}(2018)\citenamefont {Wang},
  \citenamefont {Borodin}, \citenamefont {Ding}, \citenamefont {Gobet},
  \citenamefont {Vatamanu}, \citenamefont {Fan}, \citenamefont {Gao},
  \citenamefont {Edison}, \citenamefont {Liang}, \citenamefont {Sun} \emph
  {et~al.}}]{wang2018hybrid}%
  \BibitemOpen
  \bibfield  {author} {\bibinfo {author} {\bibfnamefont {F.}~\bibnamefont
  {Wang}}, \bibinfo {author} {\bibfnamefont {O.}~\bibnamefont {Borodin}},
  \bibinfo {author} {\bibfnamefont {M.~S.}\ \bibnamefont {Ding}}, \bibinfo
  {author} {\bibfnamefont {M.}~\bibnamefont {Gobet}}, \bibinfo {author}
  {\bibfnamefont {J.}~\bibnamefont {Vatamanu}}, \bibinfo {author}
  {\bibfnamefont {X.}~\bibnamefont {Fan}}, \bibinfo {author} {\bibfnamefont
  {T.}~\bibnamefont {Gao}}, \bibinfo {author} {\bibfnamefont {N.}~\bibnamefont
  {Edison}}, \bibinfo {author} {\bibfnamefont {Y.}~\bibnamefont {Liang}},
  \bibinfo {author} {\bibfnamefont {W.}~\bibnamefont {Sun}},  \emph {et~al.},\
  }\bibfield  {title} {\enquote {\bibinfo {title} {Hybrid aqueous/non-aqueous
  electrolyte for safe and high-energy li-ion batteries},}\ }\href@noop {}
  {\bibfield  {journal} {\bibinfo  {journal} {Joule}\ }\textbf {\bibinfo
  {volume} {2}},\ \bibinfo {pages} {927--937} (\bibinfo {year}
  {2018})}\BibitemShut {NoStop}%
\bibitem [{\citenamefont {Leonard}\ \emph {et~al.}(2018)\citenamefont
  {Leonard}, \citenamefont {Wei}, \citenamefont {Chen}, \citenamefont {Du},\
  and\ \citenamefont {Ji}}]{leonard2018}%
  \BibitemOpen
  \bibfield  {author} {\bibinfo {author} {\bibfnamefont {D.~P.}\ \bibnamefont
  {Leonard}}, \bibinfo {author} {\bibfnamefont {Z.}~\bibnamefont {Wei}},
  \bibinfo {author} {\bibfnamefont {G.}~\bibnamefont {Chen}}, \bibinfo {author}
  {\bibfnamefont {F.}~\bibnamefont {Du}}, \ and\ \bibinfo {author}
  {\bibfnamefont {X.}~\bibnamefont {Ji}},\ }\bibfield  {title} {\enquote
  {\bibinfo {title} {Water-in-salt electrolyte for potassium-ion batteries},}\
  }\href@noop {} {\bibfield  {journal} {\bibinfo  {journal} {ACS Energy Lett.}\
  }\textbf {\bibinfo {volume} {3}},\ \bibinfo {pages} {373--374} (\bibinfo
  {year} {2018})}\BibitemShut {NoStop}%
\bibitem [{\citenamefont {Yang}\ \emph {et~al.}(2019)\citenamefont {Yang},
  \citenamefont {Chen}, \citenamefont {Ji}, \citenamefont {Pollard},
  \citenamefont {L{\"{u}}}, \citenamefont {Sun}, \citenamefont {Hou},
  \citenamefont {Liu}, \citenamefont {Liu}, \citenamefont {Qing}, \citenamefont
  {Wang}, \citenamefont {Borodin}, \citenamefont {Ren}, \citenamefont {Xu},\
  and\ \citenamefont {Wang}}]{Yang2019}%
  \BibitemOpen
  \bibfield  {author} {\bibinfo {author} {\bibfnamefont {C.}~\bibnamefont
  {Yang}}, \bibinfo {author} {\bibfnamefont {J.}~\bibnamefont {Chen}}, \bibinfo
  {author} {\bibfnamefont {X.}~\bibnamefont {Ji}}, \bibinfo {author}
  {\bibfnamefont {T.~P.}\ \bibnamefont {Pollard}}, \bibinfo {author}
  {\bibfnamefont {X.}~\bibnamefont {L{\"{u}}}}, \bibinfo {author}
  {\bibfnamefont {C.~J.}\ \bibnamefont {Sun}}, \bibinfo {author} {\bibfnamefont
  {S.}~\bibnamefont {Hou}}, \bibinfo {author} {\bibfnamefont {Q.}~\bibnamefont
  {Liu}}, \bibinfo {author} {\bibfnamefont {C.}~\bibnamefont {Liu}}, \bibinfo
  {author} {\bibfnamefont {T.}~\bibnamefont {Qing}}, \bibinfo {author}
  {\bibfnamefont {Y.}~\bibnamefont {Wang}}, \bibinfo {author} {\bibfnamefont
  {O.}~\bibnamefont {Borodin}}, \bibinfo {author} {\bibfnamefont
  {Y.}~\bibnamefont {Ren}}, \bibinfo {author} {\bibfnamefont {K.}~\bibnamefont
  {Xu}}, \ and\ \bibinfo {author} {\bibfnamefont {C.}~\bibnamefont {Wang}},\
  }\href {\doibase 10.1038/s41586-019-1175-6} {\enquote {\bibinfo {title}
  {{Aqueous Li-ion battery enabled by halogen conversion–intercalation
  chemistry in graphite}},}\ } (\bibinfo {year} {2019})\BibitemShut {NoStop}%
\bibitem [{\citenamefont {Dou}\ \emph {et~al.}(2019)\citenamefont {Dou},
  \citenamefont {Lu}, \citenamefont {Su}, \citenamefont {Zhang}, \citenamefont
  {Lei}, \citenamefont {Bu}, \citenamefont {Liu}, \citenamefont {Xiao},
  \citenamefont {Chen}, \citenamefont {Shi},\ and\ \citenamefont
  {Yan}}]{Dou2019}%
  \BibitemOpen
  \bibfield  {author} {\bibinfo {author} {\bibfnamefont {Q.}~\bibnamefont
  {Dou}}, \bibinfo {author} {\bibfnamefont {Y.}~\bibnamefont {Lu}}, \bibinfo
  {author} {\bibfnamefont {L.}~\bibnamefont {Su}}, \bibinfo {author}
  {\bibfnamefont {X.}~\bibnamefont {Zhang}}, \bibinfo {author} {\bibfnamefont
  {S.}~\bibnamefont {Lei}}, \bibinfo {author} {\bibfnamefont {X.}~\bibnamefont
  {Bu}}, \bibinfo {author} {\bibfnamefont {L.}~\bibnamefont {Liu}}, \bibinfo
  {author} {\bibfnamefont {D.}~\bibnamefont {Xiao}}, \bibinfo {author}
  {\bibfnamefont {J.}~\bibnamefont {Chen}}, \bibinfo {author} {\bibfnamefont
  {S.}~\bibnamefont {Shi}}, \ and\ \bibinfo {author} {\bibfnamefont
  {X.}~\bibnamefont {Yan}},\ }\bibfield  {title} {\enquote {\bibinfo {title}
  {{A sodium perchlorate-based hybrid electrolyte with high salt-to-water molar
  ratio for safe 2.5 V carbon-based supercapacitor}},}\ }\href {\doibase
  10.1016/j.ensm.2019.03.016} {\bibfield  {journal} {\bibinfo  {journal}
  {Energy Storage Materials}\ }\textbf {\bibinfo {volume} {23}} (\bibinfo
  {year} {2019}),\ 10.1016/j.ensm.2019.03.016}\BibitemShut {NoStop}%
\bibitem [{\citenamefont {Chen}\ \emph {et~al.}(2018)\citenamefont {Chen},
  \citenamefont {Goodwin}, \citenamefont {Feng},\ and\ \citenamefont
  {Kornyshev}}]{Chen2017}%
  \BibitemOpen
  \bibfield  {author} {\bibinfo {author} {\bibfnamefont {M.}~\bibnamefont
  {Chen}}, \bibinfo {author} {\bibfnamefont {Z.~A.~H.}\ \bibnamefont
  {Goodwin}}, \bibinfo {author} {\bibfnamefont {G.}~\bibnamefont {Feng}}, \
  and\ \bibinfo {author} {\bibfnamefont {A.~A.}\ \bibnamefont {Kornyshev}},\
  }\bibfield  {title} {\enquote {\bibinfo {title} {{On the temperature
  dependence of the double layer capacitance of ionic liquids}},}\ }\href
  {\doibase 10.1016/j.jelechem.2017.11.005} {\bibfield  {journal} {\bibinfo
  {journal} {J. Electroanal. Chem.}\ }\textbf {\bibinfo {volume} {819}},\
  \bibinfo {pages} {347--358} (\bibinfo {year} {2018})}\BibitemShut {NoStop}%
\bibitem [{\citenamefont {Goodwin}\ and\ \citenamefont
  {Kornyshev}(2017)}]{goodwin2017underscreening}%
  \BibitemOpen
  \bibfield  {author} {\bibinfo {author} {\bibfnamefont {Z.~A.}\ \bibnamefont
  {Goodwin}}\ and\ \bibinfo {author} {\bibfnamefont {A.~A.}\ \bibnamefont
  {Kornyshev}},\ }\bibfield  {title} {\enquote {\bibinfo {title}
  {Underscreening, overscreening and double-layer capacitance},}\ }\href@noop
  {} {\bibfield  {journal} {\bibinfo  {journal} {Electrochemistry
  Communications}\ }\textbf {\bibinfo {volume} {82}},\ \bibinfo {pages}
  {129--133} (\bibinfo {year} {2017})}\BibitemShut {NoStop}%
\bibitem [{\citenamefont {Feng}\ \emph {et~al.}(2019)\citenamefont {Feng},
  \citenamefont {Chen}, \citenamefont {Bi}, \citenamefont {Goodwin},
  \citenamefont {Postnikov}, \citenamefont {Brilliantov}, \citenamefont
  {Urbakh},\ and\ \citenamefont {Kornyshev}}]{feng2019free}%
  \BibitemOpen
  \bibfield  {author} {\bibinfo {author} {\bibfnamefont {G.}~\bibnamefont
  {Feng}}, \bibinfo {author} {\bibfnamefont {M.}~\bibnamefont {Chen}}, \bibinfo
  {author} {\bibfnamefont {S.}~\bibnamefont {Bi}}, \bibinfo {author}
  {\bibfnamefont {Z.~A.}\ \bibnamefont {Goodwin}}, \bibinfo {author}
  {\bibfnamefont {E.~B.}\ \bibnamefont {Postnikov}}, \bibinfo {author}
  {\bibfnamefont {N.}~\bibnamefont {Brilliantov}}, \bibinfo {author}
  {\bibfnamefont {M.}~\bibnamefont {Urbakh}}, \ and\ \bibinfo {author}
  {\bibfnamefont {A.~A.}\ \bibnamefont {Kornyshev}},\ }\bibfield  {title}
  {\enquote {\bibinfo {title} {Free and bound states of ions in ionic liquids,
  conductivity, and underscreening paradox},}\ }\href@noop {} {\bibfield
  {journal} {\bibinfo  {journal} {Physical Review X}\ }\textbf {\bibinfo
  {volume} {9}},\ \bibinfo {pages} {021024} (\bibinfo {year}
  {2019})}\BibitemShut {NoStop}%
\bibitem [{\citenamefont {Gebbie}\ \emph
  {et~al.}(2013{\natexlab{a}})\citenamefont {Gebbie}, \citenamefont {Valtiner},
  \citenamefont {Banquy}, \citenamefont {Fox}, \citenamefont {Henderson},\ and\
  \citenamefont {Israelachvili}}]{Gebbie2013}%
  \BibitemOpen
  \bibfield  {author} {\bibinfo {author} {\bibfnamefont {M.~A.}\ \bibnamefont
  {Gebbie}}, \bibinfo {author} {\bibfnamefont {M.}~\bibnamefont {Valtiner}},
  \bibinfo {author} {\bibfnamefont {X.}~\bibnamefont {Banquy}}, \bibinfo
  {author} {\bibfnamefont {E.~T.}\ \bibnamefont {Fox}}, \bibinfo {author}
  {\bibfnamefont {W.~A.}\ \bibnamefont {Henderson}}, \ and\ \bibinfo {author}
  {\bibfnamefont {J.~N.}\ \bibnamefont {Israelachvili}},\ }\bibfield  {title}
  {\enquote {\bibinfo {title} {{Ionic liquids behave as dilute electrolyte
  solutions}},}\ }\href {\doibase 10.1073/pnas.1307871110} {\bibfield
  {journal} {\bibinfo  {journal} {Proceedings of the National Academy of
  Sciences}\ }\textbf {\bibinfo {volume} {110}},\ \bibinfo {pages} {9674--9679}
  (\bibinfo {year} {2013}{\natexlab{a}})}\BibitemShut {NoStop}%
\bibitem [{\citenamefont {Gebbie}\ \emph {et~al.}(2015)\citenamefont {Gebbie},
  \citenamefont {Dobes}, \citenamefont {Valtiner},\ and\ \citenamefont
  {Israelachvili}}]{Gebbie2015}%
  \BibitemOpen
  \bibfield  {author} {\bibinfo {author} {\bibfnamefont {M.~A.}\ \bibnamefont
  {Gebbie}}, \bibinfo {author} {\bibfnamefont {H.~A.}\ \bibnamefont {Dobes}},
  \bibinfo {author} {\bibfnamefont {M.}~\bibnamefont {Valtiner}}, \ and\
  \bibinfo {author} {\bibfnamefont {J.~N.}\ \bibnamefont {Israelachvili}},\
  }\bibfield  {title} {\enquote {\bibinfo {title} {{Long-range electrostatic
  screening in ionic liquids}},}\ }\href@noop {} {\bibfield  {journal}
  {\bibinfo  {journal} {Proceedings of the National Academy of Sciences}\
  }\textbf {\bibinfo {volume} {112}},\ \bibinfo {pages} {7432–7437} (\bibinfo
  {year} {2015})}\BibitemShut {NoStop}%
\bibitem [{\citenamefont {Smith}, \citenamefont {Lee},\ and\ \citenamefont
  {Perkin}(2016{\natexlab{a}})}]{Smith2016}%
  \BibitemOpen
  \bibfield  {author} {\bibinfo {author} {\bibfnamefont {A.~M.}\ \bibnamefont
  {Smith}}, \bibinfo {author} {\bibfnamefont {A.~A.}\ \bibnamefont {Lee}}, \
  and\ \bibinfo {author} {\bibfnamefont {S.}~\bibnamefont {Perkin}},\
  }\bibfield  {title} {\enquote {\bibinfo {title} {{The Electrostatic Screening
  Length in Concentrated Electrolytes Increases with Concentration}},}\ }\href
  {\doibase 10.1021/acs.jpclett.6b00867} {\bibfield  {journal} {\bibinfo
  {journal} {The Journal of Physical Chemistry Letters}\ }\textbf {\bibinfo
  {volume} {7}},\ \bibinfo {pages} {2157--2163} (\bibinfo {year}
  {2016}{\natexlab{a}})}\BibitemShut {NoStop}%
\bibitem [{\citenamefont {Goodwin}, \citenamefont {Feng},\ and\ \citenamefont
  {Kornyshev}(2017)}]{goodwin2017mean}%
  \BibitemOpen
  \bibfield  {author} {\bibinfo {author} {\bibfnamefont {Z.~A.}\ \bibnamefont
  {Goodwin}}, \bibinfo {author} {\bibfnamefont {G.}~\bibnamefont {Feng}}, \
  and\ \bibinfo {author} {\bibfnamefont {A.~A.}\ \bibnamefont {Kornyshev}},\
  }\bibfield  {title} {\enquote {\bibinfo {title} {Mean-field theory of
  electrical double layer in ionic liquids with account of short-range
  correlations},}\ }\href@noop {} {\bibfield  {journal} {\bibinfo  {journal}
  {Electrochimica Acta}\ }\textbf {\bibinfo {volume} {225}},\ \bibinfo {pages}
  {190--197} (\bibinfo {year} {2017})}\BibitemShut {NoStop}%
\bibitem [{\citenamefont {Kim}\ \emph {et~al.}(2014)\citenamefont {Kim},
  \citenamefont {Kim}, \citenamefont {Choi},\ and\ \citenamefont
  {Cho}}]{kim2014ion}%
  \BibitemOpen
  \bibfield  {author} {\bibinfo {author} {\bibfnamefont {S.}~\bibnamefont
  {Kim}}, \bibinfo {author} {\bibfnamefont {H.}~\bibnamefont {Kim}}, \bibinfo
  {author} {\bibfnamefont {J.-H.}\ \bibnamefont {Choi}}, \ and\ \bibinfo
  {author} {\bibfnamefont {M.}~\bibnamefont {Cho}},\ }\bibfield  {title}
  {\enquote {\bibinfo {title} {Ion aggregation in high salt solutions: Ion
  network versus ion cluster},}\ }\href@noop {} {\bibfield  {journal} {\bibinfo
   {journal} {The Journal of chemical physics}\ }\textbf {\bibinfo {volume}
  {141}},\ \bibinfo {pages} {124510} (\bibinfo {year} {2014})}\BibitemShut
  {NoStop}%
\bibitem [{\citenamefont {Choi}\ and\ \citenamefont {Cho}(2014)}]{choi2014ion}%
  \BibitemOpen
  \bibfield  {author} {\bibinfo {author} {\bibfnamefont {J.-H.}\ \bibnamefont
  {Choi}}\ and\ \bibinfo {author} {\bibfnamefont {M.}~\bibnamefont {Cho}},\
  }\bibfield  {title} {\enquote {\bibinfo {title} {Ion aggregation in high salt
  solutions. ii. spectral graph analysis of water hydrogen-bonding network and
  ion aggregate structures},}\ }\href@noop {} {\bibfield  {journal} {\bibinfo
  {journal} {The Journal of chemical physics}\ }\textbf {\bibinfo {volume}
  {141}},\ \bibinfo {pages} {154502} (\bibinfo {year} {2014})}\BibitemShut
  {NoStop}%
\bibitem [{\citenamefont {Choi}\ and\ \citenamefont {Cho}(2015)}]{choi2015ion}%
  \BibitemOpen
  \bibfield  {author} {\bibinfo {author} {\bibfnamefont {J.-H.}\ \bibnamefont
  {Choi}}\ and\ \bibinfo {author} {\bibfnamefont {M.}~\bibnamefont {Cho}},\
  }\bibfield  {title} {\enquote {\bibinfo {title} {Ion aggregation in high salt
  solutions. iv. graph-theoretical analyses of ion aggregate structure and
  water hydrogen bonding network},}\ }\href@noop {} {\bibfield  {journal}
  {\bibinfo  {journal} {The Journal of chemical physics}\ }\textbf {\bibinfo
  {volume} {143}},\ \bibinfo {pages} {104110} (\bibinfo {year}
  {2015})}\BibitemShut {NoStop}%
\bibitem [{\citenamefont {Choi}\ \emph {et~al.}(2017)\citenamefont {Choi},
  \citenamefont {Choi}, \citenamefont {Jeon},\ and\ \citenamefont
  {Cho}}]{choi2017ion}%
  \BibitemOpen
  \bibfield  {author} {\bibinfo {author} {\bibfnamefont {J.-H.}\ \bibnamefont
  {Choi}}, \bibinfo {author} {\bibfnamefont {H.~R.}\ \bibnamefont {Choi}},
  \bibinfo {author} {\bibfnamefont {J.}~\bibnamefont {Jeon}}, \ and\ \bibinfo
  {author} {\bibfnamefont {M.}~\bibnamefont {Cho}},\ }\bibfield  {title}
  {\enquote {\bibinfo {title} {Ion aggregation in high salt solutions. vii. the
  effect of cations on the structures of ion aggregates and water
  hydrogen-bonding network},}\ }\href@noop {} {\bibfield  {journal} {\bibinfo
  {journal} {The Journal of chemical physics}\ }\textbf {\bibinfo {volume}
  {147}},\ \bibinfo {pages} {154107} (\bibinfo {year} {2017})}\BibitemShut
  {NoStop}%
\bibitem [{\citenamefont {Borodin}\ \emph {et~al.}(2017)\citenamefont
  {Borodin}, \citenamefont {Suo}, \citenamefont {Gobet}, \citenamefont {Ren},
  \citenamefont {Wang}, \citenamefont {Faraone}, \citenamefont {Peng},
  \citenamefont {Olguin}, \citenamefont {Schroeder}, \citenamefont {Ding} \emph
  {et~al.}}]{borodin2017liquid}%
  \BibitemOpen
  \bibfield  {author} {\bibinfo {author} {\bibfnamefont {O.}~\bibnamefont
  {Borodin}}, \bibinfo {author} {\bibfnamefont {L.}~\bibnamefont {Suo}},
  \bibinfo {author} {\bibfnamefont {M.}~\bibnamefont {Gobet}}, \bibinfo
  {author} {\bibfnamefont {X.}~\bibnamefont {Ren}}, \bibinfo {author}
  {\bibfnamefont {F.}~\bibnamefont {Wang}}, \bibinfo {author} {\bibfnamefont
  {A.}~\bibnamefont {Faraone}}, \bibinfo {author} {\bibfnamefont
  {J.}~\bibnamefont {Peng}}, \bibinfo {author} {\bibfnamefont {M.}~\bibnamefont
  {Olguin}}, \bibinfo {author} {\bibfnamefont {M.}~\bibnamefont {Schroeder}},
  \bibinfo {author} {\bibfnamefont {M.~S.}\ \bibnamefont {Ding}},  \emph
  {et~al.},\ }\bibfield  {title} {\enquote {\bibinfo {title} {Liquid structure
  with nano-heterogeneity promotes cationic transport in concentrated
  electrolytes},}\ }\href@noop {} {\bibfield  {journal} {\bibinfo  {journal}
  {ACS nano}\ }\textbf {\bibinfo {volume} {11}},\ \bibinfo {pages}
  {10462--10471} (\bibinfo {year} {2017})}\BibitemShut {NoStop}%
\bibitem [{\citenamefont {France-Lanord}\ and\ \citenamefont
  {Grossman}(2019)}]{france2019}%
  \BibitemOpen
  \bibfield  {author} {\bibinfo {author} {\bibfnamefont {A.}~\bibnamefont
  {France-Lanord}}\ and\ \bibinfo {author} {\bibfnamefont {J.~C.}\ \bibnamefont
  {Grossman}},\ }\bibfield  {title} {\enquote {\bibinfo {title} {Correlations
  from ion pairing and the nernst-einstein equation},}\ }\href@noop {}
  {\bibfield  {journal} {\bibinfo  {journal} {Physical review letters}\
  }\textbf {\bibinfo {volume} {122}},\ \bibinfo {pages} {136001} (\bibinfo
  {year} {2019})}\BibitemShut {NoStop}%
\bibitem [{\citenamefont {Yu}\ \emph {et~al.}(2020)\citenamefont {Yu},
  \citenamefont {Curtiss}, \citenamefont {Winans}, \citenamefont {Li},\ and\
  \citenamefont {Cheng}}]{yu2020asymmetric}%
  \BibitemOpen
  \bibfield  {author} {\bibinfo {author} {\bibfnamefont {Z.}~\bibnamefont
  {Yu}}, \bibinfo {author} {\bibfnamefont {L.~A.}\ \bibnamefont {Curtiss}},
  \bibinfo {author} {\bibfnamefont {R.~E.}\ \bibnamefont {Winans}}, \bibinfo
  {author} {\bibfnamefont {T.}~\bibnamefont {Li}}, \ and\ \bibinfo {author}
  {\bibfnamefont {L.}~\bibnamefont {Cheng}},\ }\bibfield  {title} {\enquote
  {\bibinfo {title} {Asymmetric composition of ionic aggregates and the origin
  of high correlated transference number in water-in-salt electrolytes},}\
  }\href@noop {} {\bibfield  {journal} {\bibinfo  {journal} {The Journal of
  Physical Chemistry Letters}\ } (\bibinfo {year} {2020})}\BibitemShut
  {NoStop}%
\bibitem [{\citenamefont {Lim}\ \emph {et~al.}(2018)\citenamefont {Lim},
  \citenamefont {Park}, \citenamefont {Lee}, \citenamefont {Kim}, \citenamefont
  {Kwak},\ and\ \citenamefont {Cho}}]{lim2018}%
  \BibitemOpen
  \bibfield  {author} {\bibinfo {author} {\bibfnamefont {J.}~\bibnamefont
  {Lim}}, \bibinfo {author} {\bibfnamefont {K.}~\bibnamefont {Park}}, \bibinfo
  {author} {\bibfnamefont {H.}~\bibnamefont {Lee}}, \bibinfo {author}
  {\bibfnamefont {J.}~\bibnamefont {Kim}}, \bibinfo {author} {\bibfnamefont
  {K.}~\bibnamefont {Kwak}}, \ and\ \bibinfo {author} {\bibfnamefont
  {M.}~\bibnamefont {Cho}},\ }\bibfield  {title} {\enquote {\bibinfo {title}
  {Nanometric water channels in water-in-salt lithium ion battery
  electrolyte},}\ }\href@noop {} {\bibfield  {journal} {\bibinfo  {journal}
  {Journal of the American Chemical Society}\ }\textbf {\bibinfo {volume}
  {140}},\ \bibinfo {pages} {15661--15667} (\bibinfo {year}
  {2018})}\BibitemShut {NoStop}%
\bibitem [{\citenamefont {Lewis}\ \emph {et~al.}(2020)\citenamefont {Lewis},
  \citenamefont {Zhang}, \citenamefont {Dereka}, \citenamefont {Carino},
  \citenamefont {Maginn},\ and\ \citenamefont
  {Tokmakoff}}]{lewis2020signatures}%
  \BibitemOpen
  \bibfield  {author} {\bibinfo {author} {\bibfnamefont {N.~H.}\ \bibnamefont
  {Lewis}}, \bibinfo {author} {\bibfnamefont {Y.}~\bibnamefont {Zhang}},
  \bibinfo {author} {\bibfnamefont {B.}~\bibnamefont {Dereka}}, \bibinfo
  {author} {\bibfnamefont {E.~V.}\ \bibnamefont {Carino}}, \bibinfo {author}
  {\bibfnamefont {E.~J.}\ \bibnamefont {Maginn}}, \ and\ \bibinfo {author}
  {\bibfnamefont {A.}~\bibnamefont {Tokmakoff}},\ }\bibfield  {title} {\enquote
  {\bibinfo {title} {Signatures of ion-pairing and aggregation in the
  vibrational spectroscopy of super-concentrated aqueous lithium bistriflimide
  solutions},}\ }\href@noop {} {\bibfield  {journal} {\bibinfo  {journal} {The
  Journal of Physical Chemistry C}\ } (\bibinfo {year} {2020})}\BibitemShut
  {NoStop}%
\bibitem [{\citenamefont {Molinari}\ \emph {et~al.}(2019)\citenamefont
  {Molinari}, \citenamefont {Mailoa}, \citenamefont {Craig}, \citenamefont
  {Christensen},\ and\ \citenamefont {Kozinsky}}]{molinari2019transport}%
  \BibitemOpen
  \bibfield  {author} {\bibinfo {author} {\bibfnamefont {N.}~\bibnamefont
  {Molinari}}, \bibinfo {author} {\bibfnamefont {J.~P.}\ \bibnamefont
  {Mailoa}}, \bibinfo {author} {\bibfnamefont {N.}~\bibnamefont {Craig}},
  \bibinfo {author} {\bibfnamefont {J.}~\bibnamefont {Christensen}}, \ and\
  \bibinfo {author} {\bibfnamefont {B.}~\bibnamefont {Kozinsky}},\ }\bibfield
  {title} {\enquote {\bibinfo {title} {Transport anomalies emerging from strong
  correlation in ionic liquid electrolytes},}\ }\href@noop {} {\bibfield
  {journal} {\bibinfo  {journal} {Journal of Power Sources}\ }\textbf {\bibinfo
  {volume} {428}},\ \bibinfo {pages} {27--36} (\bibinfo {year}
  {2019})}\BibitemShut {NoStop}%
\bibitem [{\citenamefont {Molinari}, \citenamefont {Mailoa},\ and\
  \citenamefont {Kozinsky}(2019)}]{molinari2019general}%
  \BibitemOpen
  \bibfield  {author} {\bibinfo {author} {\bibfnamefont {N.}~\bibnamefont
  {Molinari}}, \bibinfo {author} {\bibfnamefont {J.~P.}\ \bibnamefont
  {Mailoa}}, \ and\ \bibinfo {author} {\bibfnamefont {B.}~\bibnamefont
  {Kozinsky}},\ }\bibfield  {title} {\enquote {\bibinfo {title} {General trend
  of a negative li effective charge in ionic liquid electrolytes},}\
  }\href@noop {} {\bibfield  {journal} {\bibinfo  {journal} {The journal of
  physical chemistry letters}\ }\textbf {\bibinfo {volume} {10}},\ \bibinfo
  {pages} {2313--2319} (\bibinfo {year} {2019})}\BibitemShut {NoStop}%
\bibitem [{\citenamefont {Flory}(1941)}]{flory1941molecular}%
  \BibitemOpen
  \bibfield  {author} {\bibinfo {author} {\bibfnamefont {P.~J.}\ \bibnamefont
  {Flory}},\ }\bibfield  {title} {\enquote {\bibinfo {title} {Molecular size
  distribution in three dimensional polymers. i. gelation1},}\ }\href@noop {}
  {\bibfield  {journal} {\bibinfo  {journal} {Journal of the American Chemical
  Society}\ }\textbf {\bibinfo {volume} {63}},\ \bibinfo {pages} {3083--3090}
  (\bibinfo {year} {1941})}\BibitemShut {NoStop}%
\bibitem [{\citenamefont {Flory}(1942{\natexlab{a}})}]{flory1942constitution}%
  \BibitemOpen
  \bibfield  {author} {\bibinfo {author} {\bibfnamefont {P.~J.}\ \bibnamefont
  {Flory}},\ }\bibfield  {title} {\enquote {\bibinfo {title} {Constitution of
  three-dimensional polymers and the theory of gelation.}}\ }\href@noop {}
  {\bibfield  {journal} {\bibinfo  {journal} {The Journal of Physical
  Chemistry}\ }\textbf {\bibinfo {volume} {46}},\ \bibinfo {pages} {132--140}
  (\bibinfo {year} {1942}{\natexlab{a}})}\BibitemShut {NoStop}%
\bibitem [{\citenamefont {Stockmayer}(1943)}]{stockmayer1943theory}%
  \BibitemOpen
  \bibfield  {author} {\bibinfo {author} {\bibfnamefont {W.~H.}\ \bibnamefont
  {Stockmayer}},\ }\bibfield  {title} {\enquote {\bibinfo {title} {Theory of
  molecular size distribution and gel formation in branched-chain polymers},}\
  }\href@noop {} {\bibfield  {journal} {\bibinfo  {journal} {The Journal of
  chemical physics}\ }\textbf {\bibinfo {volume} {11}},\ \bibinfo {pages}
  {45--55} (\bibinfo {year} {1943})}\BibitemShut {NoStop}%
\bibitem [{\citenamefont {Stockmayer}(1944)}]{stockmayer1944theory}%
  \BibitemOpen
  \bibfield  {author} {\bibinfo {author} {\bibfnamefont {W.~H.}\ \bibnamefont
  {Stockmayer}},\ }\bibfield  {title} {\enquote {\bibinfo {title} {Theory of
  molecular size distribution and gel formation in branched polymers ii.
  general cross linking},}\ }\href@noop {} {\bibfield  {journal} {\bibinfo
  {journal} {The Journal of Chemical Physics}\ }\textbf {\bibinfo {volume}
  {12}},\ \bibinfo {pages} {125--131} (\bibinfo {year} {1944})}\BibitemShut
  {NoStop}%
\bibitem [{\citenamefont {Stauffer}\ and\ \citenamefont
  {Aharony}(1994)}]{stauffer1994introduction}%
  \BibitemOpen
  \bibfield  {author} {\bibinfo {author} {\bibfnamefont {D.}~\bibnamefont
  {Stauffer}}\ and\ \bibinfo {author} {\bibfnamefont {A.}~\bibnamefont
  {Aharony}},\ }\bibfield  {title} {\enquote {\bibinfo {title} {Introduction to
  percolation theory.(2nd edn), 1992},}\ }\href@noop {} {\bibfield  {journal}
  {\bibinfo  {journal} {London, Taylor and Francis.}\ } (\bibinfo {year}
  {1994})}\BibitemShut {NoStop}%
\bibitem [{\citenamefont {Tanaka}(1989)}]{tanaka1989}%
  \BibitemOpen
  \bibfield  {author} {\bibinfo {author} {\bibfnamefont {F.}~\bibnamefont
  {Tanaka}},\ }\bibfield  {title} {\enquote {\bibinfo {title} {Theory of
  thermoreversible gelation},}\ }\href@noop {} {\bibfield  {journal} {\bibinfo
  {journal} {Macromolecules}\ }\textbf {\bibinfo {volume} {22}},\ \bibinfo
  {pages} {1988--1994} (\bibinfo {year} {1989})}\BibitemShut {NoStop}%
\bibitem [{\citenamefont {Tanaka}(1990)}]{tanaka1990thermodynamic}%
  \BibitemOpen
  \bibfield  {author} {\bibinfo {author} {\bibfnamefont {F.}~\bibnamefont
  {Tanaka}},\ }\bibfield  {title} {\enquote {\bibinfo {title} {Thermodynamic
  theory of network-forming polymer solutions. 1},}\ }\href@noop {} {\bibfield
  {journal} {\bibinfo  {journal} {Macromolecules}\ }\textbf {\bibinfo {volume}
  {23}},\ \bibinfo {pages} {3784--3789} (\bibinfo {year} {1990})}\BibitemShut
  {NoStop}%
\bibitem [{\citenamefont {Tanaka}\ and\ \citenamefont
  {Stockmayer}(1994)}]{tanaka1994}%
  \BibitemOpen
  \bibfield  {author} {\bibinfo {author} {\bibfnamefont {F.}~\bibnamefont
  {Tanaka}}\ and\ \bibinfo {author} {\bibfnamefont {W.~H.}\ \bibnamefont
  {Stockmayer}},\ }\bibfield  {title} {\enquote {\bibinfo {title}
  {Thermoreversible gelation with junctions of variable multiplicity},}\
  }\href@noop {} {\bibfield  {journal} {\bibinfo  {journal} {Macromolecules}\
  }\textbf {\bibinfo {volume} {27}},\ \bibinfo {pages} {3943--3954} (\bibinfo
  {year} {1994})}\BibitemShut {NoStop}%
\bibitem [{\citenamefont {Tanaka}\ and\ \citenamefont
  {Ishida}(1995)}]{tanaka1995}%
  \BibitemOpen
  \bibfield  {author} {\bibinfo {author} {\bibfnamefont {F.}~\bibnamefont
  {Tanaka}}\ and\ \bibinfo {author} {\bibfnamefont {M.}~\bibnamefont
  {Ishida}},\ }\bibfield  {title} {\enquote {\bibinfo {title} {Thermoreversible
  gelation of hydrated polymers},}\ }\href@noop {} {\bibfield  {journal}
  {\bibinfo  {journal} {Journal of the Chemical Society, Faraday Transactions}\
  }\textbf {\bibinfo {volume} {91}},\ \bibinfo {pages} {2663--2670} (\bibinfo
  {year} {1995})}\BibitemShut {NoStop}%
\bibitem [{\citenamefont {Ishida}\ and\ \citenamefont
  {Tanaka}(1997)}]{ishida1997}%
  \BibitemOpen
  \bibfield  {author} {\bibinfo {author} {\bibfnamefont {M.}~\bibnamefont
  {Ishida}}\ and\ \bibinfo {author} {\bibfnamefont {F.}~\bibnamefont
  {Tanaka}},\ }\bibfield  {title} {\enquote {\bibinfo {title} {Theoretical
  study of the postgel regime in thermoreversible gelation},}\ }\href@noop {}
  {\bibfield  {journal} {\bibinfo  {journal} {Macromolecules}\ }\textbf
  {\bibinfo {volume} {30}},\ \bibinfo {pages} {3900--3909} (\bibinfo {year}
  {1997})}\BibitemShut {NoStop}%
\bibitem [{\citenamefont {Tanaka}(1998)}]{tanaka1998}%
  \BibitemOpen
  \bibfield  {author} {\bibinfo {author} {\bibfnamefont {F.}~\bibnamefont
  {Tanaka}},\ }\bibfield  {title} {\enquote {\bibinfo {title} {Thermoreversible
  gelation of associating polymers},}\ }\href@noop {} {\bibfield  {journal}
  {\bibinfo  {journal} {Physica A: Statistical Mechanics and its Applications}\
  }\textbf {\bibinfo {volume} {257}},\ \bibinfo {pages} {245--255} (\bibinfo
  {year} {1998})}\BibitemShut {NoStop}%
\bibitem [{\citenamefont {Tanaka}\ and\ \citenamefont
  {Ishida}(1999)}]{tanaka1999}%
  \BibitemOpen
  \bibfield  {author} {\bibinfo {author} {\bibfnamefont {F.}~\bibnamefont
  {Tanaka}}\ and\ \bibinfo {author} {\bibfnamefont {M.}~\bibnamefont
  {Ishida}},\ }\bibfield  {title} {\enquote {\bibinfo {title} {Thermoreversible
  gelation with two-component networks},}\ }\href@noop {} {\bibfield  {journal}
  {\bibinfo  {journal} {Macromolecules}\ }\textbf {\bibinfo {volume} {32}},\
  \bibinfo {pages} {1271--1283} (\bibinfo {year} {1999})}\BibitemShut {NoStop}%
\bibitem [{\citenamefont {Tanaka}(2002)}]{tanaka2002}%
  \BibitemOpen
  \bibfield  {author} {\bibinfo {author} {\bibfnamefont {F.}~\bibnamefont
  {Tanaka}},\ }\bibfield  {title} {\enquote {\bibinfo {title} {Theoretical
  study of molecular association and thermoreversible gelation in polymers},}\
  }\href@noop {} {\bibfield  {journal} {\bibinfo  {journal} {Polymer journal}\
  }\textbf {\bibinfo {volume} {34}},\ \bibinfo {pages} {479} (\bibinfo {year}
  {2002})}\BibitemShut {NoStop}%
\bibitem [{\citenamefont
  {Flory}(1942{\natexlab{b}})}]{flory1942thermodynamics}%
  \BibitemOpen
  \bibfield  {author} {\bibinfo {author} {\bibfnamefont {P.~J.}\ \bibnamefont
  {Flory}},\ }\bibfield  {title} {\enquote {\bibinfo {title} {Thermodynamics of
  high polymer solutions},}\ }\href@noop {} {\bibfield  {journal} {\bibinfo
  {journal} {The Journal of chemical physics}\ }\textbf {\bibinfo {volume}
  {10}},\ \bibinfo {pages} {51--61} (\bibinfo {year}
  {1942}{\natexlab{b}})}\BibitemShut {NoStop}%
\bibitem [{\citenamefont {Flory}(1953)}]{flory1953principles}%
  \BibitemOpen
  \bibfield  {author} {\bibinfo {author} {\bibfnamefont {P.~J.}\ \bibnamefont
  {Flory}},\ }\href@noop {} {\emph {\bibinfo {title} {Principles of polymer
  chemistry}}}\ (\bibinfo  {publisher} {Cornell University Press},\ \bibinfo
  {year} {1953})\BibitemShut {NoStop}%
\bibitem [{\citenamefont {Stockmayer}(1952)}]{stockmayer1952molecular}%
  \BibitemOpen
  \bibfield  {author} {\bibinfo {author} {\bibfnamefont {W.~H.}\ \bibnamefont
  {Stockmayer}},\ }\bibfield  {title} {\enquote {\bibinfo {title} {Molecular
  distribution in condensation polymers},}\ }\href@noop {} {\bibfield
  {journal} {\bibinfo  {journal} {Journal of Polymer Science}\ }\textbf
  {\bibinfo {volume} {9}},\ \bibinfo {pages} {69--71} (\bibinfo {year}
  {1952})}\BibitemShut {NoStop}%
\bibitem [{\citenamefont {Matsuyama}\ and\ \citenamefont
  {Tanaka}(1990)}]{matsuyama1990theory}%
  \BibitemOpen
  \bibfield  {author} {\bibinfo {author} {\bibfnamefont {A.}~\bibnamefont
  {Matsuyama}}\ and\ \bibinfo {author} {\bibfnamefont {F.}~\bibnamefont
  {Tanaka}},\ }\bibfield  {title} {\enquote {\bibinfo {title} {Theory of
  solvation-induced reentrant phase separation in polymer solutions},}\
  }\href@noop {} {\bibfield  {journal} {\bibinfo  {journal} {Physical review
  letters}\ }\textbf {\bibinfo {volume} {65}},\ \bibinfo {pages} {341}
  (\bibinfo {year} {1990})}\BibitemShut {NoStop}%
\bibitem [{\citenamefont {Gebbie}\ \emph
  {et~al.}(2013{\natexlab{b}})\citenamefont {Gebbie}, \citenamefont {Valtiner},
  \citenamefont {Banquy}, \citenamefont {Fox}, \citenamefont {Henderson},\ and\
  \citenamefont {Israelachvili}}]{gebbie2013ionic}%
  \BibitemOpen
  \bibfield  {author} {\bibinfo {author} {\bibfnamefont {M.~A.}\ \bibnamefont
  {Gebbie}}, \bibinfo {author} {\bibfnamefont {M.}~\bibnamefont {Valtiner}},
  \bibinfo {author} {\bibfnamefont {X.}~\bibnamefont {Banquy}}, \bibinfo
  {author} {\bibfnamefont {E.~T.}\ \bibnamefont {Fox}}, \bibinfo {author}
  {\bibfnamefont {W.~A.}\ \bibnamefont {Henderson}}, \ and\ \bibinfo {author}
  {\bibfnamefont {J.~N.}\ \bibnamefont {Israelachvili}},\ }\bibfield  {title}
  {\enquote {\bibinfo {title} {Ionic liquids behave as dilute electrolyte
  solutions},}\ }\href@noop {} {\bibfield  {journal} {\bibinfo  {journal}
  {Proceedings of the National Academy of Sciences}\ }\textbf {\bibinfo
  {volume} {110}},\ \bibinfo {pages} {9674--9679} (\bibinfo {year}
  {2013}{\natexlab{b}})}\BibitemShut {NoStop}%
\bibitem [{\citenamefont {Debye}\ and\ \citenamefont
  {H{\"u}ckel}(1923)}]{debye1923theory}%
  \BibitemOpen
  \bibfield  {author} {\bibinfo {author} {\bibfnamefont {P.}~\bibnamefont
  {Debye}}\ and\ \bibinfo {author} {\bibfnamefont {E.}~\bibnamefont
  {H{\"u}ckel}},\ }\bibfield  {title} {\enquote {\bibinfo {title} {The theory
  of electrolytes. i. freezing point depres-sion and related phenomena [zur
  theorie der elektrolyte. i. gefrierpunktserniedrigung und verwandte
  erscheinungen]},}\ }\href@noop {} {\bibfield  {journal} {\bibinfo  {journal}
  {Physikalische Zeitschrift}\ }\textbf {\bibinfo {volume} {24}},\ \bibinfo
  {pages} {185--206} (\bibinfo {year} {1923})}\BibitemShut {NoStop}%
\bibitem [{\citenamefont {Vincze}, \citenamefont {Valisk{\'o}},\ and\
  \citenamefont {Boda}(2010)}]{vincze2010nonmonotonic}%
  \BibitemOpen
  \bibfield  {author} {\bibinfo {author} {\bibfnamefont {J.}~\bibnamefont
  {Vincze}}, \bibinfo {author} {\bibfnamefont {M.}~\bibnamefont {Valisk{\'o}}},
  \ and\ \bibinfo {author} {\bibfnamefont {D.}~\bibnamefont {Boda}},\
  }\bibfield  {title} {\enquote {\bibinfo {title} {The nonmonotonic
  concentration dependence of the mean activity coefficient of electrolytes is
  a result of a balance between solvation and ion-ion correlations},}\
  }\href@noop {} {\bibfield  {journal} {\bibinfo  {journal} {The Journal of
  chemical physics}\ }\textbf {\bibinfo {volume} {133}},\ \bibinfo {pages}
  {154507} (\bibinfo {year} {2010})}\BibitemShut {NoStop}%
\bibitem [{\citenamefont {H{\"u}ckel}(1925)}]{huckel1925theorie}%
  \BibitemOpen
  \bibfield  {author} {\bibinfo {author} {\bibfnamefont {E.}~\bibnamefont
  {H{\"u}ckel}},\ }\bibfield  {title} {\enquote {\bibinfo {title} {Zur theorie
  konzentrierterer w{\"a}sseriger l{\"o}sungen starker elektrolyte},}\
  }\href@noop {} {\bibfield  {journal} {\bibinfo  {journal} {Phys. Z}\ }\textbf
  {\bibinfo {volume} {26}},\ \bibinfo {pages} {93--147} (\bibinfo {year}
  {1925})}\BibitemShut {NoStop}%
\bibitem [{\citenamefont {Born}(1920)}]{born1920volumen}%
  \BibitemOpen
  \bibfield  {author} {\bibinfo {author} {\bibfnamefont {M.}~\bibnamefont
  {Born}},\ }\bibfield  {title} {\enquote {\bibinfo {title} {Volumen und
  hydratationsw{\"a}rme der ionen},}\ }\href@noop {} {\bibfield  {journal}
  {\bibinfo  {journal} {Zeitschrift f{\"u}r Physik A Hadrons and Nuclei}\
  }\textbf {\bibinfo {volume} {1}},\ \bibinfo {pages} {45--48} (\bibinfo {year}
  {1920})}\BibitemShut {NoStop}%
\bibitem [{\citenamefont {Weing{\"a}rtner}(2006)}]{weingartner2006static}%
  \BibitemOpen
  \bibfield  {author} {\bibinfo {author} {\bibfnamefont {H.}~\bibnamefont
  {Weing{\"a}rtner}},\ }\bibfield  {title} {\enquote {\bibinfo {title} {The
  static dielectric constant of ionic liquids},}\ }\href@noop {} {\bibfield
  {journal} {\bibinfo  {journal} {Zeitschrift f{\"u}r Physikalische Chemie}\
  }\textbf {\bibinfo {volume} {220}},\ \bibinfo {pages} {1395--1405} (\bibinfo
  {year} {2006})}\BibitemShut {NoStop}%
\bibitem [{\citenamefont {Shilov}\ and\ \citenamefont
  {Lyashchenko}(2015)}]{shilov2015role}%
  \BibitemOpen
  \bibfield  {author} {\bibinfo {author} {\bibfnamefont {I.~Y.}\ \bibnamefont
  {Shilov}}\ and\ \bibinfo {author} {\bibfnamefont {A.~K.}\ \bibnamefont
  {Lyashchenko}},\ }\bibfield  {title} {\enquote {\bibinfo {title} {The role of
  concentration dependent static permittivity of electrolyte solutions in the
  debye--huckel theory},}\ }\href@noop {} {\bibfield  {journal} {\bibinfo
  {journal} {The Journal of Physical Chemistry B}\ }\textbf {\bibinfo {volume}
  {119}},\ \bibinfo {pages} {10087--10095} (\bibinfo {year}
  {2015})}\BibitemShut {NoStop}%
\bibitem [{\citenamefont {Smith}, \citenamefont {Lee},\ and\ \citenamefont
  {Perkin}(2016{\natexlab{b}})}]{smith2016electrostatic}%
  \BibitemOpen
  \bibfield  {author} {\bibinfo {author} {\bibfnamefont {A.~M.}\ \bibnamefont
  {Smith}}, \bibinfo {author} {\bibfnamefont {A.~A.}\ \bibnamefont {Lee}}, \
  and\ \bibinfo {author} {\bibfnamefont {S.}~\bibnamefont {Perkin}},\
  }\bibfield  {title} {\enquote {\bibinfo {title} {The electrostatic screening
  length in concentrated electrolytes increases with concentration},}\
  }\href@noop {} {\bibfield  {journal} {\bibinfo  {journal} {The journal of
  physical chemistry letters}\ }\textbf {\bibinfo {volume} {7}},\ \bibinfo
  {pages} {2157--2163} (\bibinfo {year} {2016}{\natexlab{b}})}\BibitemShut
  {NoStop}%
\bibitem [{\citenamefont {Tanaka}(2011)}]{tanaka2011polymer}%
  \BibitemOpen
  \bibfield  {author} {\bibinfo {author} {\bibfnamefont {F.}~\bibnamefont
  {Tanaka}},\ }\href@noop {} {\emph {\bibinfo {title} {Polymer physics:
  applications to molecular association and thermoreversible gelation}}}\
  (\bibinfo  {publisher} {Cambridge University Press},\ \bibinfo {year}
  {2011})\BibitemShut {NoStop}%
\bibitem [{\citenamefont {Macosko}\ and\ \citenamefont
  {Miller}(1976)}]{macosko1976}%
  \BibitemOpen
  \bibfield  {author} {\bibinfo {author} {\bibfnamefont {C.~W.}\ \bibnamefont
  {Macosko}}\ and\ \bibinfo {author} {\bibfnamefont {D.~R.}\ \bibnamefont
  {Miller}},\ }\bibfield  {title} {\enquote {\bibinfo {title} {A new derivation
  of average molecular weights of nonlinear polymers},}\ }\href@noop {}
  {\bibfield  {journal} {\bibinfo  {journal} {Macromolecules}\ }\textbf
  {\bibinfo {volume} {9}},\ \bibinfo {pages} {199--206} (\bibinfo {year}
  {1976})}\BibitemShut {NoStop}%
\bibitem [{\citenamefont {De~Groot}\ and\ \citenamefont
  {Mazur}(2013)}]{de2013non}%
  \BibitemOpen
  \bibfield  {author} {\bibinfo {author} {\bibfnamefont {S.~R.}\ \bibnamefont
  {De~Groot}}\ and\ \bibinfo {author} {\bibfnamefont {P.}~\bibnamefont
  {Mazur}},\ }\href@noop {} {\emph {\bibinfo {title} {Non-equilibrium
  thermodynamics}}}\ (\bibinfo  {publisher} {Courier Corporation},\ \bibinfo
  {year} {2013})\BibitemShut {NoStop}%
\bibitem [{\citenamefont {Krishna}\ and\ \citenamefont
  {Wesselingh}(1997)}]{krishna1997maxwell}%
  \BibitemOpen
  \bibfield  {author} {\bibinfo {author} {\bibfnamefont {R.}~\bibnamefont
  {Krishna}}\ and\ \bibinfo {author} {\bibfnamefont {J.}~\bibnamefont
  {Wesselingh}},\ }\bibfield  {title} {\enquote {\bibinfo {title} {The
  maxwell-stefan approach to mass transfer},}\ }\href@noop {} {\bibfield
  {journal} {\bibinfo  {journal} {Chemical engineering science}\ }\textbf
  {\bibinfo {volume} {52}},\ \bibinfo {pages} {861--911} (\bibinfo {year}
  {1997})}\BibitemShut {NoStop}%
\bibitem [{\citenamefont {Deen}(1998)}]{deen1998analysis}%
  \BibitemOpen
  \bibfield  {author} {\bibinfo {author} {\bibfnamefont {W.~M.}\ \bibnamefont
  {Deen}},\ }\bibfield  {title} {\enquote {\bibinfo {title} {Analysis of
  transport phenomena},}\ }\href@noop {} {\  (\bibinfo {year}
  {1998})}\BibitemShut {NoStop}%
\bibitem [{\citenamefont {Newman}\ and\ \citenamefont
  {Thomas-Alyea}(2012)}]{newman2012electrochemical}%
  \BibitemOpen
  \bibfield  {author} {\bibinfo {author} {\bibfnamefont {J.}~\bibnamefont
  {Newman}}\ and\ \bibinfo {author} {\bibfnamefont {K.~E.}\ \bibnamefont
  {Thomas-Alyea}},\ }\href@noop {} {\emph {\bibinfo {title} {Electrochemical
  systems}}}\ (\bibinfo  {publisher} {John Wiley \& Sons},\ \bibinfo {year}
  {2012})\BibitemShut {NoStop}%
\bibitem [{\citenamefont {Smith}\ and\ \citenamefont
  {Bazant}(2017)}]{smith2017multiphase}%
  \BibitemOpen
  \bibfield  {author} {\bibinfo {author} {\bibfnamefont {R.~B.}\ \bibnamefont
  {Smith}}\ and\ \bibinfo {author} {\bibfnamefont {M.~Z.}\ \bibnamefont
  {Bazant}},\ }\bibfield  {title} {\enquote {\bibinfo {title} {Multiphase
  porous electrode theory},}\ }\href@noop {} {\bibfield  {journal} {\bibinfo
  {journal} {Journal of The Electrochemical Society}\ }\textbf {\bibinfo
  {volume} {164}},\ \bibinfo {pages} {E3291--E3310} (\bibinfo {year}
  {2017})}\BibitemShut {NoStop}%
\bibitem [{\citenamefont {Thomas}, \citenamefont {Darling},\ and\ \citenamefont
  {Newman}(2002)}]{thomas2002advances}%
  \BibitemOpen
  \bibfield  {author} {\bibinfo {author} {\bibfnamefont {K.}~\bibnamefont
  {Thomas}}, \bibinfo {author} {\bibfnamefont {R.}~\bibnamefont {Darling}}, \
  and\ \bibinfo {author} {\bibfnamefont {J.}~\bibnamefont {Newman}},\
  }\bibfield  {title} {\enquote {\bibinfo {title} {Advances in lithium-ion
  batteries ed w. van schalkwijk and b},}\ }\href@noop {} {\bibfield  {journal}
  {\bibinfo  {journal} {Scrosati in}\ } (\bibinfo {year} {2002})}\BibitemShut
  {NoStop}%
\bibitem [{\citenamefont {Val{\o}en}\ and\ \citenamefont
  {Reimers}(2005)}]{valoen2005transport}%
  \BibitemOpen
  \bibfield  {author} {\bibinfo {author} {\bibfnamefont {L.~O.}\ \bibnamefont
  {Val{\o}en}}\ and\ \bibinfo {author} {\bibfnamefont {J.~N.}\ \bibnamefont
  {Reimers}},\ }\bibfield  {title} {\enquote {\bibinfo {title} {Transport
  properties of lipf6-based li-ion battery electrolytes},}\ }\href@noop {}
  {\bibfield  {journal} {\bibinfo  {journal} {Journal of The Electrochemical
  Society}\ }\textbf {\bibinfo {volume} {152}},\ \bibinfo {pages} {A882--A891}
  (\bibinfo {year} {2005})}\BibitemShut {NoStop}%
\bibitem [{\citenamefont {Nyman}, \citenamefont {Behm},\ and\ \citenamefont
  {Lindbergh}(2008)}]{nyman2008electrochemical}%
  \BibitemOpen
  \bibfield  {author} {\bibinfo {author} {\bibfnamefont {A.}~\bibnamefont
  {Nyman}}, \bibinfo {author} {\bibfnamefont {M.}~\bibnamefont {Behm}}, \ and\
  \bibinfo {author} {\bibfnamefont {G.}~\bibnamefont {Lindbergh}},\ }\bibfield
  {title} {\enquote {\bibinfo {title} {Electrochemical characterisation and
  modelling of the mass transport phenomena in lipf6--ec--emc electrolyte},}\
  }\href@noop {} {\bibfield  {journal} {\bibinfo  {journal} {Electrochimica
  Acta}\ }\textbf {\bibinfo {volume} {53}},\ \bibinfo {pages} {6356--6365}
  (\bibinfo {year} {2008})}\BibitemShut {NoStop}%
\bibitem [{\citenamefont {Lundgren}, \citenamefont {Behm},\ and\ \citenamefont
  {Lindbergh}(2015)}]{lundgren2015electrochemical}%
  \BibitemOpen
  \bibfield  {author} {\bibinfo {author} {\bibfnamefont {H.}~\bibnamefont
  {Lundgren}}, \bibinfo {author} {\bibfnamefont {M.}~\bibnamefont {Behm}}, \
  and\ \bibinfo {author} {\bibfnamefont {G.}~\bibnamefont {Lindbergh}},\
  }\bibfield  {title} {\enquote {\bibinfo {title} {Electrochemical
  characterization and temperature dependency of mass-transport properties of
  lipf6 in ec: Dec},}\ }\href@noop {} {\bibfield  {journal} {\bibinfo
  {journal} {Journal of The Electrochemical Society}\ }\textbf {\bibinfo
  {volume} {162}},\ \bibinfo {pages} {A413--A420} (\bibinfo {year}
  {2015})}\BibitemShut {NoStop}%
\bibitem [{\citenamefont {Wheeler}\ and\ \citenamefont
  {Newman}(2004)}]{wheeler2004molecular}%
  \BibitemOpen
  \bibfield  {author} {\bibinfo {author} {\bibfnamefont {D.~R.}\ \bibnamefont
  {Wheeler}}\ and\ \bibinfo {author} {\bibfnamefont {J.}~\bibnamefont
  {Newman}},\ }\bibfield  {title} {\enquote {\bibinfo {title} {Molecular
  dynamics simulations of multicomponent diffusion. 1. equilibrium method},}\
  }\href@noop {} {\bibfield  {journal} {\bibinfo  {journal} {The Journal of
  Physical Chemistry B}\ }\textbf {\bibinfo {volume} {108}},\ \bibinfo {pages}
  {18353--18361} (\bibinfo {year} {2004})}\BibitemShut {NoStop}%
\bibitem [{\citenamefont {Psaltis}\ and\ \citenamefont
  {Farrell}(2011)}]{psaltis2011comparing}%
  \BibitemOpen
  \bibfield  {author} {\bibinfo {author} {\bibfnamefont {S.}~\bibnamefont
  {Psaltis}}\ and\ \bibinfo {author} {\bibfnamefont {T.~W.}\ \bibnamefont
  {Farrell}},\ }\bibfield  {title} {\enquote {\bibinfo {title} {Comparing
  charge transport predictions for a ternary electrolyte using the
  maxwell--stefan and nernst--planck equations},}\ }\href@noop {} {\bibfield
  {journal} {\bibinfo  {journal} {Journal of The Electrochemical Society}\
  }\textbf {\bibinfo {volume} {158}},\ \bibinfo {pages} {A33--A42} (\bibinfo
  {year} {2011})}\BibitemShut {NoStop}%
\bibitem [{\citenamefont {Balu}\ and\ \citenamefont
  {Khair}(2018)}]{balu2018role}%
  \BibitemOpen
  \bibfield  {author} {\bibinfo {author} {\bibfnamefont {B.}~\bibnamefont
  {Balu}}\ and\ \bibinfo {author} {\bibfnamefont {A.~S.}\ \bibnamefont
  {Khair}},\ }\bibfield  {title} {\enquote {\bibinfo {title} {Role of
  stefan--maxwell fluxes in the dynamics of concentrated electrolytes},}\
  }\href@noop {} {\bibfield  {journal} {\bibinfo  {journal} {Soft Matter}\
  }\textbf {\bibinfo {volume} {14}},\ \bibinfo {pages} {8267--8275} (\bibinfo
  {year} {2018})}\BibitemShut {NoStop}%
\bibitem [{\citenamefont {Kraaijeveld}\ and\ \citenamefont
  {Wesselingh}(1993)}]{kraaijeveld1993negative}%
  \BibitemOpen
  \bibfield  {author} {\bibinfo {author} {\bibfnamefont {G.}~\bibnamefont
  {Kraaijeveld}}\ and\ \bibinfo {author} {\bibfnamefont {J.~A.}\ \bibnamefont
  {Wesselingh}},\ }\bibfield  {title} {\enquote {\bibinfo {title} {Negative
  maxwell-stefan diffusion coefficients},}\ }\href@noop {} {\bibfield
  {journal} {\bibinfo  {journal} {Industrial \& engineering chemistry
  research}\ }\textbf {\bibinfo {volume} {32}},\ \bibinfo {pages} {738--742}
  (\bibinfo {year} {1993})}\BibitemShut {NoStop}%
\bibitem [{\citenamefont {Wesselingh}, \citenamefont {Vonk},\ and\
  \citenamefont {Kraaijeveld}(1995)}]{wesselingh1995exploring}%
  \BibitemOpen
  \bibfield  {author} {\bibinfo {author} {\bibfnamefont {J.}~\bibnamefont
  {Wesselingh}}, \bibinfo {author} {\bibfnamefont {P.}~\bibnamefont {Vonk}}, \
  and\ \bibinfo {author} {\bibfnamefont {G.}~\bibnamefont {Kraaijeveld}},\
  }\bibfield  {title} {\enquote {\bibinfo {title} {Exploring the maxwell-stefan
  description of ion exchange},}\ }\href@noop {} {\bibfield  {journal}
  {\bibinfo  {journal} {The Chemical Engineering Journal and The Biochemical
  Engineering Journal}\ }\textbf {\bibinfo {volume} {57}},\ \bibinfo {pages}
  {75--89} (\bibinfo {year} {1995})}\BibitemShut {NoStop}%
\bibitem [{\citenamefont {LOBO}\ and\ \citenamefont
  {Quaresma}(1989)}]{lobo1989handbook}%
  \BibitemOpen
  \bibfield  {author} {\bibinfo {author} {\bibfnamefont {V.~M.}\ \bibnamefont
  {LOBO}}\ and\ \bibinfo {author} {\bibfnamefont {J.}~\bibnamefont
  {Quaresma}},\ }\bibfield  {title} {\enquote {\bibinfo {title} {Handbook of
  electrolyte solutions. b},}\ }\href@noop {} {\bibfield  {journal} {\bibinfo
  {journal} {Physical sciences data}\ }\textbf {\bibinfo {volume} {41}}
  (\bibinfo {year} {1989})}\BibitemShut {NoStop}%
\bibitem [{\citenamefont {Stoppa}, \citenamefont {Hunger},\ and\ \citenamefont
  {Buchner}(2009)}]{stoppa2009conductivities}%
  \BibitemOpen
  \bibfield  {author} {\bibinfo {author} {\bibfnamefont {A.}~\bibnamefont
  {Stoppa}}, \bibinfo {author} {\bibfnamefont {J.}~\bibnamefont {Hunger}}, \
  and\ \bibinfo {author} {\bibfnamefont {R.}~\bibnamefont {Buchner}},\
  }\bibfield  {title} {\enquote {\bibinfo {title} {Conductivities of binary
  mixtures of ionic liquids with polar solvents},}\ }\href@noop {} {\bibfield
  {journal} {\bibinfo  {journal} {Journal of Chemical \& Engineering Data}\
  }\textbf {\bibinfo {volume} {54}},\ \bibinfo {pages} {472--479} (\bibinfo
  {year} {2009})}\BibitemShut {NoStop}%
\bibitem [{\citenamefont {Li}\ \emph {et~al.}(2007)\citenamefont {Li},
  \citenamefont {Zhang}, \citenamefont {Han}, \citenamefont {Hu}, \citenamefont
  {Xie},\ and\ \citenamefont {Yang}}]{li2007effect}%
  \BibitemOpen
  \bibfield  {author} {\bibinfo {author} {\bibfnamefont {W.}~\bibnamefont
  {Li}}, \bibinfo {author} {\bibfnamefont {Z.}~\bibnamefont {Zhang}}, \bibinfo
  {author} {\bibfnamefont {B.}~\bibnamefont {Han}}, \bibinfo {author}
  {\bibfnamefont {S.}~\bibnamefont {Hu}}, \bibinfo {author} {\bibfnamefont
  {Y.}~\bibnamefont {Xie}}, \ and\ \bibinfo {author} {\bibfnamefont
  {G.}~\bibnamefont {Yang}},\ }\bibfield  {title} {\enquote {\bibinfo {title}
  {Effect of water and organic solvents on the ionic dissociation of ionic
  liquids},}\ }\href@noop {} {\bibfield  {journal} {\bibinfo  {journal} {The
  Journal of Physical Chemistry B}\ }\textbf {\bibinfo {volume} {111}},\
  \bibinfo {pages} {6452--6456} (\bibinfo {year} {2007})}\BibitemShut {NoStop}%
\bibitem [{\citenamefont {Chaban}\ \emph {et~al.}(2012)\citenamefont {Chaban},
  \citenamefont {Voroshylova}, \citenamefont {Kalugin},\ and\ \citenamefont
  {Prezhdo}}]{chaban2012acetonitrile}%
  \BibitemOpen
  \bibfield  {author} {\bibinfo {author} {\bibfnamefont {V.~V.}\ \bibnamefont
  {Chaban}}, \bibinfo {author} {\bibfnamefont {I.~V.}\ \bibnamefont
  {Voroshylova}}, \bibinfo {author} {\bibfnamefont {O.~N.}\ \bibnamefont
  {Kalugin}}, \ and\ \bibinfo {author} {\bibfnamefont {O.~V.}\ \bibnamefont
  {Prezhdo}},\ }\bibfield  {title} {\enquote {\bibinfo {title} {Acetonitrile
  boosts conductivity of imidazolium ionic liquids},}\ }\href@noop {}
  {\bibfield  {journal} {\bibinfo  {journal} {The Journal of Physical Chemistry
  B}\ }\textbf {\bibinfo {volume} {116}},\ \bibinfo {pages} {7719--7727}
  (\bibinfo {year} {2012})}\BibitemShut {NoStop}%
\bibitem [{\citenamefont {Makino}\ \emph {et~al.}(2008)\citenamefont {Makino},
  \citenamefont {Kishikawa}, \citenamefont {Mizoshiri}, \citenamefont
  {Takeda},\ and\ \citenamefont {Yao}}]{makino2008viscoelastic}%
  \BibitemOpen
  \bibfield  {author} {\bibinfo {author} {\bibfnamefont {W.}~\bibnamefont
  {Makino}}, \bibinfo {author} {\bibfnamefont {R.}~\bibnamefont {Kishikawa}},
  \bibinfo {author} {\bibfnamefont {M.}~\bibnamefont {Mizoshiri}}, \bibinfo
  {author} {\bibfnamefont {S.}~\bibnamefont {Takeda}}, \ and\ \bibinfo {author}
  {\bibfnamefont {M.}~\bibnamefont {Yao}},\ }\bibfield  {title} {\enquote
  {\bibinfo {title} {Viscoelastic properties of room temperature ionic
  liquids},}\ }\href@noop {} {\bibfield  {journal} {\bibinfo  {journal} {The
  Journal of chemical physics}\ }\textbf {\bibinfo {volume} {129}},\ \bibinfo
  {pages} {104510} (\bibinfo {year} {2008})}\BibitemShut {NoStop}%
\bibitem [{\citenamefont {Winter}\ and\ \citenamefont
  {Chambon}(1986)}]{winter1986analysis}%
  \BibitemOpen
  \bibfield  {author} {\bibinfo {author} {\bibfnamefont {H.~H.}\ \bibnamefont
  {Winter}}\ and\ \bibinfo {author} {\bibfnamefont {F.}~\bibnamefont
  {Chambon}},\ }\bibfield  {title} {\enquote {\bibinfo {title} {Analysis of
  linear viscoelasticity of a crosslinking polymer at the gel point},}\
  }\href@noop {} {\bibfield  {journal} {\bibinfo  {journal} {Journal of
  rheology}\ }\textbf {\bibinfo {volume} {30}},\ \bibinfo {pages} {367--382}
  (\bibinfo {year} {1986})}\BibitemShut {NoStop}%
\bibitem [{\citenamefont {Levy}, \citenamefont {McEldrew},\ and\ \citenamefont
  {Bazant}(2019)}]{levy2019spin}%
  \BibitemOpen
  \bibfield  {author} {\bibinfo {author} {\bibfnamefont {A.}~\bibnamefont
  {Levy}}, \bibinfo {author} {\bibfnamefont {M.}~\bibnamefont {McEldrew}}, \
  and\ \bibinfo {author} {\bibfnamefont {M.~Z.}\ \bibnamefont {Bazant}},\
  }\bibfield  {title} {\enquote {\bibinfo {title} {Spin-glass charge ordering
  in ionic liquids},}\ }\href@noop {} {\bibfield  {journal} {\bibinfo
  {journal} {Physical Review Materials}\ }\textbf {\bibinfo {volume} {3}},\
  \bibinfo {pages} {055606} (\bibinfo {year} {2019})}\BibitemShut {NoStop}%
\bibitem [{\citenamefont {Avni}, \citenamefont {Adar},\ and\ \citenamefont
  {Andelman}(2020)}]{avni2020charge}%
  \BibitemOpen
  \bibfield  {author} {\bibinfo {author} {\bibfnamefont {Y.}~\bibnamefont
  {Avni}}, \bibinfo {author} {\bibfnamefont {R.~M.}\ \bibnamefont {Adar}}, \
  and\ \bibinfo {author} {\bibfnamefont {D.}~\bibnamefont {Andelman}},\
  }\bibfield  {title} {\enquote {\bibinfo {title} {Charge oscillations in ionic
  liquids: A microscopic cluster model},}\ }\href@noop {} {\bibfield  {journal}
  {\bibinfo  {journal} {Physical Review E}\ }\textbf {\bibinfo {volume}
  {101}},\ \bibinfo {pages} {010601} (\bibinfo {year} {2020})}\BibitemShut
  {NoStop}%
\bibitem [{\citenamefont {McEldrew}\ \emph {et~al.}(2018)\citenamefont
  {McEldrew}, \citenamefont {Goodwin}, \citenamefont {Kornyshev},\ and\
  \citenamefont {Bazant}}]{mceldrew2018}%
  \BibitemOpen
  \bibfield  {author} {\bibinfo {author} {\bibfnamefont {M.}~\bibnamefont
  {McEldrew}}, \bibinfo {author} {\bibfnamefont {Z.~A.}\ \bibnamefont
  {Goodwin}}, \bibinfo {author} {\bibfnamefont {A.~A.}\ \bibnamefont
  {Kornyshev}}, \ and\ \bibinfo {author} {\bibfnamefont {M.~Z.}\ \bibnamefont
  {Bazant}},\ }\bibfield  {title} {\enquote {\bibinfo {title} {Theory of the
  double layer in water-in-salt electrolytes},}\ }\href@noop {} {\bibfield
  {journal} {\bibinfo  {journal} {The journal of physical chemistry letters}\
  }\textbf {\bibinfo {volume} {9}},\ \bibinfo {pages} {5840--5846} (\bibinfo
  {year} {2018})}\BibitemShut {NoStop}%
\bibitem [{\citenamefont {Bazant}\ \emph {et~al.}(2009)\citenamefont {Bazant},
  \citenamefont {Kilic}, \citenamefont {Storey},\ and\ \citenamefont
  {Ajdari}}]{Bazant2009a}%
  \BibitemOpen
  \bibfield  {author} {\bibinfo {author} {\bibfnamefont {M.~Z.}\ \bibnamefont
  {Bazant}}, \bibinfo {author} {\bibfnamefont {M.~S.}\ \bibnamefont {Kilic}},
  \bibinfo {author} {\bibfnamefont {B.}~\bibnamefont {Storey}}, \ and\ \bibinfo
  {author} {\bibfnamefont {A.}~\bibnamefont {Ajdari}},\ }\bibfield  {title}
  {\enquote {\bibinfo {title} {Towards an understanding of nonlinear
  electrokinetics at large voltages in concentrated solutions},}\ }\href@noop
  {} {\bibfield  {journal} {\bibinfo  {journal} {Advances in Colloid and
  Interface Science}\ }\textbf {\bibinfo {volume} {152}},\ \bibinfo {pages}
  {48--88} (\bibinfo {year} {2009})}\BibitemShut {NoStop}%
\bibitem [{\citenamefont {Storey}\ and\ \citenamefont
  {Bazant}(2012)}]{storey2012effects}%
  \BibitemOpen
  \bibfield  {author} {\bibinfo {author} {\bibfnamefont {B.~D.}\ \bibnamefont
  {Storey}}\ and\ \bibinfo {author} {\bibfnamefont {M.~Z.}\ \bibnamefont
  {Bazant}},\ }\bibfield  {title} {\enquote {\bibinfo {title} {Effects of
  electrostatic correlations on electrokinetic phenomena},}\ }\href@noop {}
  {\bibfield  {journal} {\bibinfo  {journal} {Physical Review E}\ }\textbf
  {\bibinfo {volume} {86}},\ \bibinfo {pages} {056303} (\bibinfo {year}
  {2012})}\BibitemShut {NoStop}%
\end{thebibliography}%

\end{document}